\documentclass[twocolumn]{aastex62}

\def\mic              {\hbox{$\mu\mathrm{m}$}}

\def\comm#1           {\texttt{(COMMENT: #1)}}

\usepackage{graphicx}
\usepackage{xspace}

\shorttitle{The Large Synoptic Survey Telescope}
\shortauthors{Ivezi\'{c}, Kahn, Tyson, Abel, Acosta, Allsman, Alonso, AlSayyad, Anderson, et al.}

\begin{document}

\title{LSST: from Science Drivers to Reference Design and Anticipated Data Products}

\AuthorCallLimit=999

\author[0000-0001-5250-2633]{\v{Z}eljko~Ivezi\'{c}}
\affiliation{University of Washington, Dept.\ of Astronomy, Box 351580, Seattle, WA 98195}

\author{Steven~M.~Kahn}
\affiliation{LSST Project Office, 950 N.\ Cherry Avenue, Tucson, AZ  85719}
\affiliation{Kavli Institute for Particle Astrophysics and Cosmology, SLAC National Accelerator Laboratory, Stanford University, Stanford, CA 94025}

\author[0000-0002-9242-8797]{J.~Anthony~Tyson}
\affiliation{Physics Department, University of California, One Shields Avenue, Davis, CA 95616}

\author{Bob~Abel}
\affiliation{Olympic College, 1600 Chester Ave., Bremerton, WA 98337-1699}

\author{Emily~Acosta}
\affiliation{LSST Project Office, 950 N.\ Cherry Avenue, Tucson, AZ  85719}

\author{Robyn~Allsman}
\affiliation{LSST Project Office, 950 N.\ Cherry Avenue, Tucson, AZ  85719}

\author{David~Alonso}
\affiliation{Department of Physics, University of Oxford, Denys Wilkinson Building, Keble Road, Oxford, OX1 3RH, UK}

\author{Yusra~AlSayyad}
\affiliation{Department of Astrophysical Sciences, Princeton University, Princeton, NJ 08544}

\author{Scott~F.~Anderson}
\affiliation{University of Washington, Dept.\ of Astronomy, Box 351580, Seattle, WA 98195}

\author{John~Andrew}
\affiliation{LSST Project Office, 950 N.\ Cherry Avenue, Tucson, AZ  85719}

% ...

\begin{abstract}
Major advances in our understanding of the
Universe frequently arise from dramatic improvements in our ability to accurately
measure astronomical quantities. Aided by rapid progress in information
technology, current sky surveys are changing the way we view and study the
Universe. Next-generation surveys will maintain this revolutionary
progress. We describe here the most ambitious survey currently planned in
the optical, the Large Synoptic Survey Telescope (LSST). A vast array
of science will be enabled by a single wide-deep-fast sky survey, and
LSST will have unique survey capability in the faint time domain. The LSST design is
driven by four main science themes: probing dark energy and dark matter,
taking an inventory of the Solar System, exploring the transient optical sky,
and mapping the Milky Way. LSST will be a large, wide-field ground-based system
designed to obtain repeated images covering the sky visible from
Cerro Pach\'{o}n in northern Chile. The telescope will have an 8.4 m
(6.5 m effective) primary mirror, a 9.6 deg$^2$ field of view, and a 3.2 Gigapixel
camera.  The standard observing sequence will consist of pairs of
15-second exposures in a given field, with two such visits in each
pointing in a given night to identify and constrain the orbits of
asteroids.  With these repeats, the LSST system is capable of imaging
about 10,000 square degrees of sky in a
single filter in three clear nights.   The typical 5$\sigma$
point-source depth in a single visit in $r$ will be $\sim 24.5$ (AB).
The system is designed to yield high image quality as well as superb astrometric
and photometric accuracy. The project is in the construction phase and will begin
regular survey operations by 2022. The survey area will
be contained within 30,000 deg$^2$ with $\delta<+34.5^\circ$, and will be imaged multiple times
in six bands, $ugrizy$, covering the wavelength range 320--1050 nm.
About 90\% of the observing time will be devoted to a deep-wide-fast survey mode
which will uniformly observe a 18,000 deg$^2$
region about 800 times (summed over all six bands) during the anticipated 10 years
of operations, and will yield a coadded map to $r\sim27.5$. These data will result in
databases including 20 billion galaxies and a similar number of stars, and will
serve the majority of the primary science programs. The remaining 10\% of the observing time
will be allocated to special projects such as a Very Deep and Fast time domain
survey, whose details are currently under discussion. We illustrate
how the LSST science drivers led to these choices of system
parameters, and describe the expected data products and their characteristics.
The goal is to make LSST data products  including a relational database of about 32
trillion observations of 40 billion objects available to the public and scientists around the
world -- everyone will be able to view and study a high-definition color movie of
the deep Universe.
\end{abstract}

\keywords{
  astronomical data bases: atlases, catalogs, surveys ---
  Solar System ---
  stars ---
  the Galaxy ---
  galaxies ---
  cosmology
}

\section{INTRODUCTION}

Major advances in our understanding of the Universe have historically arisen
from dramatic improvements in our ability to ``see''. We have developed
progressively larger telescopes over the past century, allowing us
to peer further into space, and further back in time. With the development of
advanced instrumentation -- imagers, spectrographs, and polarimeters -- we
have been able to parse radiation detected from distant sources over the
full electromagnetic spectrum in increasingly subtle ways.
These data have provided the detailed information needed to construct physical
models of planets, stars, galaxies, quasars, and larger structures, and to probe the
new physics of dark matter and dark energy.

Until recently, most astronomical investigations have focused on small samples
of cosmic sources or individual objects. This is because our largest telescope
facilities typically had rather small fields of view, and those with large
fields of view could not detect very faint sources. With all of our existing
telescope facilities, we have still surveyed only a small fraction of the
observable Universe (except when considering the most luminous quasars).

Over the past two decades, however, advances in technology have made it possible to
move beyond the traditional observational paradigm and to undertake large-scale
sky surveys. As vividly demonstrated by surveys such as the Sloan Digital Sky
Survey \citep[SDSS;][]{2000AJ....120.1579Y}, the Two Micron All Sky Survey \citep[2MASS;][]{2006AJ....131.1163S},
the Galaxy Evolution Explorer \citep[GALEX;][]{2005ApJ...619L...1M},
and Gaia \citep{2016A&A...595A...2G} to name but a few, sensitive and accurate
multi-color surveys over a large fraction of the sky enable an extremely broad range of
new scientific investigations. These projects, based on a synergy of advances in
telescope construction, detectors, and above all, information technology,
have dramatically impacted nearly all fields of astronomy
-- and several areas of fundamental physics. In addition, the world-wide attention
received by Sky in Google Earth\footnote{\url{https://www.google.com/sky/}}
\citep{2007arXiv0709.0752S}, the World Wide Telescope\footnote{\url{http://worldwidetelescope.org/home}},
and the hundreds of thousands of volunteers
classifying galaxies in the Galaxy Zoo project \citep{2011MNRAS.410..166L}
and its extensions demonstrate that the impact of sky surveys extends
far beyond fundamental science progress and reaches all of society.

Motivated by the evident scientific progress enabled by large sky surveys,
three nationally-endorsed reports by the U.S.\ National Academy of Sciences
\citep{NAP9839,NAP10079,NAP10432}
concluded that a dedicated ground-based wide-field imaging telescope with an effective aperture
of 6--8 meters ``is a high priority for planetary science, astronomy, and physics
over the next decade.'' The Large Synoptic Survey Telescope (LSST) described here is
such a system. Located on Cerro Pach\'on in northern Chile,
the LSST will be a large, wide-field ground-based telescope
designed to obtain multi-band images over a substantial fraction of the sky
every few nights. The survey will yield contiguous overlapping imaging of over
half the sky in six optical bands, with each sky location visited close to 1000 times over
10 years. The 2010 report ``New Worlds, New Horizons in Astronomy and Astrophysics''
by the NRC Committee for a Decadal Survey of Astronomy and
Astrophysics \citep{NAP12951}
ranked LSST as its top priority for large ground-based projects, and in May 2014 the National
Science Board approved the project for construction.
As of this writing, the LSST construction phase is close to the peak
of activity.
After initial tests with a commissioning camera and full commissioning
with the main camera, the
ten year sky survey is projected to begin in 2022.

The purpose of this paper is to provide an overall summary of the main LSST
science drivers and how they led to the current system design parameters
(\S~\ref{Sec:refdesign}), to describe the anticipated data products (\S~\ref{Sec:dataprod}),
and to provide a few examples of the science programs that LSST will enable
(\S~\ref{Sec:science}). The community involvement is discussed in \S~\ref{Sec:community},
and broad educational and societal impacts of the project in \S~\ref{Sec:impact}. Concluding
remarks are presented in \S~\ref{Sec:conclusions}. This publication will be maintained
at the arXiv.org site\footnote{\url{https://arxiv.org/abs/0805.2366}}, and will also
be available from the LSST website (\url{http://www.lsst.org}). The latest arXiv version of this paper
should be consulted and referenced for the most up-to-date information about the
LSST system.

\section{  FROM SCIENCE DRIVERS TO REFERENCE DESIGN}
\label{Sec:refdesign}

The most important characteristic that determines the speed at which a system can
survey a given sky area to a given flux limit (i.e., its depth) is its \'etendue
(or grasp), the product of its primary mirror area and the angular
area of its field of view (for a given set of observing conditions, such as
seeing and sky brightness).
The effective \'etendue for LSST will be greater than 300 m$^2$ deg$^2$, which
is more than an order of magnitude larger than that of any existing facility.
For example, the SDSS, with its 2.5-m telescope \citep{2006AJ....131.2332G} and a
camera with 30 imaging CCDs \citep{1998AJ....116.3040G}, has an effective \'etendue of
only 5.9 m$^2$ deg$^2$.

The range of scientific investigations which will be enabled by such a
dramatic improvement in survey capability is extremely broad. Guided by
the community-wide input assembled in the report of the Science Working Group of the
LSST in 2004 \citep{Document-26952}, the LSST is designed to
achieve goals set by four main science themes:

\begin{enumerate}
\item Probing Dark Energy and Dark Matter;
\item Taking an Inventory of the Solar System;
\item Exploring the Transient Optical Sky;
\item Mapping the Milky Way.
\end{enumerate}

Each of these four themes itself encompasses a variety of analyses, with
varying sensitivity to instrumental and system parameters. These themes
fully exercise the technical capabilities of the system, such as photometric
and astrometric accuracy and image quality. About 90\% of the observing time
will be devoted to a deep-wide-fast (main) survey mode. The working paradigm is that all
scientific investigations will utilize a common database constructed from an optimized
observing program (the main survey mode), such as that discussed in
\S~\ref{sec:baseline}.
Here we briefly describe these science goals and the most challenging requirements for the
telescope and instrument that are derived from those goals, which will
inform the overall system design decisions discussed below.
For a more detailed discussion, we refer the reader to the LSST Science Requirements
Document \citep{LPM-17}, the LSST Science Book
\citep[][hereafter SciBook]{2009arXiv0912.0201L},
and links to technical papers and presentations at
\url{https://www.lsst.org/scientists}.

\subsection{The Main Science Drivers }

The main science drivers are used to optimize various system parameters.
Ultimately, in this high-dimensional parameter space, there is a
manifold defined by the total project cost. The science
drivers must both justify this cost, as well as provide guidance
on how to optimize various parameters while staying within the cost envelope.

Here we summarize the dozen or so most important interlocking constraints on data
and system properties placed by the four main science themes:

\begin{enumerate}
\item  The depth of a single visit to a given field;
\item  Image quality;
\item  Photometric accuracy;
\item  Astrometric accuracy;
\item  Optimal exposure time;
\item  The filter complement;
\item  The distribution of revisit times (i.e., the cadence of observations),
                including the survey lifetime;
\item  The total number of visits to a given area of sky;
\item  The coadded survey depth;
\item  The distribution of visits on the sky, and the total sky coverage;
\item  The distribution of visits per filter; and
\item  Parameters characterizing data processing and data access
  (such as the maximum time allowed after each exposure to report
         transient sources, and the maximum allowed software
         contribution to measurement errors).
\end{enumerate}

We present a detailed discussion of how these science-driven data properties are
transformed to system parameters below.

\subsubsection{Probing Dark Energy and Dark Matter}
\label{sec:Dark_Energy}

Current models of cosmology require the existence of both dark matter and dark
energy to match observational constraints
\citep{2007ApJ...659...98R,2009ApJS..180..330K,2010MNRAS.401.2148P,2012arXiv1211.0310L,2015PNAS..11212249W}, and
references therein). Dark energy affects the cosmic history of both the Hubble expansion
and mass clustering. Distinguishing competing models for the physical
nature of dark energy, or alternative explanations involving
modifications of the General Theory of Relativity, will require
percent level measurements of both the cosmic expansion and the growth
of dark matter structure as a function of redshift.  Any given
cosmological probe is sensitive to, and thus constrains degenerate
combinations of, several cosmological and astrophysical/systematic parameters.  Therefore the most robust
cosmological constraints are the result of using interlocking combinations
of probes. The most powerful probes include weak gravitational lens cosmic shear (WL), galaxy clustering and baryon
acoustic oscillations (LSS), the mass function and clustering of clusters of galaxies,
time delays in lensed quasar and supernova systems (SL),
and photometry of type Ia supernovae (SN) -- all as functions of
redshift. Using the cosmic microwave background fluctuations as the normalization, the
combination of these probes can yield the needed precision to distinguish among models of dark
energy \citep[see e.g.,][and references therein]{2006JCAP...08..008Z}. The challenge is to turn this available precision into accuracy, by careful modeling and marginalization over a variety of systematic effects \citep[see e.g.,][]{2017MNRAS.470.2100K}.

Meanwhile, there are a number of astrophysical probes of the fundamental
properties of dark matter worth exploring, including, for example,
weak and strong lensing observations of the mass distribution in
galaxies and isolated and
merging clusters, in conjunction with dynamical and
X-ray observations \citep[see e.g.,][]{2012ApJ...747L..42D,
2013ApJ...765...24N, 2013MNRAS.430...81R}, the numbers and gamma-ray
emission from dwarf satellite galaxies (see e.g., \citealt{2014ApJ...795L..13H};
\citealt{2015ApJ...809L...4D}),  the subtle perturbations of stellar
streams in the Milky Way halo by dark matter substructure
\citep{2016MNRAS.456..602B}, and massive compact halo object
microlensing \citep{2001ApJ...550L.169A}.

Three of the primary Dark Energy probes, WL, LSS and SN,  provide unique and
independent constraints on the LSST system design (SciBook Ch.~11--15).

Weak lensing (WL) techniques can be used to map the distribution of
mass as a function of redshift and thereby trace the history of both
the expansion of the Universe and the growth of structure (e.g., \citealt{1999ApJ...514L..65H};
for recent reviews see \citealt{2015RPPh...78h6901K}; \citealt{2017arXiv171003235M}).  Measurements of cosmic shear as a function of
redshift allow determination of angular distances versus cosmic time,
providing multiple independent constraints on the nature of dark
energy.  These investigations require deep wide-area multi-color
imaging with stringent requirements on shear systematics in at least
two bands, and excellent photometry in at least five bands to measure
photometric redshifts (a requirement shared with LSS, and indeed all
extragalactic science drivers). The strongest constraints on the LSST
image quality arise from this science program. In order to control
systematic errors in shear measurement, the desired depth must be
achieved with many short exposures (allowing for systematics in the
measurement of galaxy shapes related
to the PSF and telescope pointing to be diagnosed and removed). Detailed simulations of
weak lensing techniques show that imaging over $\sim20,000$ deg$^2$ to
a 5$\sigma$ point-source depth of $r_{AB} \sim 27.5$ gives adequate
signal to measure shapes for of order 2 billion galaxies for weak
lensing.  These numbers are adequate to reach
Stage IV goals for dark energy, as defined by the Dark Energy Task
Force \citep{2006astro.ph..9591A}.
This
depth, and the corresponding deep surface brightness limit,
optimize the number of galaxies with measured shapes in ground-based
seeing, and allow their detection in significant numbers to beyond a
redshift of two.  Analyzing these data will
require sophisticated data processing techniques.  For example, rather
than simply coadding all images in a given region of sky, the
individual exposures, each with their own PSF and noise
characteristics,  should be analyzed simultaneously to optimally
measure the shapes of galaxies \citep{2008ASPC..394..107T,2011PASP..123..596J}.

Type Ia supernovae provided the first robust evidence that the expansion of the
Universe is accelerating \citep{1998AJ....116.1009R,1999ApJ...517..565P}. To fully
exploit the supernova science potential, light curves sampled in multiple
bands every few days over the course of a few months are required. This is
essential to search for systematic differences in supernova populations
(e.g., due to differing progenitor channels) which
may masquerade as cosmological effects, as well as to determine photometric
redshifts from the supernovae themselves. Unlike other cosmological probes,
even a single object gives information on the relationship between
redshift and distance.  Thus a large
number of SN across the sky allows one to search for any dependence
of dark energy properties on direction, which
would be an indicator of new physics. The results from this method can be compared
with similar measures of anisotropy from the combination of WL and LSS
\citep{2009ApJ...690..923Z}.
Given the expected SN flux distribution
at the redshifts where dark energy is important, the
single visit depth should be at least $r\sim24$. Good image quality is
required to separate SN photometrically from
their host galaxies. Observations in at least five photometric bands will allow
proper K-corrected light curves to be measured over a range of
redshift.  Carrying out these K-corrections requires that the
calibration of the relative offsets in photometric zero points between filters and
the system response functions, especially near the edges of
bandpasses, be accurate to about 1\% \citep{2007ApJ...666..694W},
similar to the requirements from photometric redshifts of galaxies. Deeper data
($r>26$) for small areas of the sky can extend the discovery of SN to a mean
redshift of 0.7 (from $\sim0.5$ for the main survey), with some objects beyond $z\sim$1
\citep[][SciBook Ch.~11]{2004AAS...20510818G,2004AAS...20510820P}. The added statistical leverage
on the ``pre-acceleration'' era ($z\ga1$) would improve constraints on the properties of
dark energy as a function of redshift.

Finally, there will be powerful cross checks and complementarities with other planned or
proposed surveys, such as Euclid \citep{2011arXiv1110.3193L} and WFIRST
\citep{2015arXiv150303757S}, which will provide
wide-field optical-IR imaging from space;
DESI \citep{2013arXiv1308.0847L}
and PFS \citep{2014PASJ...66R...1T}, which will measure
spectroscopic BAO with
millions of galaxies; and SKA\footnote{\url{https://www.skatelescope.org}} (radio).
Large survey volumes are key to probing dynamical dark energy models (with sub-horizon
dark energy clustering or anisotropic stresses). The cross-correlation
of the three-dimensional
mass distribution -- as probed by neutral hydrogen in CHIME \citep{2014SPIE.9145E..4VN}, HIRAX \citep{2016SPIE.9906E..5XN}
or SKA, or galaxies in DESI and PFS -- with the gravitational growth
probed by tomographic shear in LSST will be a complementary way to constrain dark energy
properties beyond simply characterizing its equation of state and to test the underlying theory of gravity.
Current and future ground-based CMB experiments, such as Advanced ACT \citep{2016SPIE.9910E..14D},
SPT-3G \citep{2014SPIE.9153E..1PB}, Simons Observatory, and CMB Stage-4 \citep{2016arXiv161002743A}, will also offer invaluable opportunities for
cross-correlations with secondary CMB anisotropies.

\subsubsection{Taking an Inventory of the Solar System}

The small-body populations in the Solar System, such as asteroids, trans-Neptunian objects (TNOs)
and comets, are remnants of its early assembly. The history of accretion, collisional grinding, and
perturbation by existing and vanished giant planets is preserved in the orbital elements and size
distributions of those objects. Cataloging the orbital parameters, size distributions, colors and light
curves of these small-body populations requires a large number of observations in multiple filters,
and will lead to insights into planetary formation and evolution by providing the basis and constraints
for new theoretical models. In addition, collisions in the main asteroid belt between Mars and Jupiter
still occur, and occasionally eject objects on orbits that may place them on a collision course with Earth.
Studying the properties of main belt asteroids at sub-kilometer sizes is important for linking the near-Earth
Object (NEO) population with its source in the main belt. About 20\% of NEOs, the potentially hazardous
asteroids (PHAs), are in orbits that pass sufficiently close to Earth's orbit, to within 0.05 AU, that perturbations
on time scales of a century can lead to the possibility of collision. In December 2005,
the U.S.\ Congress directed\footnote{For details see \url{http://neo.jpl.nasa.gov/neo/report2007.html}} NASA to
implement a survey that would catalog 90\% of NEOs with diameters larger than 140 meters by 2020.

Discovering and linking objects in the Solar System moving with a wide range of apparent velocities (from
several degrees per day for NEOs to a few arc seconds per day for the most distant TNOs) places strong
constraints on the cadence of observations, requiring closely spaced pairs of observations (two or preferably
three times per lunation) in order to link detections unambiguously and derive orbits (SciBook Ch.~5). Individual
exposures should be shorter than about 30 seconds to minimize the effects of trailing for the majority of
moving objects. The images must be well sampled to enable accurate astrometry, with absolute accuracy of at
least 0.1 arcsec in order to measure orbital parameters of TNOs with enough precision to constrain theoretical
models and enable prediction of occultations. The photometry should be
better than 1--2\% to measure asteroids' colors and thus determine
their types.  The different filters
should be observed over a short time span to reduce apparent
variations in color due to changes in observing geometry, but should
be repeated over many lunations in order to determine phase curves and allow shape modeling.

The Congressional mandate can be fulfilled with a 10-meter-class
telescope equipped with a multi-gigapixel camera, and a sophisticated
and robust data processing system \citep{2007IAUS..236..353I}. The images should reach a depth of at
least 24.5 (5$\sigma$ for point sources) in the $r$ band to reach high
completeness down to the 140 m mandate for NEOs.  Such an instrument
would probe the $\sim$100 m size range at main-belt distances, and
discover rare distant TNOs such as Sedna \citep{2004ApJ...617..645B}
and 2012 VP113 \citep{2014Natur.507..471T}.

\subsubsection{Exploring the Transient Optical Sky}

Recent surveys have shown the power of measuring variability of
celestial sources for
studying gravitational lensing, searching for supernovae, determining
the physical properties of gamma-ray burst sources, discovering
gravitational wave counterparts, probing the structure of active
galactic nuclei, studying variable star populations, discovering
exoplanets, and many other subjects at the forefront of astrophysics
\citep[SciBook Ch.~8;][]{2009PASP..121.1395L,2012IAUS..285..141D,2014ApJ...784...45R}.

Time-domain science has diverse requirements for transient and
variable phenomena that are physically and phenomenologically
heterogeneous.  It requires large area coverage to enhance the probability
of detecting rare events; good image quality to enable differencing of
images, especially in crowded fields; good time sampling, necessary to
distinguish different types of variables and to infer
their properties (e.g., determining the intrinsic peak luminosity of Type Ia
supernovae requires measuring their light curve shape); accurate
color information to classify variable objects;
long term
persistent observations to characterize slow-evolving transients
(e.g., tidal disruption events, super luminous supernovae at high
redshift, and luminous blue variables); and rapid data reduction,
classification, and reporting to the community to allow immediate
follow-up with spectroscopy, further optical photometry, and imaging in other
wavebands.

Wide area, dense temporal coverage to deep limiting magnitudes will
enable the discovery and analysis of rare and exotic objects such as
neutron stars and black hole binaries, novae and stellar flares,
gamma-ray bursts and X-ray flashes, active galactic nuclei, stellar
disruptions by black holes \citep{2011Sci...333..203B,2012Natur.485..217G},
and possibly new classes of transients, such as binary mergers of
supermassive black holes \citep{2008ApJ...682..758S},
chaotic
eruptions on stellar surfaces \citep{2011ApJ...741...33A}, and, further yet, completely
unexpected phenomena.

Such a survey would likely detect microlensing by stars and compact objects in
the Milky Way, but also in the Local Group and perhaps beyond \citep{2008A&A...478..755D}.
Given the duration of the LSST it will also be possible
to detect the parallax microlensing signal of intermediate mass black holes and
measure their masses \citep{1992ApJ...392..442G}. It would open the possibility of
discovering populations of binaries and planets via transits
\citep[e.g.,][]{2006Natur.439..437B,2010arXiv1009.3048D,2013ApJ...768..129C,2014ApJ...780...54B},
as well as obtaining spectra of lensed stars in distant galaxies.

A deep and persistent survey will discover precursors of explosive and
eruptive transients, generate large samples of transients whose study
has thus far been limited by small sample size (e.g., different
subtypes of core collapse SN, \citealt{2014ApJS..213...19B}.)

Time series ranging between one minute and ten years cadence should be
probed over a significant fraction of the sky. The survey's cadence
will be sufficient, combined with the large coverage, to
serendipitously catch very short-lived events, such as eclipses in
ultra-compact double degenerate binary systems \citep{2005AJ....130.2230A},
to constrain the properties of fast faint transients (such as optical
flashes associated with gamma-ray bursts; \citealt{2008AN....329..284B}), to
detect electromagnetic counterparts to gravitational wave sources
\citep{2013ApJ...767..124N,2018ApJ...852L...3S} and to further constrain the
properties of new classes of transients discovered by programs such as
the Deep Lens Survey \citep{2004ApJ...611..418B}, the Catalina Real-time
Transient Survey \citep{2009ApJ...696..870D}, the Palomar Transient Factory
\citep{2009PASP..121.1395L}, and the Zwicky Transient Factory \citep{2014htu..conf...27B}. Observations
over a decade will enable the study of long period variables, intermediate mass
black holes, and quasars \citep{2007ApJ...659..997K,2010ApJ...721.1014M,2014MNRAS.439..703G,2016JCAP...11..042C}.

The next frontier
in this field will require measuring the colors of fast transients,
and probing variability at faint magnitudes. Classification of transients in
close-to-real time will require access to the full photometric history
of the objects, both before and after the transient event
\citep[e.g.,][]{2011BASI...39..387M}.

\subsubsection{Mapping the Milky Way}

A major challenge in extragalactic cosmology today concerns the formation of structure on sub-galactic scales, where
baryon physics becomes important, and the nature of dark matter may manifest itself in observable ways \citep[e.g.][]{2015PNAS..11212249W}.
The Milky Way and its environment provide a unique dataset for understanding the detailed processes that
shape galaxy formation and for testing the small-scale predictions of
our standard cosmological model. New insights into the nature and
evolution of the Milky Way will require wide-field surveys to constrain
its structure and accretion history.  Further insights into the stellar
populations that make up the Milky Way can be gained with a comprehensive census of the stars
within a few hundred pc of the Sun.

Mapping the Galaxy requires large area coverage, excellent image
quality to maximize photometric and astrometric accuracy,
especially in crowded fields, photometric precision of at least 1\% to
separate main sequence and
giant stars \cite[e.g.,][]{2003ApJ...586..195H} as well as to identify variable
stars such as RR Lyrae \citep{2010ApJ...708..717S,2011ApJ...728..106S},
and astrometric precision of about 10 mas per observation to enable parallax and proper motion measurements
(SciBook Ch.~6,7). In order to probe the halo out to its presumed edge at $\sim100$ kpc \citep{2004ASPC..327..104I}
with main-sequence stars, the total coadded depth must reach $r > 27$, with a similar depth in the $g$ band.
The metallicity distribution of stars can be studied photometrically in the Sgr tidal stream
\cite[e.g., see][]{2003ApJ...599.1082M,2007ApJ...670..346C} and other halo substructures
($\sim 30$ kpc, \citealt{2007Natur.450.1020C}), yielding new insights into how
they formed.  Our ability to measure these metallicities is limited by
the coadded depth in the $u$ band; to probe the outer parts of the
stellar halo, one must reach
$u\sim24.5$. To detect RR Lyrae stars beyond the Galaxy's tidal radius at $\sim 300$ kpc, the single-visit depth must
be $r \sim  24.5$.

In order to measure the tangential velocity of stars at a distance of 10 kpc, where the halo dominates over the disk, to
within 10 km s$^{-1}$ (comparable with the accuracy of
large-scale radial velocity surveys), the proper motion
accuracy should be 0.2 mas yr$^{-1}$ or better. This is the same accuracy as will be delivered by the Gaia mission\footnote{\url{http://sci.esa.int/gaia/}} \citep{2001A&A...369..339P,2012Ap&SS.341...31D} at its faint limit ($r \sim 20$).
In order to measure distances to solar neighborhood stars out to a distance of 300 pc (the thin disk scale height),
with geometric distance accuracy of at least 30\%, trigonometric parallax measurements accurate to 1 mas ($1\sigma$)
are required over 10 years. To achieve the required proper motion and parallax accuracy with an assumed astrometric
accuracy of 10 mas per observation per coordinate, approximately 1,000
separate observations are required. This requirement for a large
number of observations is similar to that from minimizing
systematics in weak lensing observations (\S~\ref{sec:Dark_Energy}).

\begin{figure}
\includegraphics[width=1.0\hsize,clip]{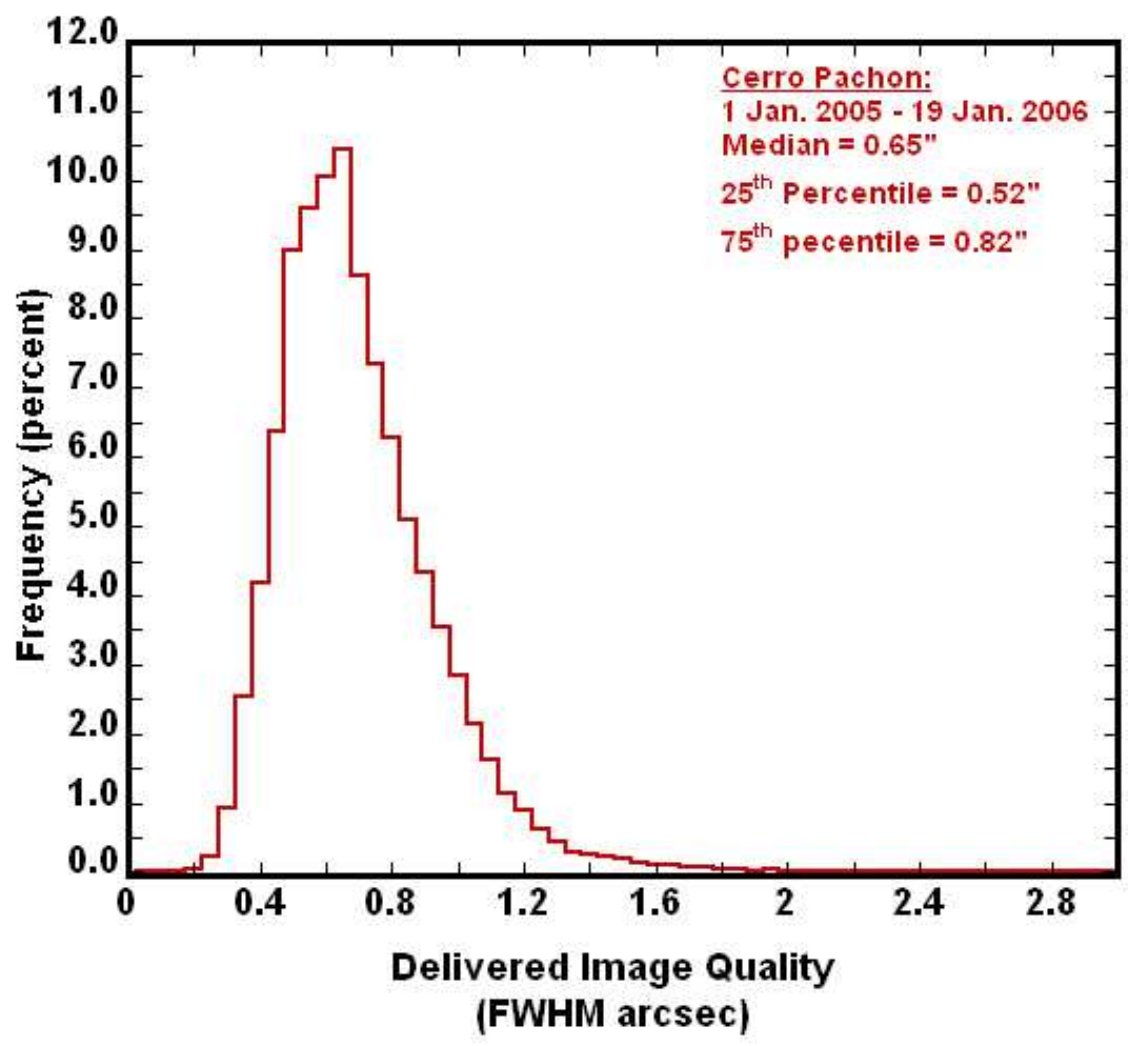}
\caption{
The image quality distribution measured at the Cerro Pach\'{o}n site using
a differential image motion monitor (DIMM) at $\lambda$ = 500 nm, and corrected
using an outer scale parameter of 30 m over an 8.4 m aperture. For details
about the outer scale correction see \citet{2002PASP..114.1156T}. The observed distribution
is well described by a log-normal distribution, with the parameters shown in
the figure.}
\label{Fig:seeing}
\end{figure}

\subsubsection{A Summary and Synthesis of Science-driven Constraints on Data Properties}

The goals of all the science programs discussed above
(and many more, of course) can be accomplished by satisfying the
minimal constraints listed below. For a more elaborate listing
of various constraints, including detailed specification of
various probability density distribution functions, please see the LSST Science
Requirements Document \citep{LPM-17}
and the LSST Science Book \citep{2009arXiv0912.0201L}.

\begin{enumerate}
\item  \textit{The single visit depth} should reach $r\sim24.5$. This limit is
   primarily driven by the search for NEOs, variable sources (e.g., SN,
   RR Lyrae stars), and by proper motion and trigonometric parallax
   measurements for stars. Indirectly, it is also driven by the
   requirements on the coadded survey depth and the minimum number of
   exposures required by WL science.  We plan to split a single visit
   into two exposures of equal length to identify and remove cosmic
   rays.
\item  \textit{Image quality} should maintain the limit set by the
     atmosphere (the median free-air seeing is 0.65 arcsec in the $r$ band
     at the chosen site, see Fig.~\ref{Fig:seeing}),
     and not be degraded appreciably by the hardware. In addition to stringent
     constraints from weak lensing, good image quality is driven by the
     required survey depth for point sources and by image differencing
     techniques.
\item  \textit{Photometric repeatability} should achieve 5 mmag precision
     at the bright end, with zeropoint stability across the sky of 10 mmag
     and band-to-band calibration errors not larger than 5 mmag.
     These requirements are driven by the need for high photometric redshift accuracy,
     the separation of stellar populations, detection of low-amplitude variable
     objects (such as eclipsing planetary systems), and the search for
     systematic effects in type Ia supernova light curves.
\item  \textit{Astrometric precision} should maintain the limit set by
     the atmosphere, of about 10 mas per visit at the bright end
     (on scales below 20 arcmin). This precision is driven by the desire to
     achieve a proper motion accuracy of 0.2 mas yr$^{-1}$ and parallax accuracy of
     1.0 mas over the course of a 10-year survey (see \S~\ref{sec:astrom}).
\item  \textit{The single visit exposure time}
should be less than about a minute
    to prevent trailing of fast moving objects and to aid control
    of various systematic effects induced by the atmosphere. It should
    be longer than $\sim$20 seconds to avoid significant efficiency losses due to
    finite readout, slew time, and read noise.  As described above, we
    are planning to split each visit into two exposures.
\item  \textit{The filter complement} should include at least six filters
    in the wavelength range limited by atmospheric absorption and
    silicon detection efficiency (320--1050 nm), with roughly
    rectangular filters and no large gaps in the coverage, in order
    to enable robust and accurate photometric redshifts and stellar typing. An
    SDSS-like $u$ band \citep{1996AJ....111.1748F} is extremely important for separating
    low-redshift quasars from hot stars, and for estimating the metallicities of
    F/G main sequence stars. A bandpass with an effective wavelength of
    about 1 micron  would enable studies of sub-stellar objects, high-redshift
    quasars (to redshifts of $\sim$7.5), and regions of the Galaxy that are obscured
    by interstellar dust.
\item  \textit{The revisit time distribution} should enable determination of
   orbits of Solar System objects and sample SN light curves every few days,
   while accommodating constraints set by proper motion and trigonometric
   parallax measurements.
\item  \textit{The total number of visits} of any given area of sky, when accounting for all
   filters, should be of the order of 1,000, as mandated by WL
   science, the search for NEOs, and proper motion and
   trigonometric parallax measurements. Studies of transient sources
   also benefit from a large number of visits.
\item  \textit{The coadded survey depth} should reach
    $r\sim27.5$, with sufficient signal-to-noise ratio in other bands
    to address both extragalactic and Galactic science drivers.
\item  \textit{The distribution of visits per filter} should enable
   accurate photometric redshifts, separation of stellar populations,
   and sufficient depth to enable detection of faint extremely red
   sources (e.g., brown dwarfs and high-redshift quasars). Detailed simulations of
   photometric redshift uncertainties
   suggest roughly similar number of visits among bandpasses
   (but because the system throughput and atmospheric properties are
    wavelength dependent, the achieved depths are different in different
    bands). The adopted time allocation
   (see Table~\ref{tab:baseline}) includes a slight preference to the $r$ and $i$ bands because of their
   dominant role in star/galaxy separation and weak lensing measurements.
\item  \textit{The distribution of visits on the sky} should extend over
   at least $\sim$18,000 deg$^2$ to obtain the required number of galaxies
   for WL studies, with attention paid to include ``special''
   regions such as the Ecliptic and Galactic planes, and the Large and Small
   Magellanic Clouds (if in the Southern Hemisphere).  For comparison,
   the full area that can be observed at airmass less than 2.0 from
   any mid-latitude site is about 30,000 deg$^2$.
\item  \textit{Data processing, data products and data access} should
  result in data products that approach the statistical uncertainties
  in the raw data; i.e., the processing must be close to optimal.
To enable fast and efficient response to
   transient sources, the processing latency for variable sources should be less than a minute,
   with a robust and accurate preliminary characterization
  
   of all reported variables.
\end{enumerate}

Remarkably, even with these joint requirements, none of the
individual science programs is severely over-designed, i.e., despite
their significant scientific diversity, these programs are highly
compatible in terms of desired data characteristics. Indeed, any one
of the four main science drivers could be removed, and the remaining
three would still yield very similar requirements for most system
parameters. As a result, the LSST system can adopt a highly
efficient survey strategy in which \textit{a single dataset serves most science
programs} (instead of science-specific surveys executed in series).
One can view this project as \textit{massively parallel astrophysics}.
The vast majority (about 90\%) of the observing time will be devoted to
a deep-wide-fast survey mode of the sort we have just described, with
the remaining 10\%
allocated to special programs which will also address multiple science
goals. Before describing these surveys in detail, we discuss the main
system parameters.

\begin{deluxetable}{l|l}[t]
\tablecaption{The LSST Baseline Design and Survey Parameters\label{tab:baseline}}
\tablehead{
\colhead{Quantity} & \colhead{Baseline Design Specification}
}
\startdata
Optical Config.                           &  3-mirror modified Paul-Baker        \\
Mount Config.                            &  Alt-azimuth          \\
Final f-ratio, aperture                 &  f/1.234, 8.4 m                \\
Field of view, \'etendue              &  9.6 deg$^2$,   319 m$^2$deg$^2$     \\
Plate Scale                                  &  50.9 $\mu$m/arcsec (0.2'' pix)  \\
Pixel count                                  &  3.2 Gigapix  \\
Wavelength Coverage                   &  320 -- 1050 nm, $ugrizy$             \\
Single visit depths, design\tablenotemark{a}  &  23.9, 25.0, 24.7, 24.0, 23.3, 22.1    \\
Single visit depths, min.\tablenotemark{b}       &  23.4, 24.6, 24.3, 23.6, 22.9, 21.7    \\
Mean number of visits\tablenotemark{c}          &  56, 80, 184, 184, 160, 160               \\
Final (coadded) depths\tablenotemark{d}         &  26.1, 27.4, 27.5, 26.8, 26.1, 24.9     \\
\enddata
\tablenotetext{a}{Design specification from the Science Requirements Document \citep[SRD;][]{LPM-17} for 5$\sigma$ depths
for point sources in the $ugrizy$ bands, respectively. The listed
values are expressed on the AB magnitude
scale, and correspond to point sources and fiducial zenith observations (about 0.2 mag loss of depth
is expected for realistic airmass distributions, see Table~\ref{tab:eqparams} for more details).}
\tablenotetext{b}{Minimum specification from the Science Requirements Document for 5$\sigma$ depths.}
\tablenotetext{c}{An illustration of the distribution of the number of visits as a function of bandpass,
taken from Table 24 in the SRD.}
\tablenotetext{d}{Idealized depth of coadded images, based on design specification for 5$\sigma$ depth and
the number of visits in the penultimate row (taken from Table 24 in the SRD).}
\end{deluxetable}

\subsection{The Main System Design Parameters}

Given the minimum science-driven constraints on the data properties listed
in the previous section, we now discuss how they are translated into
constraints on the main system design parameters: the aperture size,
the survey lifetime, the optimal exposure time, and the filter complement.

\subsubsection{The Aperture Size }
\label{Sec:apSize}
The product of the system's \'etendue and the survey lifetime, for given
observing conditions, determines
the sky area that can be surveyed to a given depth.
The
LSST field-of-view area is maximized to its practical limit, $\sim$10 deg$^2$,
determined by the requirement that the delivered image quality be dominated
by atmospheric seeing at the chosen site (Cerro Pach\'{o}n in Northern Chile).
A larger field-of-view would lead to unacceptable deterioration of the
image quality. This constraint leaves the primary mirror diameter and survey lifetime
as free parameters. The adopted survey lifetime of 10 years is a compromise
between a shorter time that leads to an excessively large and expensive mirror (15 m for a
3 year survey and 12 m for a 5 year survey) and not as effective proper motion
measurements, and a smaller telescope that would require more time to complete the
survey, with the associated increase in operations cost.

The primary mirror size is a function of the required survey depth and the
desired sky coverage. By and large, the anticipated science outcome scales
with the number of detected sources. For practically all astronomical source
populations, in order to maximize the number of detected sources, it is more
advantageous to maximize the area first, and then
the detection depth\footnote{
If the total exposure time is doubled and used to double the survey area,
the number of sources increases by a factor of two. If the survey
area is kept fixed, the increased exposure time will result in
$\sim$0.4 mag deeper data (see eq.~\ref{m5}). For cumulative source
counts described by $\log(N) = C + k*m$, the number of sources
will increase by more than a factor of two only if $k>0.75$.
Apart from $z<2$ quasars, practically all populations
have $k$ at most 0.6 (the Euclidean value), and faint stars
and galaxies have $k<0.5$. For more details, please see \citet{2003AJ....125.2740N}.}.
For this reason, the sky area for the main survey is
maximized to its practical limit, 18,000 deg$^2$, determined by the
requirement to avoid airmasses less than 1.5,
which would substantially
deteriorate the image quality and the survey depth (see eq.~\ref{m5}).

With the adopted field-of-view area, the sky coverage and the survey lifetime
fixed, the primary mirror diameter is fully driven by the required survey
depth. There are two depth requirements: the final (coadded) survey depth,
$r\sim27.5$, and the depth of a single visit, $r\sim24.5$. The two
requirements are compatible if the number of visits is several hundred
per band, which is in good agreement with independent science-driven
requirements on the latter.

\begin{figure}[t]
\includegraphics[width=\hsize,clip]{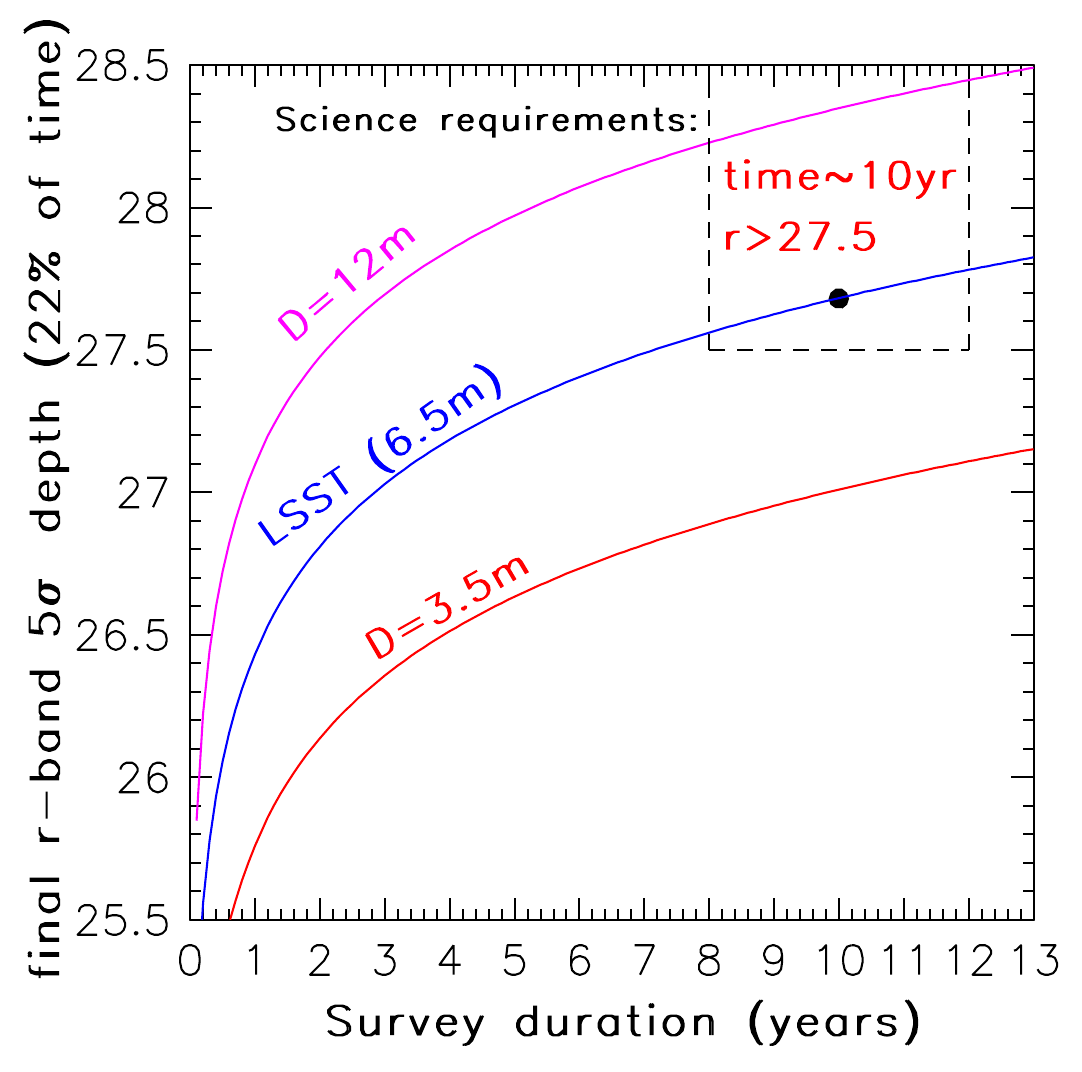}
\caption{The coadded depth in the $r$ band (AB magnitudes) vs. the effective aperture and
the survey lifetime. It is assumed that 22\% of the total observing time (corrected for
weather and other losses) is allocated for the $r$ band, and that the ratio of
the surveyed sky area to the field-of-view area is 2,000.}
\label{Fig:coaddDepth}
\end{figure}

The required coadded survey depth provides a direct constraint,
independent of the details of survey execution such as the exposure time per visit,
on the minimum effective primary mirror diameter of 6.4 m, as illustrated in
Fig.~\ref{Fig:coaddDepth}.

\subsubsection{The Optimal Exposure Time }

The single visit depth depends on both the primary mirror diameter and the
chosen exposure time, $t_\mathrm{vis}$. In turn, the exposure time
determines the time interval to revisit a given sky position and the total
number of visits, and each of these quantities has its own science
drivers. We summarize these simultaneous constraints in terms of the
single-visit exposure time:
\begin{itemize}
\item  The single-visit exposure time should not be longer than about a minute to
         prevent trailing of fast Solar System moving objects, and to enable efficient
         control of atmospheric systematics.
\item  The mean revisit time (assuming uniform cadence) for a given position
         on the sky, $n$, scales as
\begin{equation}
  n = \left( {t_\mathrm{vis} \over 10  \, \mathrm{sec}} \right)
      \left( { A_\mathrm{sky} \over 10,000  \, \mathrm{deg}^2} \right)
      \left( {10 \, \mathrm{deg}^2 \over  A_\mathrm{FOV}} \right) \mathrm{days},
\end{equation}
where two visits per night are assumed (required for efficient detection of
Solar System objects, see below), and the losses for realistic observing conditions
have been taken into account (with the aid of the Operations Simulator described below).
Science drivers such as supernova light curves and moving objects in the Solar System require
that $n<4$ days, or equivalently $t_{vis} < 40$ seconds for the nominal values
of $A_{sky} $ and $A_{FOV}$.
\item  The number of visits to a given position on the sky, $N_{visit}$,
with losses for realistic observing conditions taken into account,
is given by
\begin{equation}
      N_{visit} = \left( {3000 \over n} \right)
                    \left( { T \over 10 \, \mathrm{yr}} \right).
\end{equation}
The requirement $N_{visit}>800$ again implies that $n<4$ and
$t_{vis} < 40$ seconds if the survey lifetime, $T$ is about 10 years.
\item  These three requirements place a firm upper limit on the
optimal visit exposure time of $t_{vis} < 40$ seconds. Surveying
efficiency (the ratio of open-shutter time to the total
time spent per visit) considerations place a lower limit on
$t_{vis}$ due to finite detector read-out and telescope slew time (the longest
acceptable read-out time is set to 2 seconds, the shutter open-and-close
time is 2 seconds, and the slew and settle time is set to 5 seconds, including
the read-out time for the second exposure in a visit):
\begin{equation}
      \epsilon = \left( {t_{vis} \over t_{vis} + 9 \, \mathrm{sec}}\right).
\end{equation}
To maintain efficiency losses below $\sim$30\% (i.e., at least below the
limit set by the weather patterns), and to minimize the read noise
impact, $t_{vis} > 20$ seconds is required.
\end{itemize}

\begin{figure}[t]
\includegraphics[width=1.0\hsize,clip]{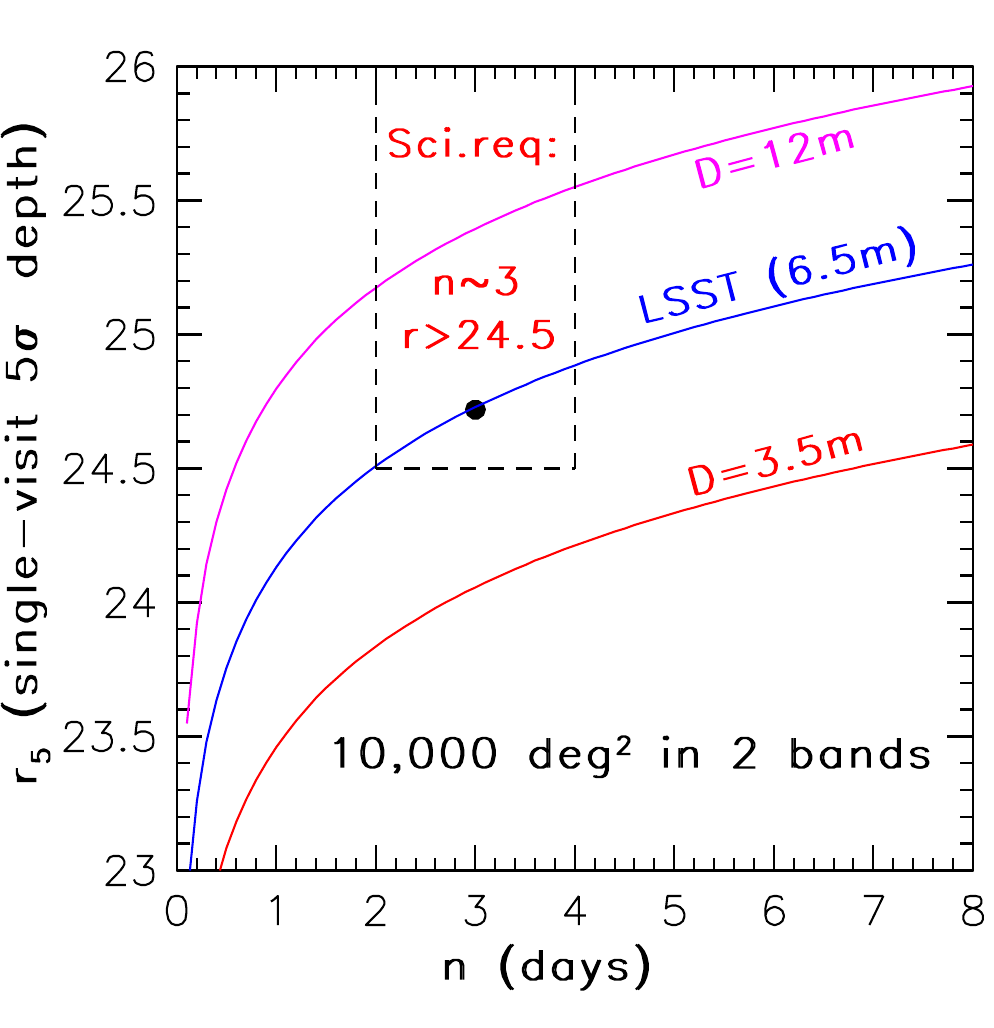}
\caption{The single-visit depth in the $r$ band (5$\sigma$ detection for
point sources, AB magnitudes) vs. revisit time, $n$ (days), as a function of
the effective aperture size. With a coverage of 10,000 deg$^2$ in two bands,
the revisit time directly constrains the visit exposure time, $t_{vis}=10\,n$
seconds. In addition to direct constraints on optimal exposure time, $t_{vis}$
is also driven by requirements on the revisit time, $n$, the total number of visits
per sky position over the survey lifetime, $N_{visit}$, and the survey efficiency,
$\epsilon$ (see eqs.1-3). Note that these constraints result in a fairly narrow range of
allowed $t_{vis}$ for the main deep-wide-fast survey.}
\label{Fig:singleDepth}
\end{figure}

Taking these constraints simultaneously into account, as summarized in
Fig.~\ref{Fig:singleDepth},
yielded the following reference design:
\begin{enumerate}
\item A primary mirror effective diameter of $\sim$6.5 m. With the adopted optical
design, described below, this effective diameter corresponds to a geometrical diameter
of $\sim$8 m. Motivated by characteristics of the existing equipment at the
Steward Mirror Laboratory, which fabricated the primary mirror, the adopted
geometrical diameter is set to 8.4 m.
\item A visit exposure time of 30 seconds (using two 15 second exposures
to efficiently reject cosmic rays; the possibility of a single exposure per visit,
to improve observing efficiency, will be investigated during the commissioning phase),
yielding $\epsilon=77$\%.
\item A revisit time of 3 days on average for 10,000 deg$^2$ of sky,
  with two visits per night.
\end{enumerate}

To summarize, the chosen primary mirror diameter is the \textit{minimum}
diameter that simultaneously satisfies the depth ($r\sim24.5$ for single visit and
$r\sim27.5$ for coadded depth) and cadence (revisit time of 3--4 days,
with 30 seconds per visit) constraints described above.

\subsection{System Design Trade-offs}

We note that the Pan-STARRS project \citep{2002SPIE.4836..154K,2010SPIE.7733E..0EK}, with similar science
goals as LSST, envisions a distributed aperture design, where the total
system \'etendue is
a sum of \'etendue values for an array of small 1.8 m telescopes\footnote{The
first of these telescopes, PS1, has been operational for some time \citep{2016arXiv161205560C}, and
has an \'etendue 1/24$^{th}$ that of LSST. }.
Similarly, the LSST system could perhaps be made as two smaller copies with
6m mirrors, or 4 copies with 4m mirrors, or 16 copies with 2m mirrors. Each
of these clones would have to have its own 3 Gigapixel camera (see below), and
given the added risk and complexity (e.g., maintenance, data processing), the monolithic
design seems advantageous for a system with such a large \'etendue as LSST.

It is informative to consider the tradeoffs that would be required
for a system with a smaller aperture, if the science requirements were
to be maintained. For this comparison, we consider a four-telescope version of
the Pan-STARRS survey (PS4). With an \'etendue about 6 times smaller
than that of LSST (effective diameters of 6.4 m and 3.0 m, and a field-of-view area
of 9.6 deg$^2$ vs. 7.2 deg$^2$), and all observing conditions being equal,
the PS4 system could in principle use a cadence identical to that of LSST. The
main difference in the datasets would be a faint limit shallower by about
1 mag in a given survey lifetime. As a result, for Euclidean populations the
sample sizes would go down by a factor of 4, while for populations of
objects with a shallower slope of the number-magnitude relation (e.g.,
galaxies around redshift of 1) the samples would be smaller by a factor 2--3.
The distance limits for nearby sources, such as Milky Way stars, would drop to
60\% of their corresponding LSST values, and the NEO completeness level mandated by
the U.S.\ Congress would not be reached.

If instead the survey coadded depth were to be maintained, then the survey sky
area would have to be 6 times smaller ($\sim$3,500 deg$^2$). If the
survey single-visit depth were to be maintained, then the exposure
time would have to be about 6 times longer (ignoring the slight difference
in the field-of-view area and simply scaling by the \'etendue ratio),
resulting in non-negligible trailing losses for Solar System objects,
and either
i) a factor of six smaller sky area observed within $n=3$ days, or
ii) the same sky area revisited every $n=18$ days.
Given these conflicts, one solution would be to split the observing time and
allocate it to individual specialized programs (e.g., large sky area vs.
deep coadded data vs. deep single-visit data vs. small $n$ data, etc.),
as is being done by the PS1 Consortium\footnote{More information about
Pan-STARRS is available from \url{http://pswww.ifa.hawaii.edu/pswww/}.}.

In summary,
\textit{given the science requirements as stated here, there is a
minimum \'etendue of $\sim$300 deg$^2$m$^2$ which enables our seemingly
disparate science goals to be addressed with a single dataset.}
A system with a smaller \'etendue would require separate specialized surveys
to address the science goals, which results in a loss of surveying
efficiency\footnote{The converse is also true: for every \'etendue
there is a set of optimal science goals that such a system can
address with a high efficiency.}. The LSST is designed to reach this
minimum \'etendue for the science goals stated in its Science Requirements
Document.

\subsection{  The Filter Complement }

\begin{figure}
\includegraphics[width=1.0\hsize,clip]{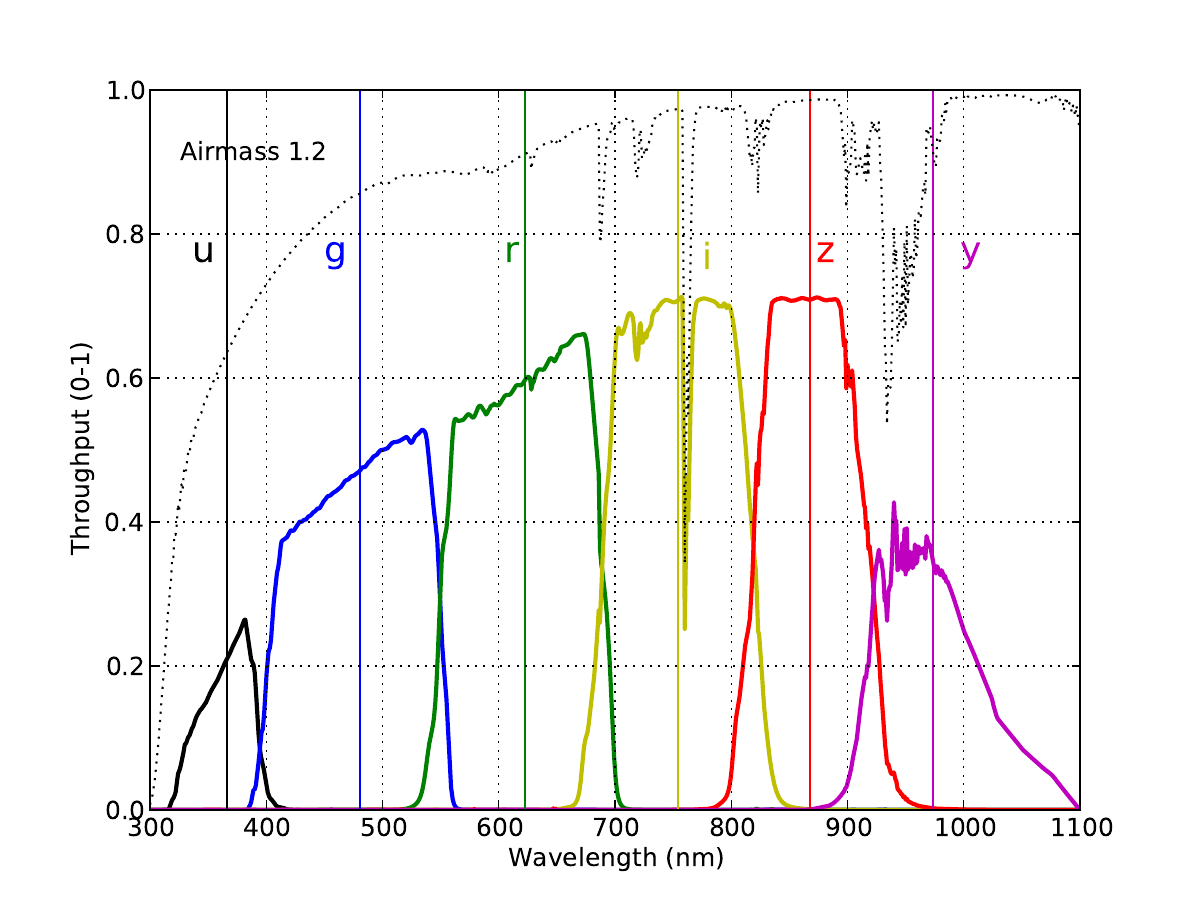}
\caption{The LSST bandpasses. The vertical axis shows the total throughput. The computation
includes the atmospheric transmission (assuming an airmass of 1.2,
dotted line), optics, and the detector sensitivity.}
\label{Fig:filters}
\end{figure}

The LSST filter complement ($ugrizy$, see Fig.~\ref{Fig:filters}) is modeled after the Sloan
Digital Sky Survey
(SDSS) system \citep{1996AJ....111.1748F} because of its demonstrated success in a wide
variety of applications, including photometric redshifts of galaxies \citep{2003ApJ...595...59B},
separation of stellar populations \citep{1998ApJS..119..121L,2003ApJ...586..195H},
and photometric selection of quasars \citep{2002AJ....123.2945R,2012ApJS..199....3R}. The extension of the
SDSS system to longer wavelengths
(the $y$ band at $\sim$1 micron) is driven by the increased effective redshift
range achievable with the LSST due to deeper imaging, the desire to study sub-stellar
objects, high-redshift quasars, and regions of the Galaxy that are obscured by
interstellar dust, and
the scientific opportunity enabled by modern CCDs with high quantum efficiency
in the near infrared.

The chosen filter complement corresponds to a design ``sweet spot''. We have
investigated the possibility of replacing the $ugrizy$ system with a
filter complement that includes only five filters. For example, each filter
width could be increased by 20\% over the same wavelength range (neither a
shorter wavelength range, nor gaps in the wavelength coverage are desirable
options), but this option is not satisfactory. Placing the red edge of the $u$
band blueward of the Balmer break allows optimal separation of stars and
quasars, and the telluric water absorption feature at 9500\,\AA\
effectively defines the blue edge of the $y$ band. Of the remaining four
filters ($griz$), the $g$ band is already quite wide. As a last option, the
$riz$ bands could be redesigned as two wider bands. However, this option is also
undesirable because the $r$ and $i$ bands are the primary bands for weak
lensing studies and for star/galaxy separation, and chromatic atmospheric
refraction would worsen the point spread function for a wider bandpass.

\begin{figure}
\includegraphics[width=1.0\hsize,clip]{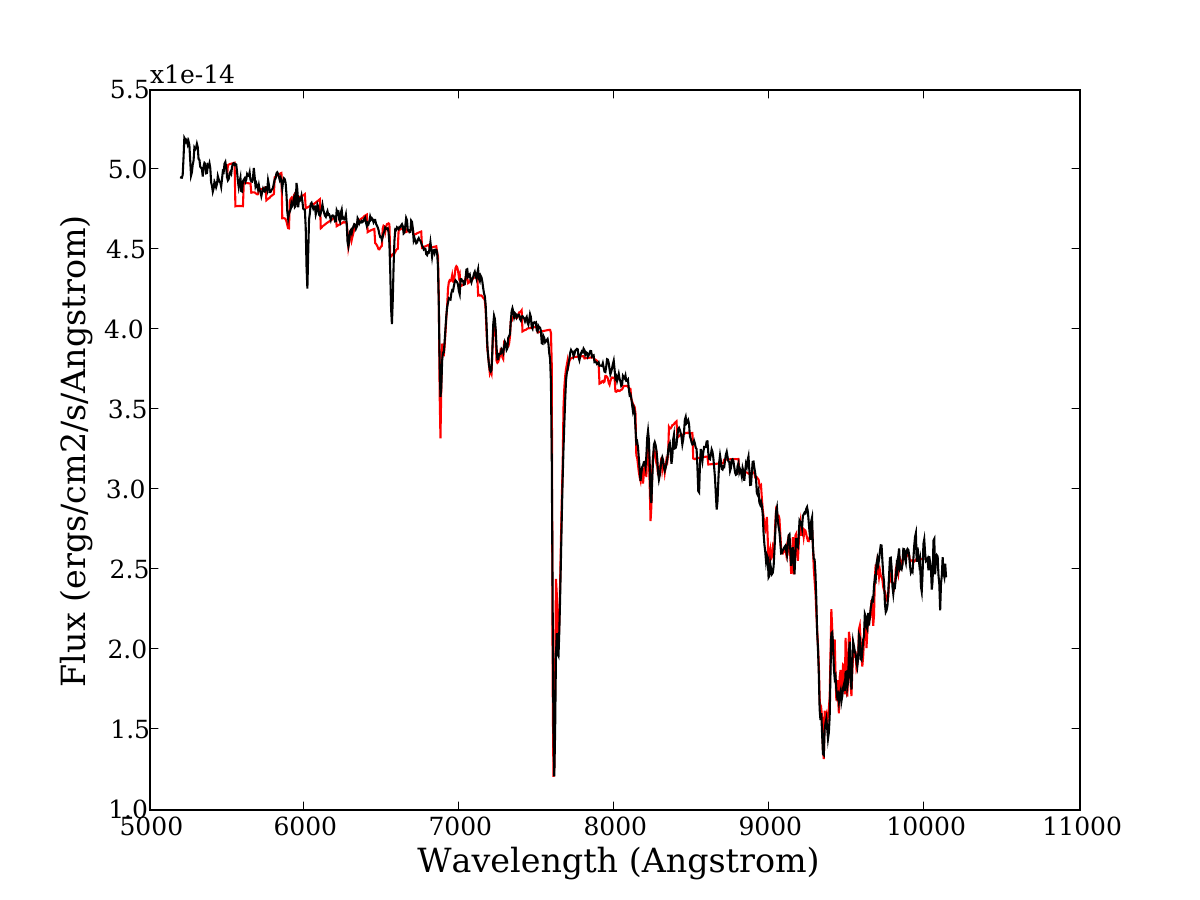}
\caption{An example of determination of the atmospheric opacity by
simultaneously fitting a three-parameter stellar model SED \citep{1979ApJS...40....1K} and
six physical parameters of a sophisticated atmospheric model \citep[MODTRAN,][]{1999SPIE.3866....2A}
to an observed F-type stellar spectrum ($F_\lambda$). The black
line is the observed spectrum and the red line is the best fit. Note that the
atmospheric water feature around 0.9--1.0 $\mu$m is exquisitely well fit.
The components of the best-fit atmospheric opacity are shown in
Fig.~\ref{Fig:modtran2}. Adapted from \citet{2010ApJ...720..811B}.}
\label{Fig:modtran1}
\end{figure}

\subsection{ The Calibration Methods }

\begin{figure}
\includegraphics[width=1.0\hsize,clip]{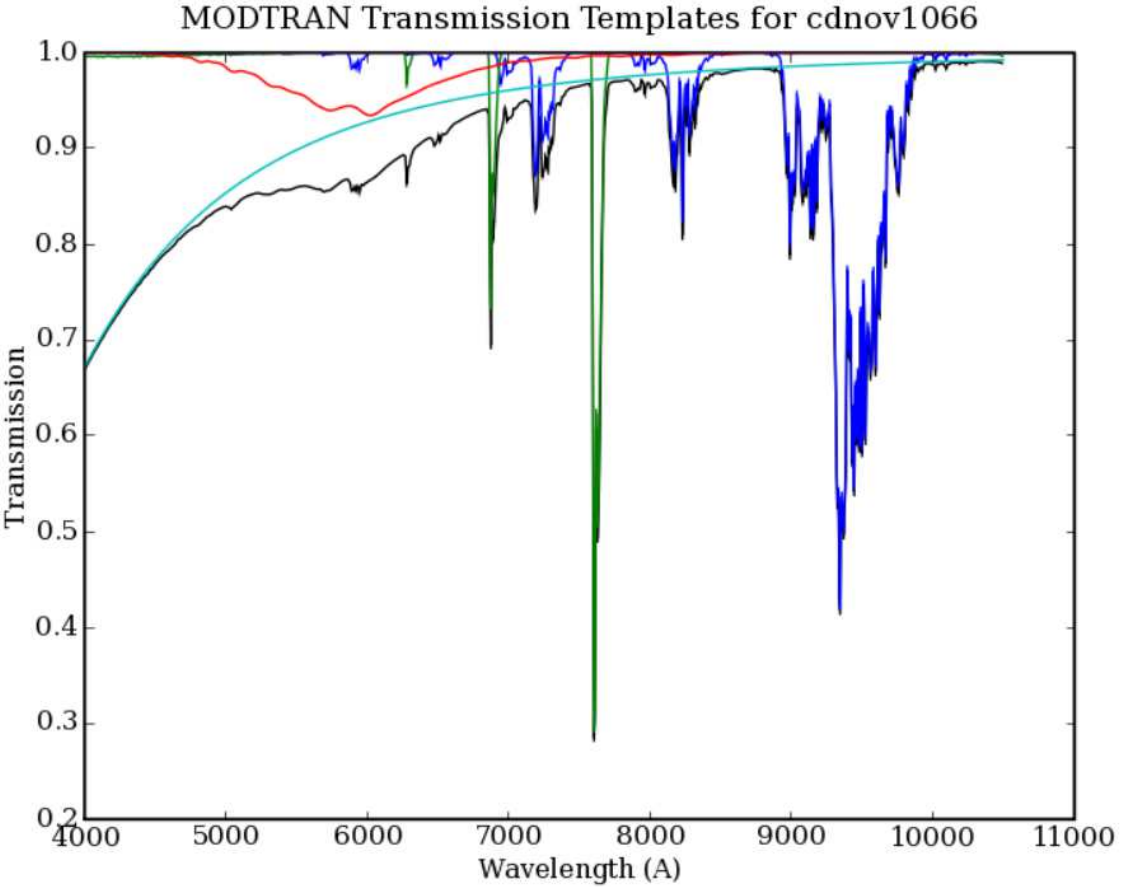}
\caption{The components of the best-fit atmospheric opacity used to
model the observed stellar spectrum shown in Fig.~\ref{Fig:modtran1}.
The atmosphere model \citep[MODTRAN,][]{1999SPIE.3866....2A} includes six
components: water vapor (blue), oxygen and other trace molecules
(green), ozone (red), Rayleigh scattering (cyan), a gray term
with a transmission of 0.989 (not shown) and an aerosol contribution
proportional to $\lambda^{-1}$ and extinction of 1.3\% at $\lambda$=0.675 \mic\
(not shown). The black line shows all six components combined.
Adapted from \citet{2010ApJ...720..811B}.}
\label{Fig:modtran2}
\end{figure}

Precise determination of the point spread function across each image,
accurate photometric and astrometric calibration, and continuous monitoring
of system performance and observing conditions will be needed to reach the
full potential of the LSST mission. Extensive precursor data including the
SDSS dataset and our own data obtained using telescopes close to
the LSST site of Cerro Pach\'{o}n (e.g., the SOAR and Gemini South telescopes),
as well as telescopes of similar aperture (e.g., Subaru), indicate that the
photometric and astrometric accuracy will be limited not by our instrumentation
or software, but rather by atmospheric effects.

The overall photometric calibration philosophy \citep{2006ApJ...646.1436S} is to measure explicitly, at 1 nm resolution, the
instrumental sensitivity as a function of wavelength using light from a monochromatic source injected
into the telescope pupil. The dose of delivered photons is measured using a calibration photodiode whose quantum
efficiency is known to high accuracy. In addition, the LSST system will explicitly measure the atmospheric transmission
spectrum associated with each image acquired. A
dedicated 1.2-meter auxiliary calibration telescope will obtain spectra of
standard stars in LSST fields, calibrating the atmospheric throughput
as a function of wavelength  \citep[][see Figs.~\ref{Fig:modtran1} and \ref{Fig:modtran2}]{2007PASP..119.1163S}.
The LSST auxiliary telescope will take
data at lower spectral resolution ($R \sim 150$) but wider spectral
coverage (340nm --- 1.05$\mu$m) than shown in these figures, using a
slitless spectrograph and an LSST corner-raft CCD.
Celestial spectrophotometric standard stars can be used as a separate means of photometric calibration, albeit only through the
comparison of band-integrated fluxes with synthetic photometry calculations.

A similar calibration process has been undertaken by the Dark Energy
Survey (DES) team, which has been approaching a calibration
precision of 5 mmag \citep{2018AJ....155...41B}.

SDSS, PS1, and DES data
taken in good photometric conditions have approached the LSST
requirement of 1\% photometric calibration
\citep{2008ApJ...674.1217P,2012ApJ...756..158S,2018AJ....155...41B}, although measurements with ground-based telescopes
typically produce data with errors a factor of two or so larger. Analysis of
repeated SDSS scans obtained in varying observing conditions demonstrates that data
obtained in
non-photometric conditions can also be calibrated with
sufficient accuracy \citep{2007AJ....134..973I}, as long as high-quality
photometric data also exist in the region.
The LSST calibration plan builds on this experience gained from the SDSS and other surveys.

The planned calibration process decouples the establishment of a stable and uniform internal
relative calibration from the task of assigning absolute optical flux to
celestial objects.

Celestial sources will be used to refine the internal photometric system and
to monitor stability and uniformity of the photometric data. We expect to use \citet{2016A&A...595A...2G} photometry, utilising
the \textit{BP} and \textit{RP} photometric measurements as well as the \text{G} magnitudes; for a subset
of stars (\textit{e.g.} F-subdwarfs) we expect to be able to transfer this rigid photometric system above
the atmosphere to objects observed by LSST.
There will be
$>$100 main-sequence stars with $17<r<20$ per detector (14$\times$14 arcmin$^2$)
even at high Galactic latitudes. Standardization of photometric scales will be
achieved through direct observation of stars with well-understood spectral
energy distributions (SEDs), in conjunction with the in-dome calibration system and the atmospheric transmission spectra.

Astrometric calibration will be based on the results from the Gaia mission \citep{2016A&A...595A...2G}, which will provide
numerous high-accuracy astrometric standards in every LSST field.

\subsection{The LSST  Reference Design}

We briefly describe the reference design for the main LSST system components.
Detailed discussion of the flow-down from science requirements to system
design parameters, and extensive system engineering analysis can be
found in the LSST Science Book (Ch.~2--3).

\begin{figure}
\includegraphics[width=1.0\hsize,clip]{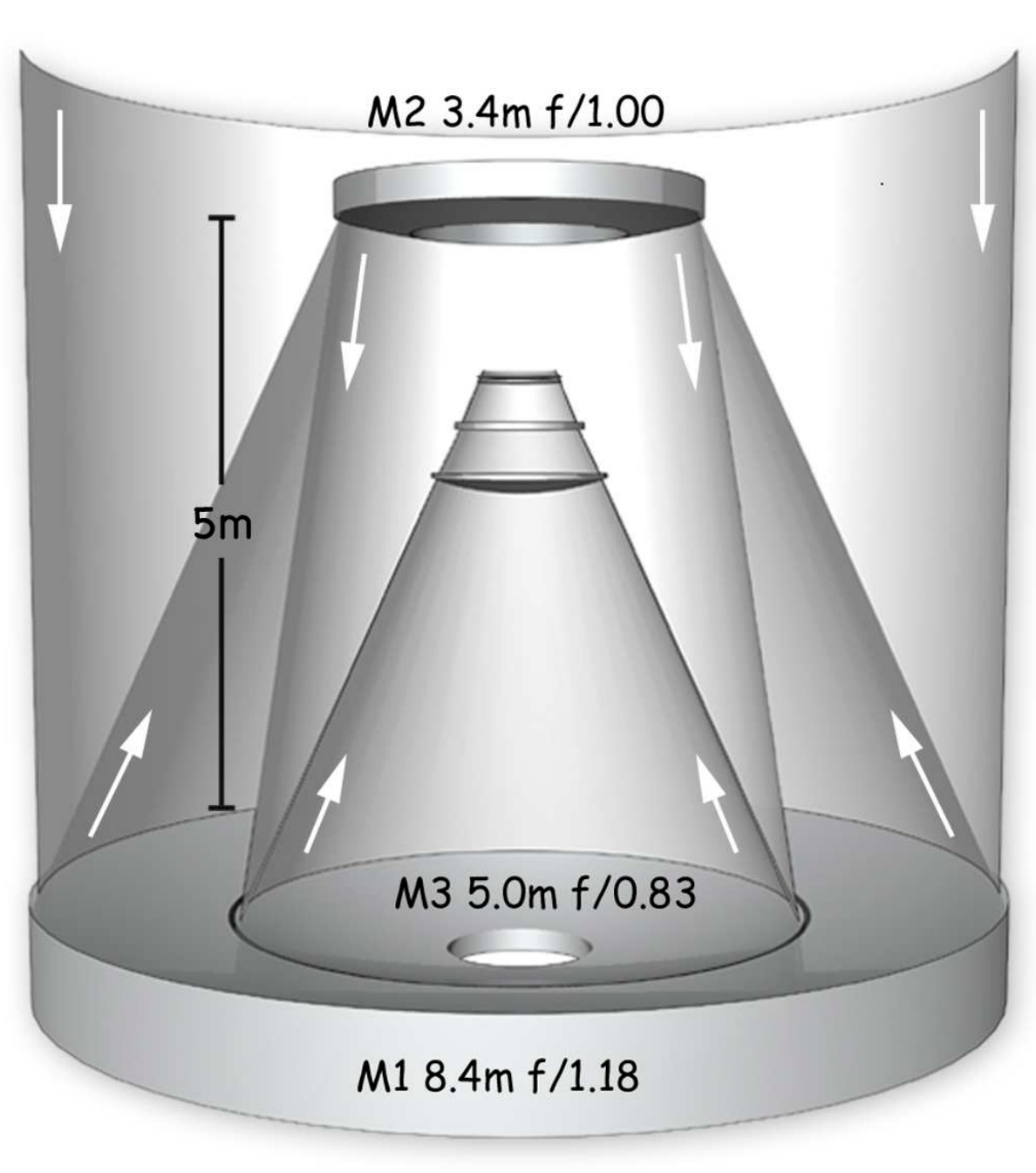}
\caption{The LSST baseline optical design (modified three-mirror
  Paul-Baker) with its unique
monolithic mirror: the primary and tertiary mirrors are positioned such
that they form a continuous compound surface, allowing them to be polished
from a single substrate.}
\label{Fig:optics}
\end{figure}

\begin{figure}
\includegraphics[width=1.0\hsize,clip]{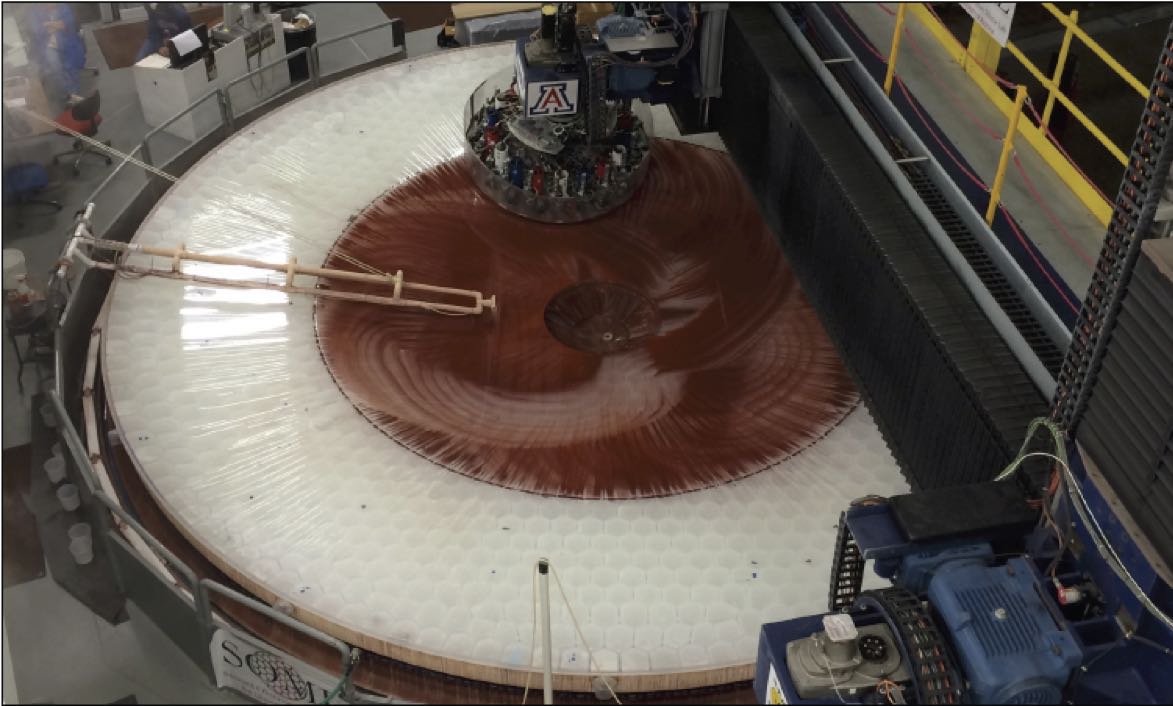}
\caption{The polishing of the primary-tertiary mirror pair at the Richard F.\ Caris Mirror Lab at the University of Arizona Tucson. }
\label{Fig:polishing}
\end{figure}

\begin{figure}
\includegraphics[width=1.0\hsize,clip]{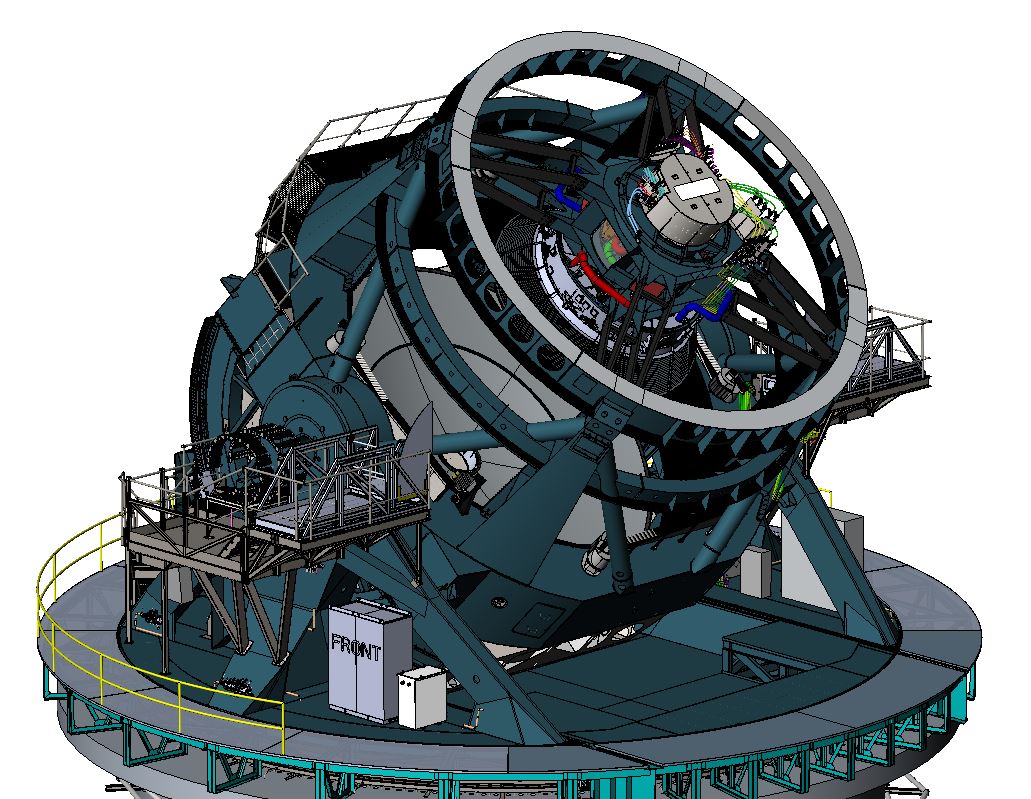}
\caption{The baseline design for the
LSST telescope.  The small focal ratio allows for a very squat
telescope, and thus a very stiff structure.  }
\label{Fig:telescope}
\end{figure}

\subsubsection{ Telescope and Site}

The large LSST \'etendue is achieved in a novel three-mirror design
\citep[modified Paul-Baker Mersenne-Schmidt system;][]{2000ASPC..195...81A} with a very fast $f$/1.234 beam. The optical
design has been optimized to yield a large field of view (9.6 deg$^2$),
with seeing-limited image quality, across a wide wavelength band (320--1050
nm). Incident light is collected by an annular primary mirror, having
an outer diameter of 8.4 m and inner diameter of 5.0 m, creating an effective filled aperture of
$\sim$6.4 m in diameter once vignetting is taken into account. The
collected light is reflected to a 3.4 m convex secondary, then onto
a 5 m concave tertiary, and finally  into the three refractive lenses of the camera (see Fig.~\ref{Fig:optics}).
In broad terms, the primary-secondary mirror pair acts as a beam condenser, while the aspheric portion of
the secondary and tertiary mirror acts as a Schmidt camera.  The three-element refractive optics of the camera
correct for the chromatic aberrations induced by the necessity of a thick dewar window and flatten the
focal surface.  During design optimization, the primary and tertiary mirror surfaces were placed such that the primary's
inner diameter coincides with the tertiary's outer diameter, thus making it possible to fabricate the mirror pair from a
single monolithic blank using spin-cast borosilicate technology. The secondary mirror is fabricated from
a thin 100 mm thick meniscus substrate, made from Corning's ultra-low expansion material. All
three mirrors will be actively supported to control wavefront distortions
introduced by gravity and environmental stresses on the telescope.
The primary-tertiary mirror was cast and polished
by the Richard F. Caris Mirror Lab at the University of Arizona in Tucson
before being inspected and accepted by LSST in April 2015
\citep{2016SPIE.9906E..0LA}. The  primary-tertiary mirror cell was
fabricated by CAID in Tucson and is undergoing acceptance tests. The
integration of the actuators and final tests with the mirror is
scheduled for early 2018.

\begin{figure}
\includegraphics[width=1.0\hsize,clip]{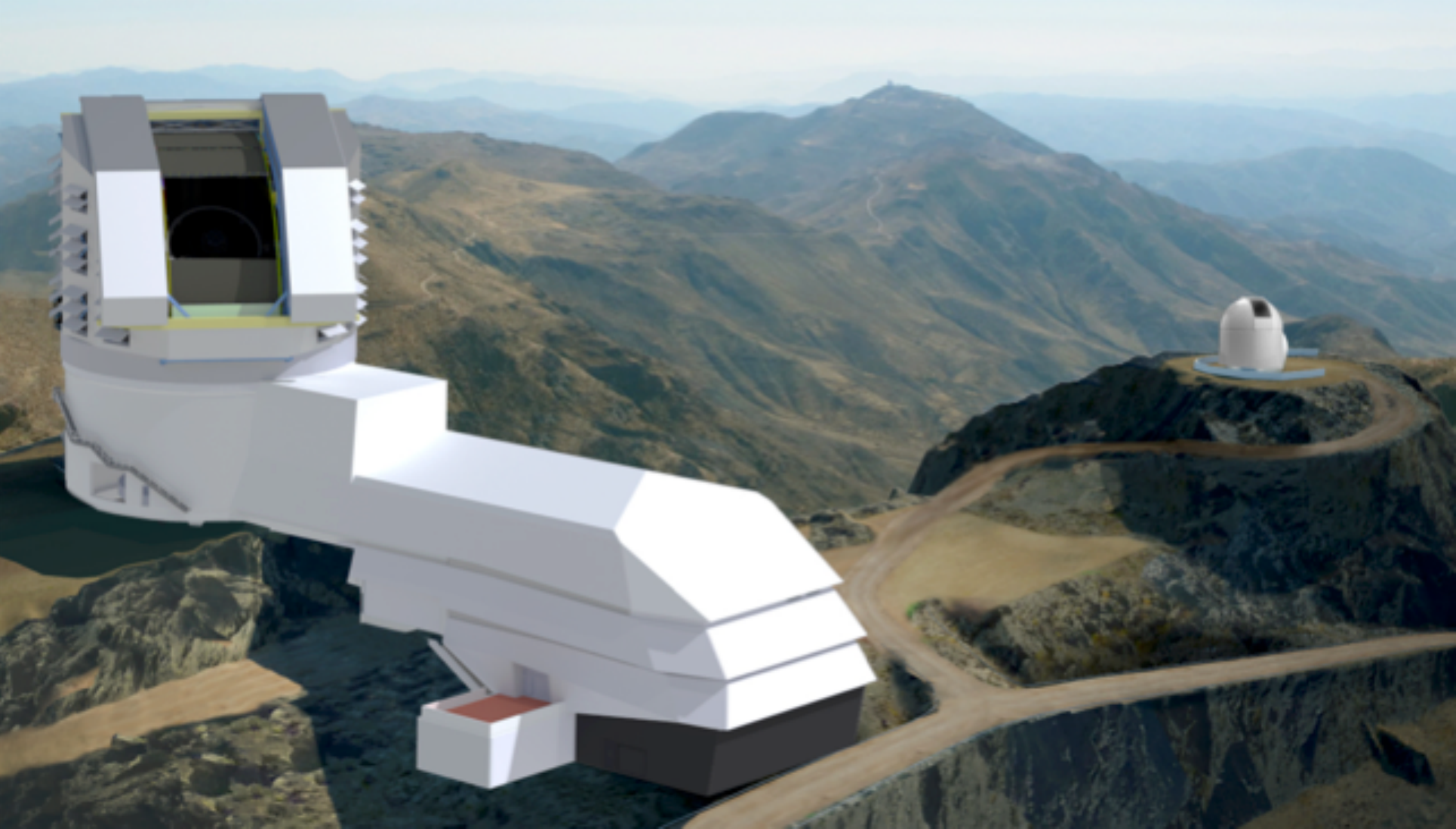}
\includegraphics[width=1.0\hsize,clip]{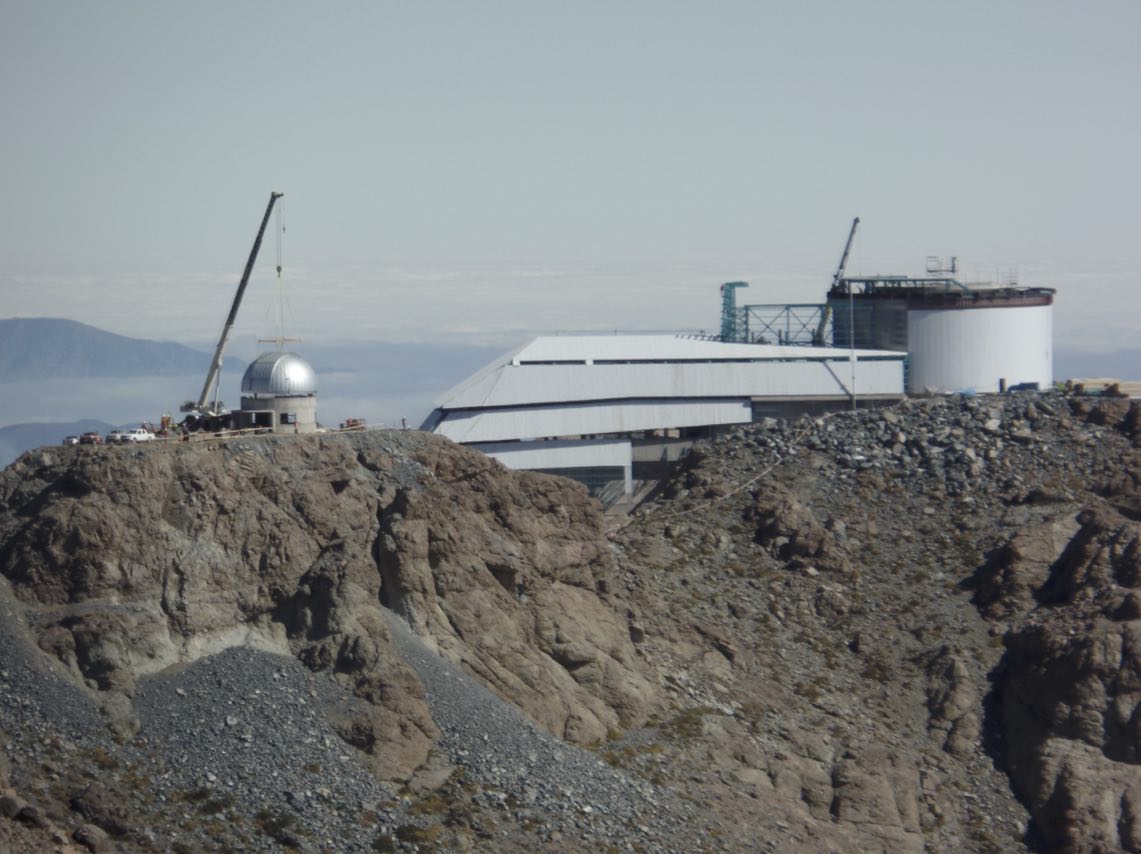}
\caption{Top: artist's rendering of the dome enclosure
with the attached summit support building on Cerro Pach\'{o}n. The LSST auxiliary
calibration telescope is shown on an adjacent rise to the right.
Bottom: Photograph of the LSST Observatory as of Summer 2017. Note the
different perspective from the artist's rendering.  The main LSST
telescope building is on the right, waiting for the dome to be
installed. The auxiliary telescope building is on the left with its
dome being installed.}
\label{Fig:observatory}
\end{figure}

The LSST Observing Facility (Fig.~\ref{Fig:observatory}),
consisting of the telescope enclosure and summit support building, is being constructed atop Cerro Pach\'{o}n in northern Chile,
sharing the ridge with the Gemini South and SOAR telescopes\footnote{Coordinates listed in older versions
of this paper were incorrect. We thank E. Mamajek for pointing out this error to us.}
(latitude: S 30$^\circ$ 14$\arcmin$ 40.68$\arcsec$; longitude: W 70$^\circ$ 44$\arcmin$ 57.90$\arcsec$; elevation: 2652 m;
\citealt{2012arXiv1210.1616M}).  The telescope enclosure houses a compact, stiff
telescope structure (see Fig.~\ref{Fig:telescope}) atop a 15 m high concrete pier
with a fundamental frequency of 8 Hz, that is crucial for achieving the required fast slew-and-settle times.  The height of the pier was set to place the telescope above the degrading
effects of the turbulent ground layer.  Capping the telescope
enclosure is a 30 m diameter dome with extensive ventilation to reduce
dome seeing
and to maintain a uniform thermal environment over the course of the night.  Furthermore, the summit support
building has been oriented with respect to the prevailing winds to shed its turbulence away from the
telescope enclosure.  The summit support building includes a coating chamber for recoating the three LSST mirrors and
clean room facilities for maintaining and servicing the camera.

\subsubsection{ Camera }

\begin{figure}[t!]
\includegraphics[width=1.05\hsize,clip]{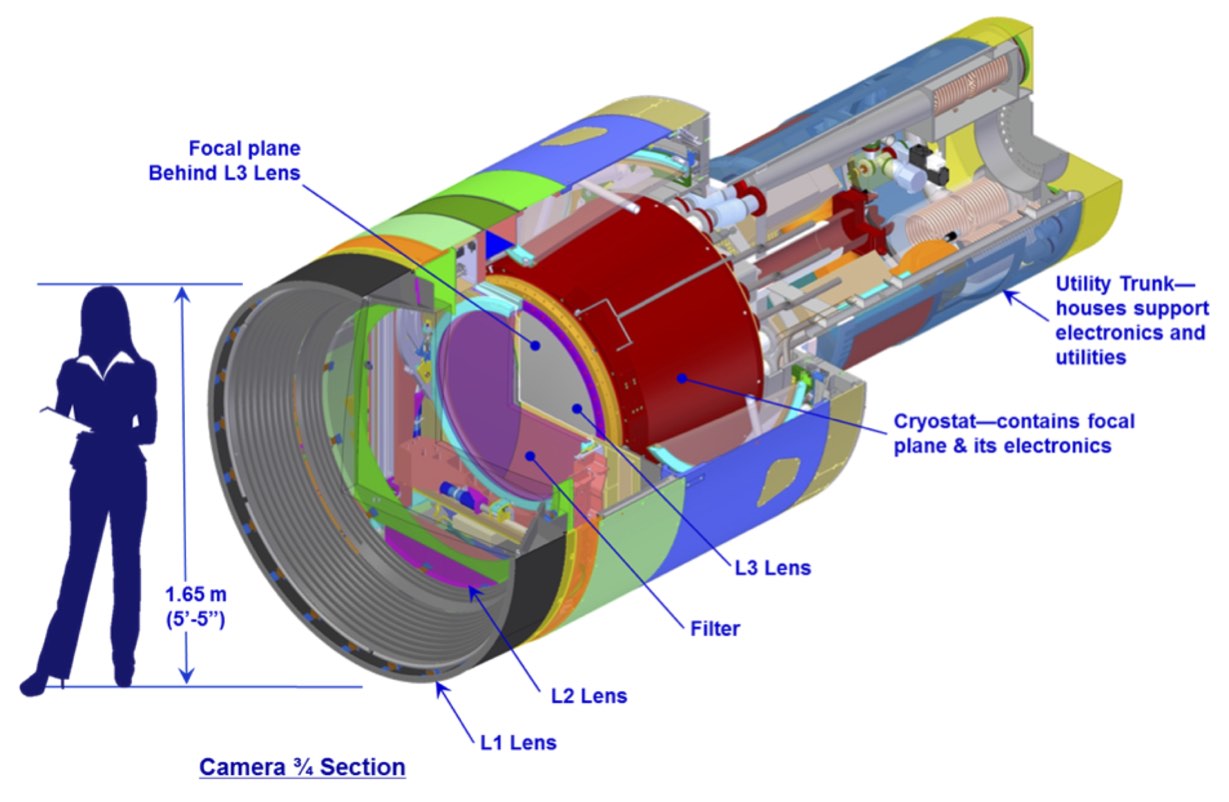}
\caption{A cutaway view of LSST camera. Not shown are the shutter, which is positioned between the filter and lens L3, and the filter exchange system.}
\label{Fig:camera}
\end{figure}

\begin{figure}[ht]
\includegraphics[width=1.0\hsize,clip]{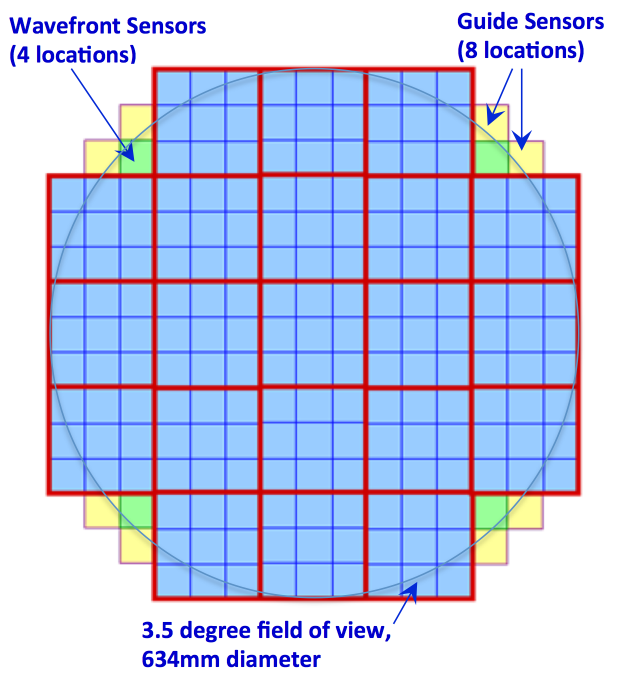}
\caption{The LSST Camera focal plane array. Each cyan square represents one
4K$\times$ 4K pixel sensor. Nine sensors are assembled into a
raft; the 21 rafts are outlined in red. There are 189 science sensors, for a total of 3.2 gigapixels. Also shown are the four corner rafts, where the guide sensors and wavefront sensors are located.}
\label{Fig:fov}
\end{figure}

\begin{figure}[ht]
\includegraphics[width=1.\hsize,clip]{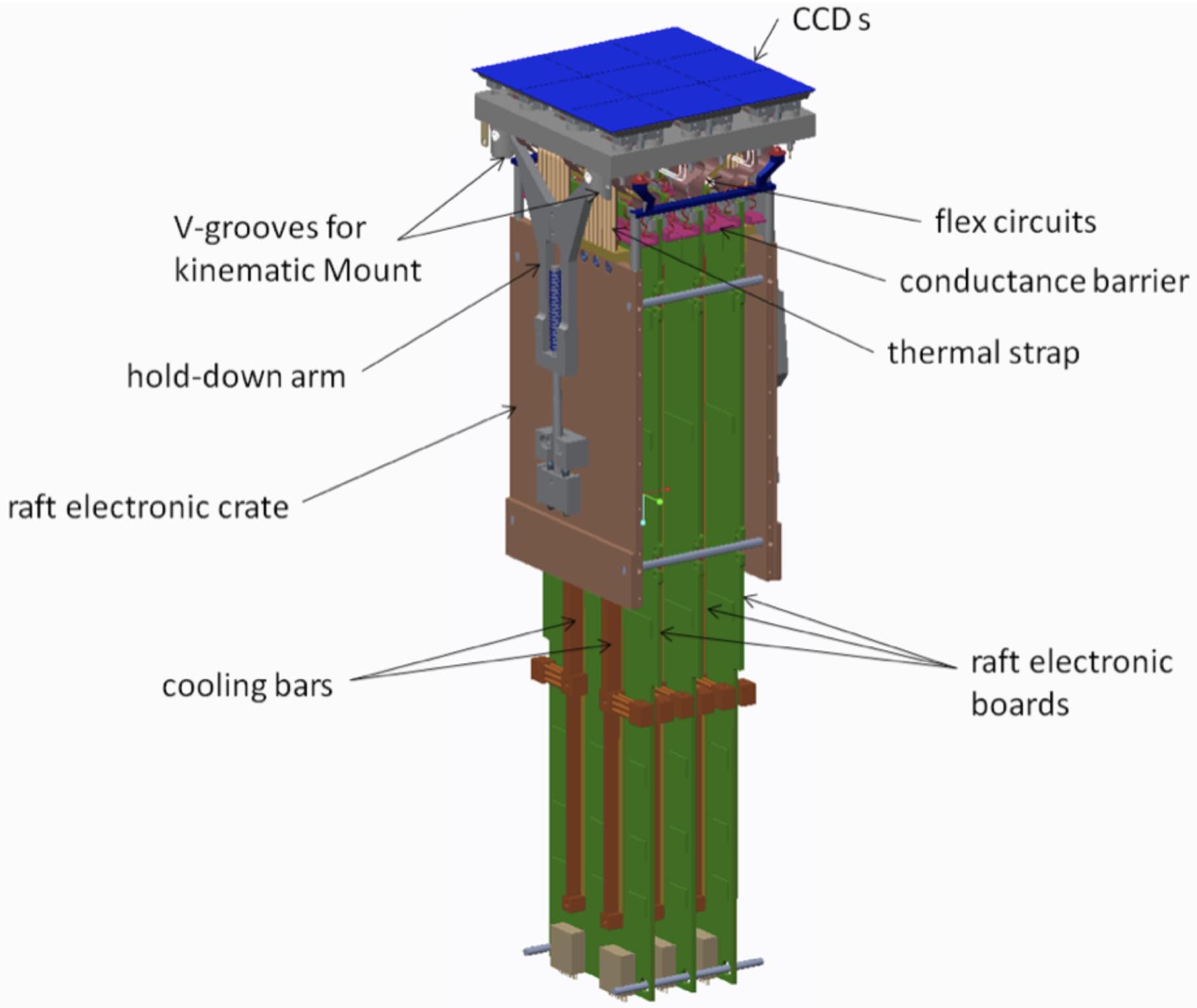}
\caption{The LSST Camera raft module, corresponding to the red squares
in Fig.~\ref{Fig:fov}, with 9 sensors, integrated electronics,
and thermal connections. Raft modules are designed to be replaceable.}
\label{Fig:raft}
\end{figure}

The LSST camera provides a 3.2 Gigapixel flat focal plane array, tiled by 189
4K$\times$4K CCD science sensors with 10 $\mu$m pixels (see Figs.~\ref{Fig:camera}
and \ref{Fig:fov}). This pixel count is a direct consequence of sampling the
9.6 deg$^2$ field-of-view (0.64\,m diameter) with 0.2$\times$0.2 arcsec$^2$
pixels (Nyquist sampling in the best expected seeing of $\sim$0.4 arcsec).
The sensors are deep depleted high resistivity silicon back-illuminated devices with
a highly segmented architecture that enables the entire array to be read in 2 seconds.
The detectors are grouped into 3$\times$3 rafts (see Fig.~\ref{Fig:raft}); each
contains its own dedicated electronics. The rafts are mounted on a silicon carbide
grid inside a vacuum cryostat, with a custom thermal control system that maintains
the CCDs at an operating temperature of around 173 K. The entrance window to the
cryostat is the third (L3) of the three refractive lenses in the camera. The other
two lenses (L1 and L2) are mounted in an optics structure at the front of the camera
body, which also contains a mechanical shutter, and a carousel assembly that holds
five large optical filters. The sixth optical filter can
replace any of the five via a procedure accomplished during daylight hours.

Each of the 21 rafts will host 3 front end electronic boards (REB) operating in the cryostat
(at $-10^\circ$ C), that read in parallel a total of 9$\times$16 segments per CCD (144 video
channels reading one million pixels each). This very high parallelization is the key to allow
for a fast readout (2 seconds) of the entire focal plane. To reach this performance with a
reasonably-sized board, a special low-noise ($<$3 electrons), low-crosstalk between channels
($<$0.02\%) and low-power dissipation (25 mW/channel) Analog Signal Processing Integrated
Circuit (ASPIC),  hosting 8 channels per chip, has been developed, which is able to read the
CCDs with a linearity better than 0.1\% \citep{1748-0221-12-03-C03017}.

\subsubsection{ Data Management }
\label{sec:dm}

The rapid cadence and scale of the LSST observing program will produce
approximately 15 TB per night of raw imaging data\footnote{For
  comparison, the volume of all imaging data collected over a decade
  by the SDSS-I/II projects
  and published in SDSS Data Release 7 \citep{2009ApJS..182..543A} is
  approximately 16 TB.} (about 20 TB with calibration exposures). As with
all large modern surveys, the large data volume, the
real-time aspects, and the complexity of processing involved requires
that the survey itself take on the task of fully reducing the data.
The data collected by the LSST
system will be automatically reduced to scientifically useful catalogs
and images by the LSST Data Management
\citep[DM;][]{2015arXiv151207914J} system.
\\

The detailed outputs of the LSST Data Management system are described
in \S~\ref{Sec:dp}.  The principal functions of the system are to:
\begin{itemize}
\item Process, in real time, the incoming stream of images generated
  by the camera system during observing by archiving raw images,
  generating alerts to new sources or sources whose properties have
  changed, and updating the relevant catalogs (Prompt  products;
  \S~\ref{Sec:dp}).
\item Process each night's data during the day and determine or refine
  orbits for all asteroids found in the imaging.
\item Periodically process the accumulated survey data to provide a
  uniform photometric and astrometric calibration, measure the
  properties of all detected objects, and characterize objects based
  on their time-dependent behavior. The results of such a processing
  run form a \emph{Data Release} (DR), which is a static,
  self-consistent dataset suitable for use in performing scientific
  analyses of LSST data and publication of the results (the data
  release products; \S~\ref{Sec:dp}). We are planning two data
  releases covering the first year of full operations, and annual data
  releases thereafter.
\item Facilitate the creation of data
  products generated by the science community, by providing suitable software,
  application programming interfaces (APIs),
and computing infrastructure at the LSST data access centers.
\item Make all LSST data available through an interface that utilizes
community-based standards   to the maximum possible extent. Provide
  enough processing, storage, and network bandwidth to enable user
  analyses of the data without the need for petabyte-scale data
  transfers.
\end{itemize}

Over the ten years of LSST operations and 11 data releases, this processing will result in a cumulative \emph{processed} data size
approaching 500 petabytes (PB) for imaging, and over 50 PB for the
catalog databases. The final data release catalog database alone is expected
to be approximately 15 PB in size.
\\

\begin{figure*}
\hskip 0.01in
\includegraphics[width=1.0\hsize,clip]{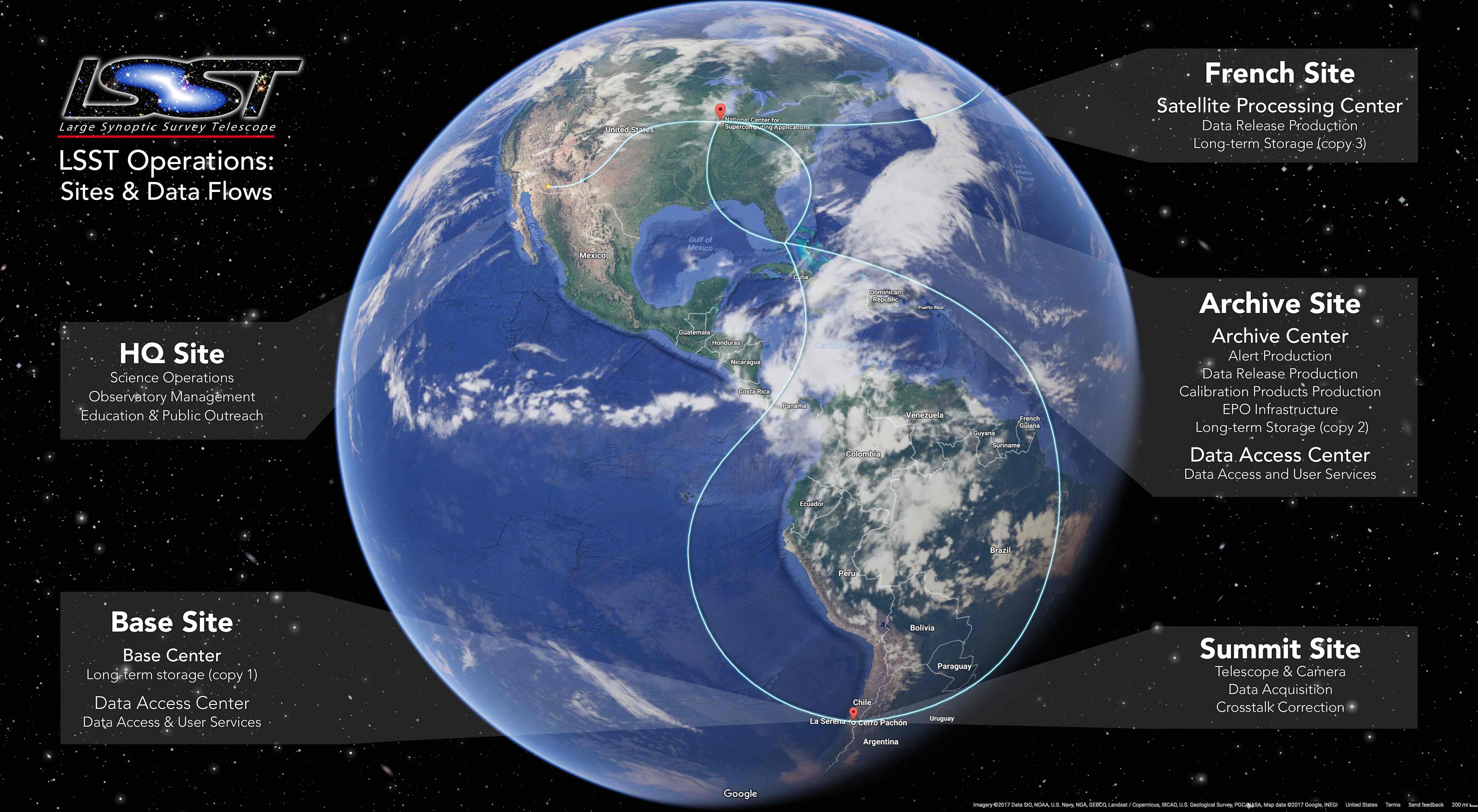}
\caption{The LSST data flow from the mountain facilities in
Chile to the data access center and processing center in the U.S., and
the satellite processing center in France.}
\label{Fig:DM2}
\end{figure*}

The DM system will span four key facilities on three continents: the Summit
Facility on Cerro Pach\'on in Chile (where the initial detector cross-talk correction
will be performed); the Base Facility in La Serena, Chile (which will serve as a retransmission
for data uploads to North America, as well as the Data Access Center for the Chilean community);
the Data Processing and Archiving Facility at the National Center for Supercomputing Applications
(NCSA) in Champaign-Urbana, IL; and the Satellite Processing Facility at CC-IN2P3 in Lyon, France.
All real-time data processing and half the data release product processing will take place at the
Data Processing and Archiving Facility, which will also serve as the Data Access Center for the US
community. The other half of the data release processing will be done at CC-IN2P3, which
will also have the role of ``Long-term Storage'' facility.

The data will be transported between the centers over existing and new high-speed optical fiber
links from South America to the U.S.\ (see Fig.~\ref{Fig:DM2}). The data processing center demands
stable, well-tested technology to ensure smooth operations. Hence, while LSST is making a novel
use of advances in information technology, it is not pushing the expected technology to the limit,
reducing the overall risk to the project.

\subsubsection{The LSST software stack}
\label{sec:dmstack}

The \emph{LSST Software Stack} is the data processing and analysis
system developed by the LSST Project to enable LSST survey data
reduction and delivery. It comprises
all science pipelines needed to accomplish LSST data processing tasks
(e.g., calibration, single frame processing, coaddition, image
differencing, multi-epoch measurement, asteroid orbit determination,
etc.), the necessary data
access and orchestration middleware, as well as the database and user
interface components.

Algorithm development for the LSST software builds on the expertise
and experience of prior large astronomical surveys (including SDSS,
Pan-STARRS, DES,
SuperMACHO, ESSENCE,  DLS, CFHTLS, and UKIDSS). The pipelines written
for these surveys have demonstrated that it is possible to carry out
largely autonomous data
reduction of large datasets, automated detection of sources and
objects, and the
extraction of scientifically useful characteristics of those objects.
While firmly footed in this prior history, the LSST software stack has
largely been written anew, for reasons of performance, extendability, and
maintainability. All LSST codes have been designed and implemented
following software engineering best practices, including modularity, clear definition
of interfaces, continuous integration,
utilization of unit testing, and a single set of documentation and coding
standards \citep{2018SPIE10707-10J}. The primary implementation language is Python and, where
necessary for performance reasons, C\texttt{++}\footnote{All components implemented
in C\texttt{++} have been wrapped and exposed as Python modules to the rest of the system.
Typical users should not have to work directly with the C\texttt{++} layer.}.

\begin{figure}
\includegraphics[width=1.0\hsize,clip]{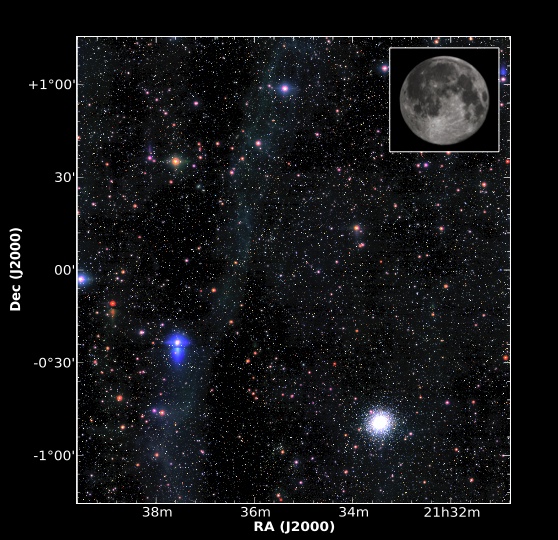}
\caption{
A small region in the vicinity of globular cluster M2, taken from a coadd of SDSS Stripe 82 data produced with LSST software stack prototypes. The coaddition employs a novel ``background matching'' technique that improves background estimation and preserves the diffuse structures in the resulting coadd.}
\label{Fig:DMStripe82}
\end{figure}

The LSST data management software has been prototyped for over eight
years. Besides processing simulated LSST data
(\S~\ref{sec:imsim}), it has been used to process images from CFHTLS \citep{2012SPIE.8448E..0MC}
and SDSS \citep{2009ApJS..182..543A}. As an example,
Fig.~\ref{Fig:DMStripe82} shows a small region in the vicinity of M2
taken from a large coaddition of SDSS Stripe 82 data, generated with LSST
software stack prototypes \citep{DMTN-035}.
\\

\begin{figure}
\includegraphics[width=1.0\hsize,clip]{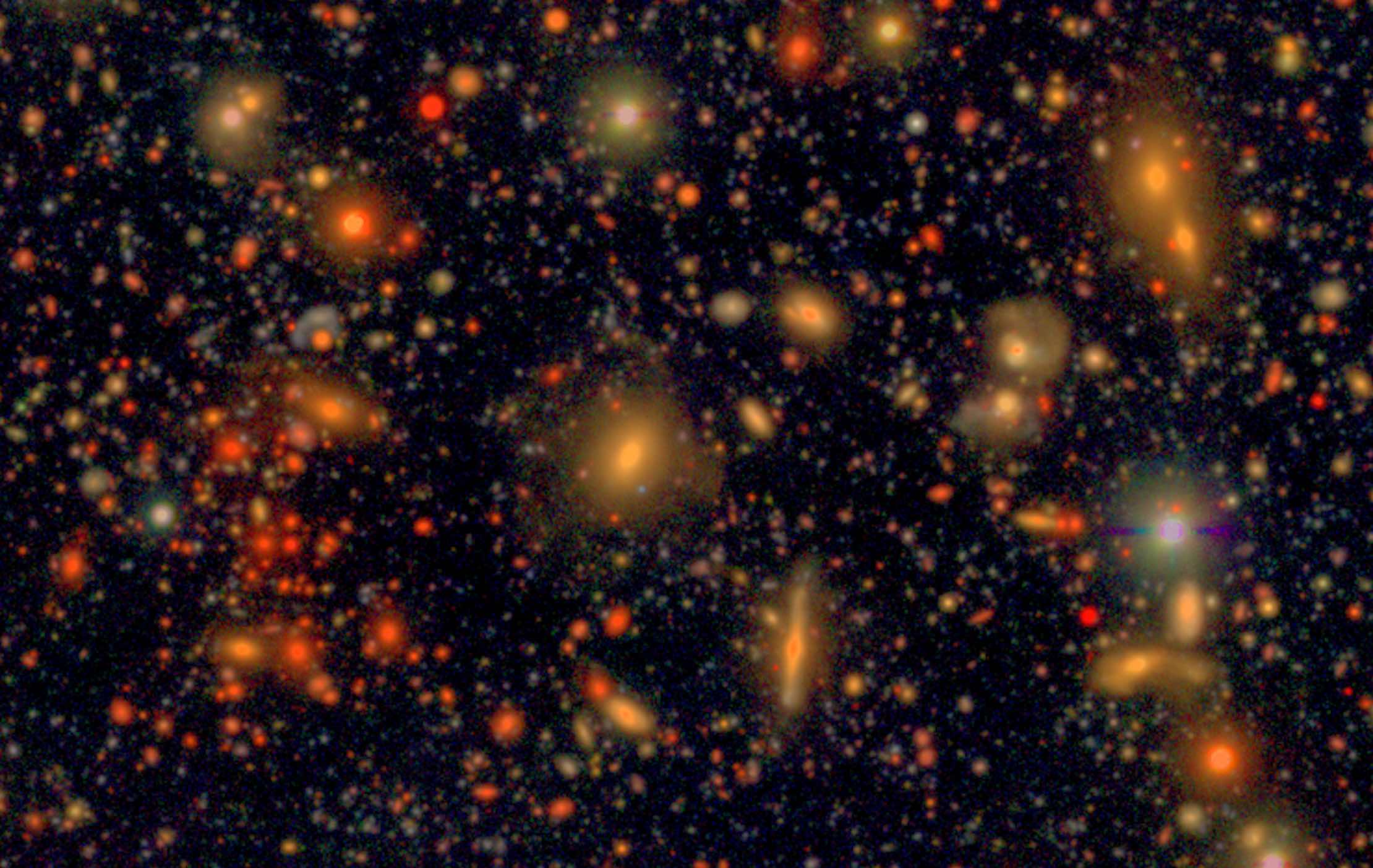}
\caption{
  A small portion, $4' \times 6'$, of the HSC \textit{gri} imaging of
  the COSMOS field.  The limiting magnitude is about 27.5, roughly
  equivalent to 10-year LSST depth.
  }
\label{Fig:HSC_cosmos}
\end{figure}

Other than when prohibited by licensing, security, or other similar
considerations, the LSST makes all newly developed source code, and especially
that pertaining to scientific algorithms, public.  Our primary goals in
publicizing the code are to simplify reproducibility of LSST data products
and to provide insight into algorithms used to create them.  Achieving these goals requires
that the source code is not only available, but appropriately documented at all
levels.
Given that, most of the LSST software stack is licensed under the terms of the GNU General
Public License (GPL), Version 3, and can be found at \url{https://github.com/lsst}.
The documentation for the LSST Science Pipelines components of the stack is available at
\url{https://pipelines.lsst.io}.

The LSST Software Stack may be of interest and (re)used beyond the LSST project (e.g.,
by other survey projects, or by individual LSST end-users).  Enabling
or supporting such applications goes beyond LSST’s construction
requirements; however, when developing the LSST codes we strongly
prefer design choices that enable future generalization.  As an example of such
re-use, a pipeline derived from the present-day LSST software stack prototypes
has been used to reduce data taken with the HSC camera \citep{2018PASJ...70S...1M} on
Subaru as part of the large SSP survey (\url{http://hsc.mtk.nao.ac.jp/ssp/survey};
\citet{2018PASJ...70S...4A,2018PASJ...70S...5B}; see Fig.~\ref{Fig:HSC_cosmos}).

\subsubsection{The LSST database design: Qserv}
\label{sec:Qserv}

The scale of the LSST data release catalogs, in combination with desired targets for user concurrency
and query response times, present some engineering challenges. The LSST project has been developing
\emph{Qserv}, a shared-nothing MPP (massively parallel processing) database system, to meet these needs
\citep{Wang:2011:QDS:2063348.2063364, 2014era..conf30303B}.
Catalog data within Qserv is spatially partitioned, and hosted on shard servers running on dedicated hardware
resources within the LSST Data Facility.  The shard servers locally leverage conventional RDBMS (relational
database management system) technologies, running behind custom front-end codes which handle query
analysis, rewrite, distribution, and result aggregation.  The Qserv shard servers also provide a facility for
cross-user synchronization of full-table scans in order to provide predictable query response times when
serving many users concurrently. More details about Qserv can be found in
the LSST document LDM-135 \citep{LDM-135}.

\subsection{Simulating the LSST System}

Throughout its design, construction and commissioning, the LSST needs
to be able to demonstrate that it can achieve the requirements laid out
in the Science Requirements Documents (SRD) given its design and
as-delivered components, that the system can be calibrated to the
required level of fidelity, that the data management software can
extract the appropriate astrophysical signals, and that this can be
achieved with sufficient efficiency such that the telescope can
complete its primary objectives within a ten-year survey.

Realizing these objectives requires that the project can characterize
the performance of the LSST including the performance of the
opto-mechanical systems, the response of the detectors and their
electronics, and the capabilities of the analysis software. A
simulation framework provides such a capability; delivering a virtual
prototype LSST against which design decisions, optimizations
(including descoping), and trade studies can be evaluated \citep{2014SPIE.9150E..14C}.

The framework underlying the LSST simulations is designed to be
extensible and scalable (i.e., capable of being run on a single
processor or across many-thousand core compute clusters). It comprises
four primary components: a simulation of the survey scheduler
(\S~\ref{sec:opsim}),
databases of simulated astrophysical catalogs of stars, galaxies,
quasars and Solar System objects (\S~\ref{sec:catalogs}), a system for generating observations
based on the pointing of the telescope, and a system for generating
realistic LSST images of a given area of sky
(\S~\ref{sec:imsim}). Computationally intensive routines are written
in C/C\texttt{++} with the overall framework and database interactions
using $Python$.  The purpose of this design is to enable the
generation of a wide range of data products for use by the
collaboration; from all-sky catalogs used in simulations of the LSST
calibration pipeline, to studies of the impact of survey cadence on
recovering variability, to simulated images of a single LSST focal
plane.

\subsubsection{ The LSST Operations Simulator \label{sec:opsim}}

The LSST Operations Simulator \citep{2014SPIE.9150E..15D} was developed to enable a
detailed quantitative analysis of the various science tradeoffs described in
this paper. It contains detailed models of site conditions, hardware and
software performance, and an algorithm for scheduling observations which will,
eventually, drive the largely robotic observatory.
Observing conditions include a model for seeing derived from an extensive body
of on-site MASS/DIMM (Multi-Aperture Scintillation Sensor and Differential
Image Motion Monitor) measurements obtained during site selection and
characterization (see Fig.~\ref{Fig:seeing}). It not only reproduces the
observed seeing distribution, but includes
the auto-correlation spectrum of seeing with time over intervals from minutes
to seasons. Weather data are taken from ten years of hourly measurements at
nearby Cerro Tololo.
Thus the simulator correctly represents the variation of limiting
magnitude between pairs of observations used to detect NEOs and the
correlation between, for example, seasonal weather patterns and observing
conditions at any given point on the sky.  In addition, down time for
observatory maintenance is also included.

The signal-to-noise ratio of each
observation is determined using a sky background model which includes the dark
sky brightness in each filter, the effects of seeing and atmospheric
transparency, and a detailed model for scattered light from the Moon and/or
twilight at each observation \citep{2016SPIE.9910E..1AY}. The time taken to move from one observation to
the next is given by a detailed model of the camera, telescope, and dome. It
includes such effects as the acceleration/deceleration profiles employed in
moving the telescope, the dome, and the wind screen,
the time needed to damp vibrations excited by each slew,
cable wrap, the time taken for active optics lock and correction as a function of
slew distance, and the time for filter changes and focal plane readout.

Observations are scheduled by a ranking algorithm. After a given exposure, all
possible next observations are assigned a score which depends upon their locations, times,
and filters according to a set of scientific requirements which can vary with
time and location. For example, if an ecliptic field has been observed in the
$r$ band, the score for another $r$-band observation of the same field will
initially be quite low, but it will rise in time to peak about an hour after
the first observation, and decline thereafter. This algorithm results in
observations being acquired as pairs roughly an hour apart, which enables
efficient association of NEO detections. To ensure uniform
sky coverage, fields with fewer previous observations will be scored more
highly than those which have already been observed more frequently.

Once all possible next observations have been scored for scientific
priority, their scores are modified according to observing conditions
(e.g., seeing, airmass, and sky brightness) and to criteria such as
slew time to move from the current position, time required to
change filters, etc. The highest-ranked observation is then performed,
and the cycle repeats. The result of a simulator run is a detailed
history of which locations on the sky were observed when, in what
filter, and with what sky background, seeing and other observing
conditions.  It takes a few days to produce a decade-long simulation
using an average PC.

Results of the simulated surveys can be visualized and analyzed using
a Python-based package called the Metrics Analysis Framework
\citep[MAF;][]{2014SPIE.9149E..0BJ}. MAF provides tools to analyze the properties of a
survey (e.g.\ the distribution of airmasses) through the creation of
functions or metrics that are applied to OpSim outputs. These metrics
can express the expected technical performance of the survey, such as
the number of visits per field or the integrated depth after 10 years,
as well as the science capabilities or a survey, such as the number of
supernovae detected or the number of supernovae with sufficient
observations to have a well-characterized light curve.

\subsubsection{Catalog Generation}
\label{sec:catalogs}

The simulated astronomical catalogs \citep[CatSim;][]{2014SPIE.9150E..14C} are
stored in an SQL database. This base catalog is queried using
sequences of observations derived from the Operations Simulator. Each
simulated pointing provides a position and time of the observation
together with the appropriate sky conditions (e.g., seeing, moon phase
and angle, sky brightness and sky transparency). Positions of sources
are propagated to the time of observation (including proper motions
for stars and orbits for Solar System sources). Magnitudes and source
counts are derived using the atmospheric and filter response functions
appropriate for the airmass of the observation and after applying
corrections for source variability.  The resulting catalogs are then
formatted to be output to users, or to be fed into an image
simulator.

The current version of the LSST simulation framework incorporates
galaxies derived from an N-body simulation of a $\Lambda$CDM
cosmology, quasars/AGNs, stars that match the observed stellar
distributions within our Galaxy, asteroids generated from simulations
of our Solar System, and a 3-D model for Galactic extinction.  Stellar
sources are based on the Galactic structure models of \citet{2008ApJ...673..864J}
and include thin-disk, thick-disk, and halo star
components. The distribution and colors of the stars match those
observed by SDSS. Each star in the simulation is matched to a template
spectral energy distribution (SED). \citet{1993sssp.book.....K} model spectra are
used to represent main-sequence F, G, and K stars as well as RGB
stars, blue horizontal branch stars, and RR Lyrae variables.  SEDs for
white dwarf stars are taken from \citet{1995PASP..107.1047B}.  SEDs for M,
L, and T dwarfs are generated from a combination of spectral models
and stacks of spectra from the SDSS
\citep[e.g.,][]{2005ApJ...623.1115C,2007AJ....133..531B,2006ApJ...640.1063B,1989A&A...217..187P,2010ApJ...714L..98K}.
The adopted metallicity for each star is based
on a model from \citet{2008ApJ...684..287I}, and proper motions are
based on the kinematic model of \citet{2010ApJ...716....1B}.  Light curve
templates are assigned to a subset of the stellar population so that
variability may also be simulated. This assignment and variability are
matched to variability trends observed by the Kepler satellite, and
augmented by
simulated distributions of RR-Lyrae and Cepheids. For Galactic reddening, a
value of $E(B-V)$ is assigned to each star using the three-dimensional
Galactic model of \citet{2005AJ....130..659A}. To provide consistency with
the modeling of extragalactic fluxes in the simulations, the dust model in the Milky Way integrated
to 100 kpc is re-normalized to match the \citet{1998ApJ...500..525S} dust maps.

Galaxy catalogs are derived from the Millennium simulations of
\citet{2006MNRAS.366..499D}.  These models extend pure dark matter N-body
simulations to include gas cooling, star formation, supernovae and
AGN, and are designed to reproduce the observed colors, luminosities,
and clustering of galaxies as a function of redshift. To generate the
LSST simulated catalogs, a light cone, covering redshifts $0<z<6$, was
constructed from 58 simulation snapshots 500\,$h^{-1}$Mpc on a side. This
light cone extends to a depth of approximately $r=28$ and covers a
4.5$^\circ$$\times$4.5$^\circ$ footprint on the sky. Replicating this
catalog across the sky simulates the full LSST footprint. As with the
stellar catalog, an SED is fit to the colors of each source using
\citet{2003MNRAS.344.1000B} spectral synthesis models. These fits are
undertaken separately for the bulge and disk components and, for the
disk, include inclination-dependent reddening. Morphologies are
modeled using two S\'ersic profiles. The bulge-to-disk ratio and disk
scale lengths are taken from \citet{2006MNRAS.366..499D}. Half-light radii
for bulges are estimated using the empirical absolute-magnitude
vs. half-light radius relation given by \citet{2009MNRAS.397.1254G}.
Comparisons between the redshift and number-magnitude
distributions of the simulated catalogs with those derived from deep
imaging and spectroscopic surveys showed that the \citet{2006MNRAS.366..499D}
models under-predict the density of sources at faint magnitudes
and high redshifts. To correct for these effects, sources are cloned
in magnitude and redshift space until their densities reflect the
average observed properties.

Quasar/AGN catalogs are generated using the \citet{2007A&A...472..443B}
luminosity function for $M_B < -15$.
Their observed SEDs are generated using a composite
rest-frame spectrum derived from SDSS data by \citet{2001AJ....122..549V}.
The host galaxy is selected to have the closest match to
the preferred stellar mass and color at the AGN's redshift, following
the results from \citet{2010ApJ...720..368X}.  Each galaxy hosts at most one
AGN, and no explicit distinction is made between low-luminosity AGN and
quasars that dramatically outshine their host galaxies. The light
curve for each AGN is generated using a damped random walk model and
prescriptions given by \citet{2010ApJ...721.1014M}.

Asteroids are simulated using the Solar System models of \citet{2007AAS...211.4721G}.
They include: Near Earth Objects (NEOs), Main Belt
Asteroids, the Trojans of Mars, Jupiter, Saturn, Uranus, and Neptune,
Trans Neptunian Objects, and Centaurs. Spectral energy distributions
are assigned using the C and S type asteroids of \citet{2009Icar..202..160D}.
Positions for the 11 million asteroids in the simulation
are stored within the base catalog (sampled once per night for the ten
year duration of the LSST survey). We generate
accurate ephemerides of all asteroids falling within a given LSST
point using the $OpenOrb$ software package \citep{2009M&PS...44.1853G}.
With typically 8000 sources per LSST field of view, this
procedure significantly reduces the computational resources
required to simulate asteroid ephemerides.

\subsubsection{Image Simulations}
\label{sec:imsim}

The framework described above provides a parametrized view of the sky
above the atmosphere. Images are simulated using two packages: GalSim
\citep{2015A&C....10..121R}, and Phosim \citep{2015ApJS..218...14P}. Galsim is a
modular and open-source package that provides a library for simulating
stars and galaxies through a range of modern astronomical
telescopes. Point-spread-functions (PSFs) are treated as either
analytic functions or modeled from ray-traced optics. Convolutions by
the PSF can be applied to parameterized galaxy profiles (e.g.\
S\'ersic profiles) or to directly observed images. Operations are
applied in Fourier space to enable an effective trade-off between
speed of simulation and accuracy. GalSim is written in C++ with a
Python API and is integrated within the LSST CatSim framework.

Phosim is an open-source package that simulates images by drawing
photons from the spectral energy distribution of each source (scaled
to the appropriate flux density based on the apparent magnitude of a
source and accounting for the spatial distribution of light for
extended sources). Each photon is ray-traced through the atmosphere,
telescope and camera to generate a CCD image. The atmosphere is
modeled using a Taylor frozen screen approximation (with the
atmosphere described by six layers). The density fluctuations within
these screens are described by a Kolmogorov spectrum with an outer
scale (typically 10\,m to 200\,m). All screens move during an exposure,
with velocities derived from NOAA measurements of the wind velocities
above the LSST site in Chile.  Typical velocities are on the order of
20 m s$^{-1}$, and are found to have a seasonable dependence that is
modeled when generating the screens. Each photon's trajectory is
altered due to refraction as it passes through each screen.

\begin{figure*}
\centerline{\includegraphics[width=0.98\hsize,clip]{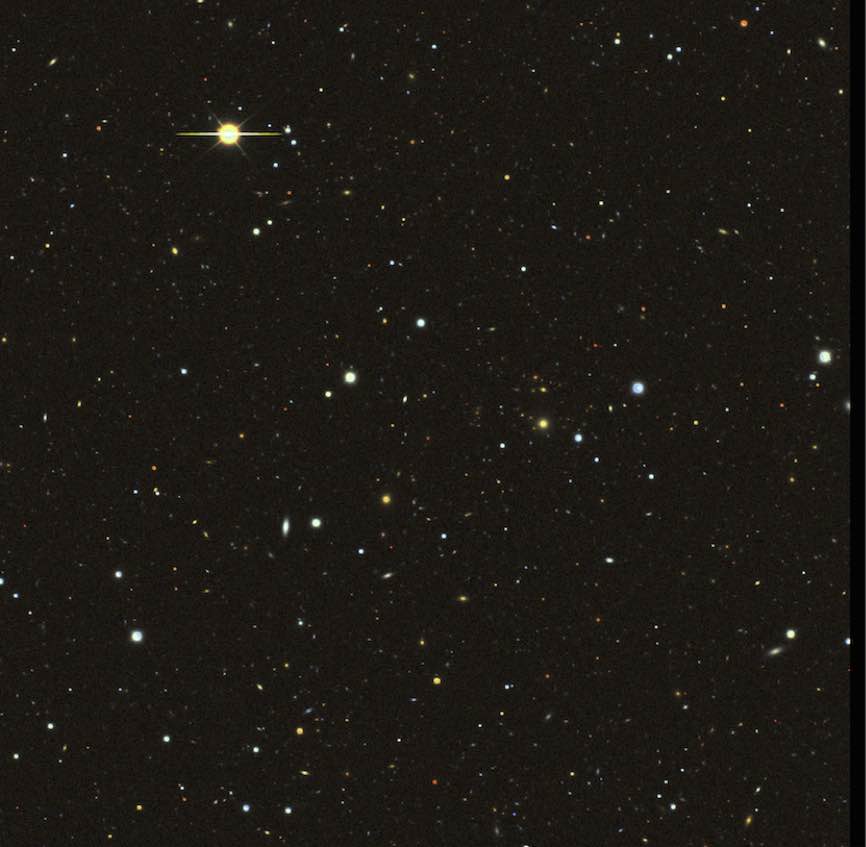}}
\caption{ A simulated image of a single LSST CCD using PhoSim
  (covering a $13.3\times13.3$ arcmin$^2$ region of the sky). The
  image is a color composite \citep{2004PASP..116..133L} from a set of 30
  second $gri$ visits.}
\label{Fig:ImSimExample}
\end{figure*}

After the atmospheric refraction, the photons in PhoSim are reflected
and refracted by the optical surfaces within the telescope and
camera. The mirrors and lenses are simulated using geometric optics
techniques in a fast ray-tracing algorithm and all optical surfaces
include a spectrum of perturbations based on design tolerances. Each
optic moves according to its six degrees of freedom within tolerances
specified by the LSST system. Fast techniques for finding intercepts
on the aspheric surface and altering the trajectory of a photon by
reflection or wavelength-dependent refraction have been implemented to
optimize the efficiency of the simulated images. Wavelength and
angle-dependent transmission functions are incorporated within each of
these techniques, including simulation of the telescope spider.

Both GalSim and PhoSim model the propagation of photons through the
silicon of the detector. The conversion probability, refraction as a
function of wavelength and temperature, and charge diffusion within
the silicon are modeled for all photons. Photons are pixelated and the
readout process simulated including blooming, charge saturation,
charge transfer inefficiency, gain and offsets, hot pixels and
columns, the dependence of the image size on intensity (a.k.a. the
``brighter-fatter'' effect), and QE variations.

An example of a simulated LSST image using PhoSim is shown in
Fig.~\ref{Fig:ImSimExample}.

\section{  ANTICIPATED DATA PRODUCTS AND THEIR CHARACTERISTICS    }
\label{Sec:dataprod}

The LSST observing strategy is designed to maximize the scientific
throughput by minimizing slew and other downtime and by making appropriate
choices of the filter bands given the real-time weather conditions.
Using simulated surveys produced with the Operations Simulator described in \S~\ref{sec:opsim},
we illustrate predictions of LSST performance with two examples.

\subsection{ The Baseline LSST Surveys }
\label{sec:baseline}

The fundamental basis of the LSST concept is to scan the sky deep, wide, and
fast, and to obtain a dataset that simultaneously satisfies the majority
of the science goals. We present here a specific realization, the
so-called ``universal cadence'', which yields the main deep-wide-fast
survey and meets our core science goals.  However, at this writing,
there is a vigorous discussion of cadence plans in the LSST community,
exploring variants and alternatives that enhance various specific
science programs, while maintaining the science requirements described
in the SRD.

The main deep-wide-fast survey
will use about 90\% of the observing time. The remaining 10\% of the observing
time will be used to obtain improved coverage of parameter space such as
very deep ($r\sim26$) observations, observations with very short revisit
times ($\sim$1 minute), and observations of ``special'' regions such as the
Ecliptic plane, Galactic plane, and the Large and Small Magellanic Clouds.

\subsubsection{ The Main Deep-Wide-Fast Survey and its extensions}

\begin{figure}
\includegraphics[width=1.0\hsize,clip]{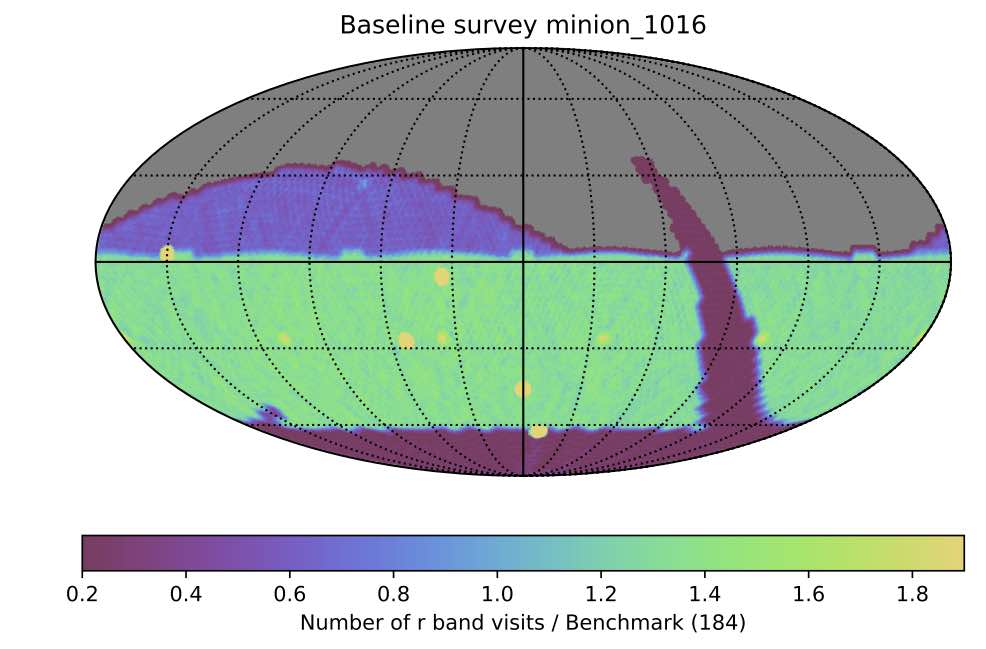}
\caption{The distribution of the $r$ band visits on the sky for a simulated
realization of the baseline cadence. The sky is shown in the equal-area Mollweide
projection in equatorial coordinates (the vernal equinoctial point is in the center, and
the right ascension is increasing from right to left). The number of visits for
a 10-year survey, normalized to the SRD design value of 184, is color-coded according
to the legend. The three regions with smaller number of visits than the main survey
(``mini-surveys'') are the Galactic plane (arc on the right), the region around the
South Celestial Pole (bottom), and the so-called ``northern Ecliptic region'' (upper left;
added in order to increase completeness for moving objects). Deep drilling fields, with
a much higher number of visits than the main survey, are also visible as small circles.
The fields were dithered on sub-field scales and pixels with angular resolution of
$\sim$30 arcmin were used to evaluate and display the coverage.}
\label{Fig:rbandSky}
\end{figure}

The observing strategy for the main survey will be optimized for the homogeneity
of depth and number of visits. In times of good seeing and at low airmass, preference
is given to $r$-band and $i$-band observations. As often as possible, each field will be
observed twice, with visits separated by 15--60 minutes. This strategy will provide motion
vectors to link detections of moving objects in the Solar System, and fine-time sampling
for measuring short-period variability. The ranking criteria also ensure that the
visits to each field are widely distributed in position angle on the sky and
rotation angle of the camera in order to minimize systematic effects in galaxy shape
determination.

The universal cadence provides most of LSST's power for detecting Near Earth
Objects (NEO) and Kuiper Belt Objects (KBOs) and naturally
incorporates the southern half of the ecliptic
within its 18,000 square degrees, with a declination cut of about
$\delta = +2^\circ$.  Additional coverage of a crescent within 10
degrees of the Northern
ecliptic plane would sample the full azimuthal distribution of KBOs,
crucial for understanding the different dynamical families in which
they fall.
Thus, we plan to extend
 the universal cadence to this region using the
$r$ and $i$ filters only, along
with more relaxed limits on airmass and seeing. Relaxed limits on airmass and
seeing are also adopted for $\sim$700 deg$^2$ around the South Celestial
Pole, allowing coverage of the Large and Small Magellanic Clouds
(Fig.~\ref{Fig:rbandSky}).

Finally, the universal cadence proposal excludes observations at low
Galactic latitudes, where the high stellar
density leads to a confusion limit at much brighter magnitudes than those
attained in the rest of the survey. Within this region, the Galactic plane
proposal provides 30 observations in each of the six filters, distributed
roughly logarithmically in time (it may not be necessary to use the
$u$ and $g$ filters for this heavily extincted region).

The resulting sky coverage for the LSST baseline cadence (known internally as
{\it minion\_1016}), based on detailed operations simulations, is shown for the
$r$ band in Fig.~\ref{Fig:rbandSky}. The anticipated total number of visits
for a ten-year LSST survey  is about 2.45 million ($\sim$4.9 million 15-second long
exposures, summing over the six filters). The per-band allocation of
these visits is shown in Table~\ref{tab:baseline}.

The baseline universal cadence is by no means the definitive plan for the entire
survey. Rather, it represents a proof of concept that it is indeed possible to
design a observing strategy which addresses a wide variety of science goals in a nearly
optimal way. With input and engagement of the community, we are
undertaking a vigorous and systematic research effort to explore
the enormously large parameter space of possible surveys (see \citealt{2017arXiv170804058L}).
The scientific commissioning period will be used to test the usefulness of various observing
modes and to explore alternative strategies.

\subsubsection{ Mini-surveys and Deep Drilling Fields}
\label{Sec:minisurveys}

Although the uniform treatment of the sky provided by the universal cadence
proposal can satisfy the majority of LSST scientific goals, roughly 10\%
of the time will be allocated to other strategies that significantly enhance the
scientific return.  These surveys aim to extend the parameter space accessible
to the main survey by going deeper or by employing different time/filter
sampling.  We have already discussed three examples of such
mini-surveys: the Northern Ecliptic Spur to improve completeness of
the asteroid and KBO population, the Southern Celestial Cap to extend
the survey footprint to the South Pole (thus providing coverage of the
Magellanic Clouds), and the Galactic Plane survey to include low
Galactic latitude fields.

As an additional  example of a mini-survey, consider a program that
uses one hour of
observing time per night to observe a single pointing (9.6 deg$^2$) to
substantially greater depth in individual visits. Accounting for
read-out time and filter changes, it could obtain about 50 consecutive
15-second exposures in each of four filters in an hour. If a field is visited
every two days over four months, about 600 deg$^2$ can be observed with this
cadence over 10 years. Taking weather into account, the selected fields would
each have on average about 40 hour-long sequences of 200 exposures each. Each
15-second exposure in a sequence would have an equivalent 5$\sigma$ depth of
$r\sim24$, and each filter subsequence when coadded would be 2 magnitudes
deeper than the main survey visits ($r\sim26.5$). When all 40 sequences and
the main survey visits are coadded, they would extend the depth to $r\sim28$.

This data set would be excellent for a wide variety of science programs. The
individual sequences would be sensitive to 1\% variability on sub-minute time
scales, allowing discovery of planetary eclipses and of interstellar scintillation
effects, expected when the light of a background star propagates through a
turbulent gas medium \citep{2003A&A...412..105M, 2011A&A...525A.108H}.
If these fields were selected
at Galactic latitudes of $|b|\sim30$ deg, they would include about 10 million
stars with $r<21$ observed with signal-to-noise ratio above 100 in each visit.
When subsequences from a given night were coadded, they would
provide dense time sampling to a faint limit of $r\sim26.5$
and would enable deep searches
for SN, trans-Neptunian objects, and other faint transient, moving and
variable sources.  For example, the SN sample
would be extended to redshifts of $z\sim1.2$, with more densely sampled light
curves than obtained from the universal cadence. Such sequences would also
serve as excellent tests of our photometric calibration procedures.

The LSST has already selected four distant extragalactic survey fields\footnote{For
details, see \url{https://www.lsst.org/News/enews/deep-drilling-201202.html}}
that the project guarantees to observe as Deep Drilling Fields with deeper coverage
and more frequent temporal sampling than provided by the standard LSST observing
pattern. These fields (Elias S1, XMM-LSS, Extended Chandra Deep Field-South, and
COSMOS) are  well-studied survey fields with substantial existing multiwavelength
coverage and other positive attributes. These four fields are only the first chosen
for deep-drilling observations.  The project plans a community call
for white papers suggesting additional deep drilling fields and other
specialized observing cadences.

\begin{deluxetable}{|r|r|r|r|r|r|r|}
\tablecaption{The Parameters From Eqs.~\ref{ggg} and \ref{m5} \label{tab:eqparams}}
\tablehead{
\colhead{} & \colhead{$u$}  &   \colhead{$g$}   & \colhead{$r$}   &  \colhead{$i$}  & \colhead{$z$}  & \colhead{$y$}
}
\startdata
$m_\mathrm{sky}$\tablenotemark{a}  & 22.99	& 22.26	 & 21.20	& 20.48	& 19.60	& 18.61	 \\
$\theta$\tablenotemark{b}       	 & 0.81	& 0.77	& 0.73	& 0.71	& 0.69	& 0.68	 \\
$\theta_{eff}$\tablenotemark{c}        & 0.92	& 0.87	& 0.83	& 0.80	& 0.78	& 0.76	 \\
$\gamma$\tablenotemark{d}  	 & 0.038	& 0.039	& 0.039	& 0.039	& 0.039	& 0.039	 \\
$k_m$\tablenotemark{e}   	 	 & 0.491	& 0.213	& 0.126	& 0.096	& 0.069	& 0.170	 \\
$C_m$\tablenotemark{f}   	         & 23.09	& 24.42	& 24.44	& 24.32	& 24.16	& 23.73	 \\
$m_5$\tablenotemark{g}    		 & 23.78	& 24.81	& 24.35	& 23.92	& 23.34	& 22.45	 \\
$\Delta C^{\infty}_m$\tablenotemark{h}   & 0.62	& 0.18	& 0.10	& 0.07	& 0.05	& 0.04	 \\
$\Delta C_m(2)$\tablenotemark{i}          & 0.23	& 0.08	& 0.05	& 0.03	& 0.02	& 0.02	 \\
$\Delta m_5$\tablenotemark{j}              & 0.21	& 0.16	& 0.14	& 0.13	& 0.13	& 0.14	 \\
\enddata
\tablenotetext{^a}{The expected median zenith sky brightness at Cerro Pach\'on (AB mag arcsec$^{-2}$).}
\tablenotetext{^b}{The expected delivered median zenith seeing
  (FWHM, arcsec). The seeing approximately scales with airmass, $X$, as $X^{0.6}$.}
\tablenotetext{^c}{The effective zenith seeing (arcsec) used for $m_5$ computation.}
\tablenotetext{^d}{The band-dependent parameter from Eq.~\ref{ggg}.}
\tablenotetext{^e}{Adopted atmospheric extinction.}
\tablenotetext{^f}{The band-dependent parameter from Eq.~\ref{m5}.}
\tablenotetext{^g}{The typical 5$\sigma$ depth for point sources at zenith, assuming exposure time of
          2$\times$15\,sec, and observing conditions as listed. For larger
          airmass the 5$\sigma$ depth is brighter; see the bottom row.}
\tablenotetext{^h}{The loss of depth due to instrumental noise (assuming 9\,e$^-$ per pixel and readout,
       and two readouts per visit).}
\tablenotetext{^i}{Additive correction to $C_m$ when exposure time is doubled from its fiducial value
          to 60\,sec.}
\tablenotetext{j}{The loss of depth at airmass of $X=1.2$ due to seeing degradation
                 and increased atmospheric extinction.}
\end{deluxetable}

\subsection{  Detailed Analysis of Simulated Surveys  }

As examples of analysis enabled by the Operations Simulator
(\S~\ref{sec:opsim}), we describe
determination of the completeness of the LSST NEO sample, and estimation
of errors expected for trigonometric parallax and proper motion measurements.
In both examples, the conclusions crucially depend on the assumed
accuracy of the photometry and astrometry, as we now describe.

\subsubsection{  Expected Photometric Signal-to-Noise Ratio }

The output of operations simulations is a data stream consisting of
a position on the sky and the time of observation, together with
observing conditions such as seeing and sky brightness. The expected
photometric error in magnitudes (roughly the inverse of the
signal-to-noise ratio) for a single visit can be written as
\begin{equation}
         \sigma_1^2 = \sigma_{sys}^2 + \sigma_{rand}^2,
\end{equation}
where $\sigma_{rand}$ is the random photometric error and $\sigma_{sys}$ is
the systematic photometric error (due to, e.g., imperfect
modeling of the point spread function, but not including uncertainties
in the
absolute photometric zeropoint). The calibration system and procedures
are designed to maintain $\sigma_{sys}<0.005$ mag. Based on
SDSS experience \citep{2007AJ....134.2236S}, the random photometric error for
point sources, as
a function of magnitude, is well described\footnote{Eq.~\ref{ggg} can
be derived from $\sigma_{rand}=N/S$, where $N$ is noise and $S$ is signal,
and by assuming that $N^2 = N_o^2 + \alpha S$. The constants $N_o$ and
$\alpha$ can be expressed in terms of a single unknown constant $\gamma$
by using the condition that $\sigma_{rand}=0.2$ for $m=m_5$.} by
\begin{equation}
\label{ggg}
  \sigma_{rand}^2 = (0.04-\gamma)\, x + \gamma \, x^2 \,\,\, \mathrm{(mag^2),}
\end{equation}
with $x \equiv 10^{0.4\,(m-m_5)}$. Here $m_5$ is the 5$\sigma$ depth (for
point sources) in a given band, and $\gamma$ depends on the sky
brightness, readout noise, etc.
Detailed determination of the system throughput yields the values of $\gamma$
listed in Table~\ref{tab:eqparams}. The 5$\sigma$ depth for point sources is determined from
\begin{eqnarray}
\label{m5}
  m_5 = C_m + 0.50\,(m_{sky}-21) + 2.5\,\log_{10}(0.7/\theta_{eff}) +  \nonumber \\
        + 1.25\,\log_{10}(t_{vis}/30) - k_m(X-1) \phantom{xxxxx}
\end{eqnarray}
where $m_{sky}$ is the sky brightness (AB mag arcsec$^{-2}$), $\theta_{eff}$ is
the seeing (in arcsec), $t_{vis}$ is the exposure time (seconds),
$k$ is the atmospheric extinction coefficient, and $X$ is airmass. Here
the seeing corresponds to the ``effective'' seeing computed from
the seeing FWHM following the procedure described in \cite{2016SPIE.9911E..18A}.
The seeing FWHM in each band is listed in
the second row of Table~\ref{tab:eqparams}, and the effective seeing is listed in the
third row of Table~\ref{tab:eqparams}.

The constants $C_m$ depend on the overall throughput of the instrument
and are computed using our current best throughput estimates for
optical elements and sensors. The resulting $C_m$ values are listed in Table~\ref{tab:eqparams}
and in all six bands they imply single visit depths $m_5$ (also listed
in Table~\ref{tab:eqparams}) that lie between the minimum and design
specification values from the Science Requirements
Document listed in Table~\ref{tab:baseline}.
The differences in performance between LSST and, for example, SDSS
follow directly from these relations\footnote{SDSS data
typically reach a 5$\sigma$ depth for point sources of $r=22.5$
with an effective aperture of $D=2.22$ m, an exposure time of $t_{vis}=54$
sec, the median $r$ band sky brightness of $r_{sky}=20.9$ mag arcsec$^{-2}$,
the median seeing of $\theta=1.5$ arcsec, and the median airmass of $X=1.3$.
In comparison, the LSST loses 0.32 mag in depth due to shorter exposures,
and gains 1.17 mag due to larger aperture, 0.83 mag due to better
seeing, and 0.20 mag due to fainter sky, for a net gain of $\sim$1.9 mag.}.

The structure of eq.~\ref{m5} nicely illustrates decoupling between the system
sensitivity which is fully absorbed into $C_m$ and observing conditions
specified by $m_{sky}$, $\theta$, $t_{vis}$, $k_m$
and $X$. The computation of $C_m$ listed in Table~\ref{tab:eqparams} assumed instrumental noise of
9 e$^-$ per pixel and per readout, whose effect on $m_5$ is
significant only in the $u$ band.
 This loss of depth due to instrumental noise, $\Delta C^{\infty}_m$,
is listed in Table~\ref{tab:eqparams}; it also corresponds to an additive correction to $C_m$ when the
exposure time $t_{vis} \rightarrow \infty$. To predict $5\sigma$ depths for
exposure time $\tau$ times longer than the fiducial $t_{vis} = 30$ sec., the
following correction should be added to the values of $C_m$ listed in Table~\ref{tab:eqparams}:
\begin{equation}
\label{eq:DCm}
 \Delta C_m(\tau) = \Delta C^\infty_m - 1.25\,\log_{10}\left[1 + {10^{(0.8 \, \Delta C^\infty_m)} - 1 \over \tau}  \right].
\end{equation}
By definition, $\Delta C_m(\tau=1)=0$. Again, this effect is only substantial in the $u$
band, as demonstrated by the values of $\Delta C_m(\tau = 2)$ listed in Table~\ref{tab:eqparams}.

The loss of depth at the airmass of $X=1.2$ due to seeing degradation
and increased atmospheric extinction is listed in the last row in Table~\ref{tab:eqparams}. Note
that the limiting depth predictions are uncertain by about 0.1--0.2 mag
due to unpredictable solar
activity (which influences the night sky brightness,
\citealt{2008A&A...481..575P}).

\subsubsection{   The NEO Completeness Analysis    }
\label{Sec:NEOc}
Detailed analyses of the LSST completeness for PHAs and NEOs are
described in \citet{2018Icar..303..181J}, \citet{2017AJ....154...12V,2017AJ....154...13V}, and \citet{2016AJ....151..172G}.
After accounting for differences in their input assumptions and models, each of these independent
works calculates a completeness value which is consistent within a few percent.
Here we briefly summarize the LSST project analysis carried out in \citet{2018Icar..303..181J}; this
approach is roughly the same for each of the studies mentioned above.

To assess the LSST completeness for PHAs, the PHA
population is represented by a sample of orbits taken from the Solar
System model of \citet{2007AAS...211.4721G}.
The simulated baseline survey is used to determine which PHAs are present in
each exposure and at what signal-to-noise ratio they were observed. In
addition to  seeing, atmospheric transparency, and sky background effects
(see eq.~\ref{m5}), the signal-to-noise computation takes into account losses
due to non-optimal detection filters and object trailing. Using mean asteroid reflectance
spectra \citep{2009Icar..202..160D}, combined with the LSST bandpasses,
we calculate expected magnitudes and colors, assuming all PHAs are C type asteroids, of
$V-m = (-1.53, -0.28, 0.18, 0.29, 0.30, 0.30)$ for $m=(u, g, r, i, z, y)$ to transform
standard $V$ band magnitudes to the magnitudes expected in each filter \citep{2001AJ....122.2749I}.
Due to very red $V-u$ colors, and the relatively bright limiting magnitude in the $y$
band, the smallest objects are preferentially detected in the $griz$ bands.
The correction for trailing is implemented by subtracting from the right-hand
side of eq.~\ref{m5}
\begin{eqnarray}
 \Delta m_5^\mathrm{trailing} & = &1.25\,\log_{10}\left(1+0.42 \, x^2 \right) \\
   x & = & {v \,t_{vis} \over 24 \, \theta},
\end{eqnarray}
where the object's velocity, $v$, is expressed in deg.~day$^{-1}$.
For the nominal exposure time of 30 seconds and $\theta=0.7$ arcsec, the loss of limiting
magnitude is 0.04 mag for $v=0.25$ deg.~day$^{-1}$, typical for objects in the main
asteroid belt, and 0.46 mag for $v=1.0$ deg.~day$^{-1}$, typical of PHAs passing
near Earth. PHAs are characterized by their ``absolute magnitude''
$H$, i.e., their apparent magnitude if they were placed 1 AU from
both the Sun and the Earth, with a phase angle of $0^\circ$.  For a
given albedo, $H$ scales directly with diameter of the asteroid.  The PHA
orbits are cloned over an $H$ magnitude distribution with $dN/dH =
10^{\alpha \, H}$,
with $\alpha=0.33$, in order to evaluate completeness as a function of $H$.

An object is considered to be discovered if the object was detected on at least three nights within a
window of 15 days, with a minimum of two visits per night. The same criterion has been used in NASA studies,
and is confirmed as reliable by a detailed analysis of orbital linking and orbit determination using the Moving
Object Processing System (MOPS) code \citep{2017AJ....154...12V,2017AJ....154...13V,2005AAS...20712102J} developed by the
Pan-STARRS project (and adopted by LSST in a collaborative effort with Pan-STARRS). The MOPS software
system and its algorithms are significantly more advanced than anything previously
fielded for this purpose to date. Realistic MOPS simulations show
$>$99\% linking efficiency across all classes of Solar System objects \citep{2013PASP..125..357D},
and at least 93\% efficiency for NEOs \citep{2017AJ....154...12V,2017AJ....154...13V}.

\begin{figure}
\includegraphics[width=1.\hsize,clip]{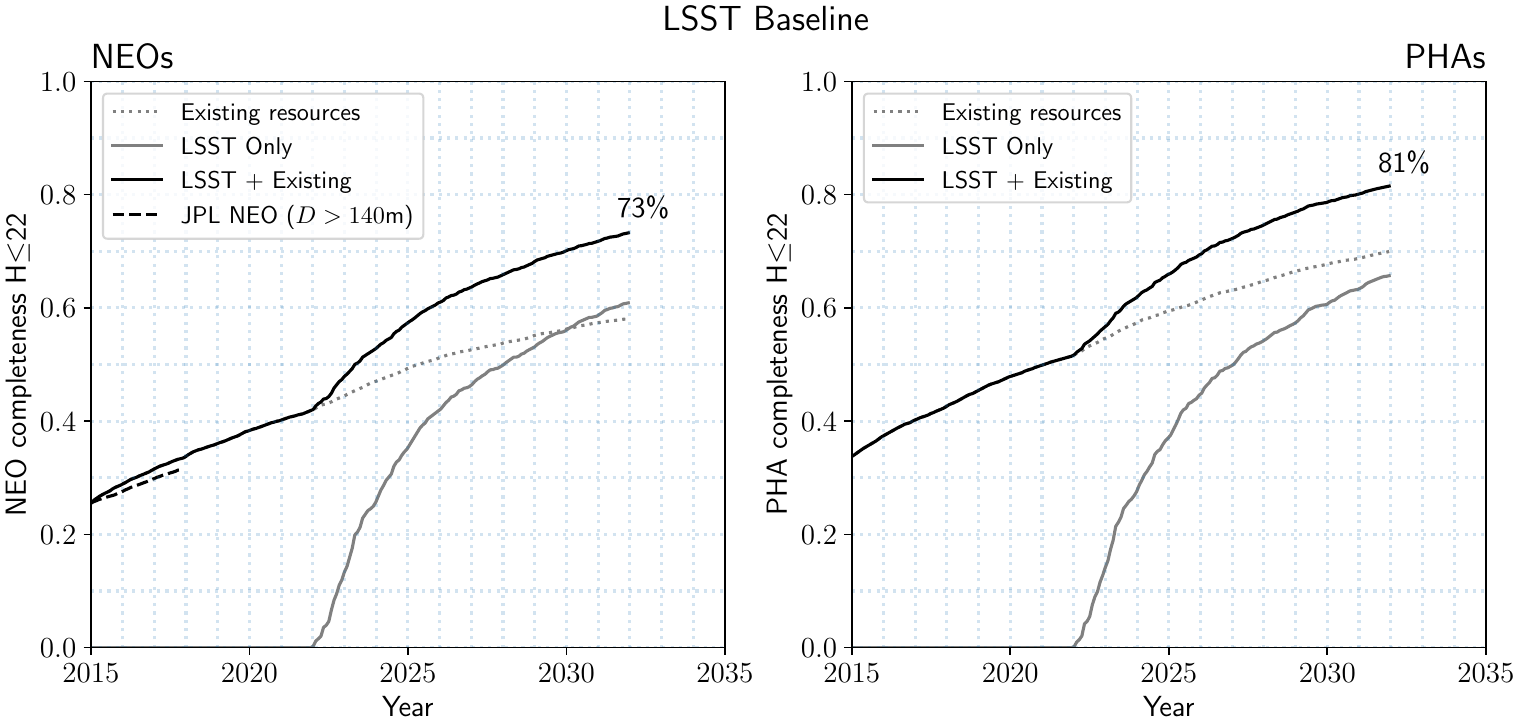}
\includegraphics[width=1.\hsize,clip]{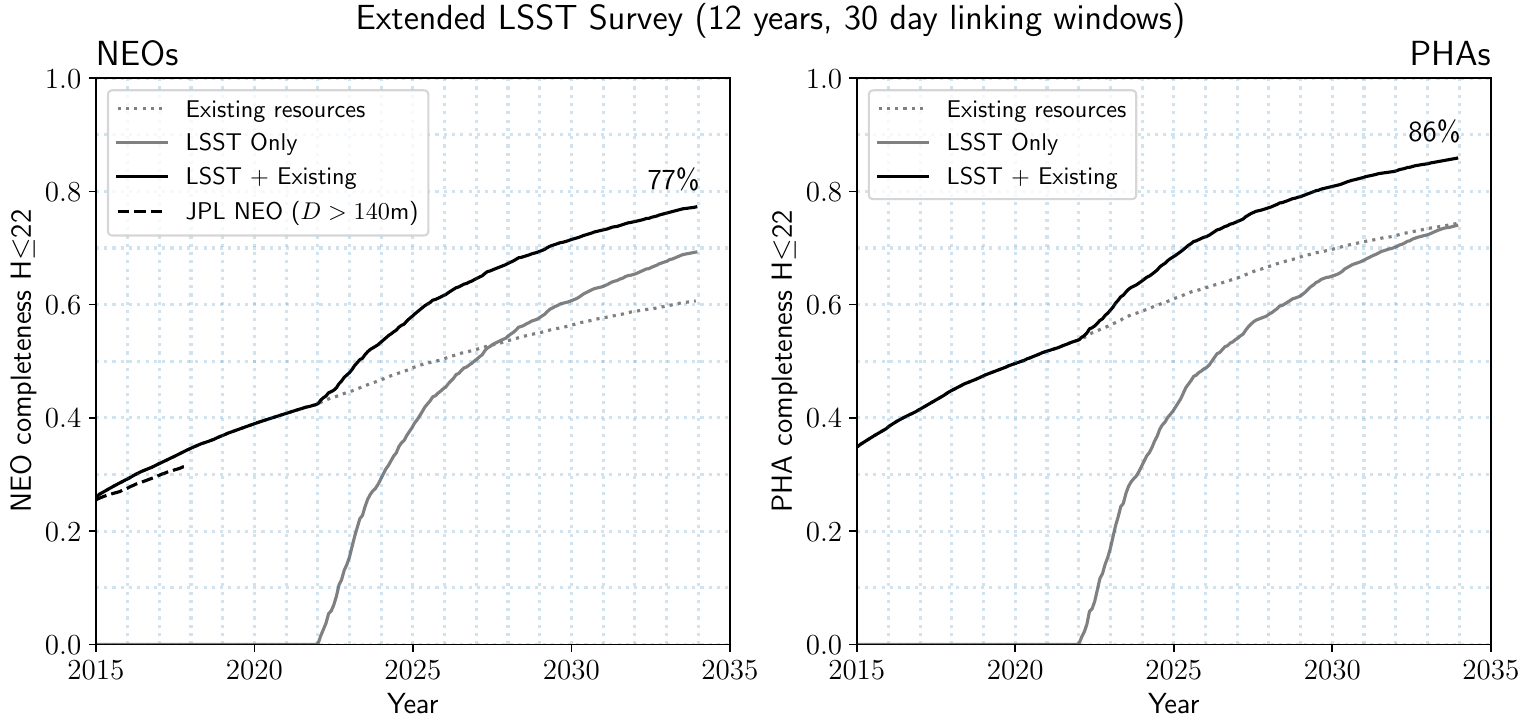}
\caption{Cumulative completeness of the LSST survey for NEOs (left in each panel) and PHAs (right in each panel)
brighter than a given absolute magnitude, $H\le22$ (related to the size of the object and albedo;
$H$=22 mag is equivalent to a typical 140\,m asteroid). The top panel illustrates cumulative completeness
for the LSST baseline cadence and MOPS configuration. In the baseline, LSST alone would discover 66\% of the PHAs
with $H\le22$ (61\% of NEOs); LSST combined with current and on-going
surveys
can discover 81\% of PHAs (73\% of NEOs). The bottom panel illustrates cumulative
completeness when LSST is operated for 12 years, with extra visits around the ecliptic, and when the MOPS linking
window is increased to 30 days from the baseline 15. In this case, LSST alone could discover 74\% of the PHAs with
$H\le22$ (69\% of NEOs); LSST combined with existing resources could discover 86\% of PHAs (77\% of NEOs).
}
\label{Fig:Cneo}
\end{figure}

The LSST baseline cadence discovers 66\% of PHAs and 61\% of NEOs with $H\leq22$ (equivalent to $D\ge140$\,m)
after 10 years of operations \citep{2018Icar..303..181J}.  This cadence spends 6\% of the total
observing time on NEO-optimized observations north of $\delta = +5^\circ$, and MOPS links objects with windows of 15 days.
The baseline survey cumulative completeness as a function of time for objects with $H\le22$ is shown in the upper panel of Fig.~\ref{Fig:Cneo},
both with and without including contributions from current and
on-going surveys. These figures are likely to be uncertain at the level of $\pm5\%$ due
to uncertainties in the orbital distribution of the true population, the size distribution, uncertain distributions of shape
(and thus light curve variations) and surface properties (thus colors and albedo), plus variations in survey cadence due to
weather, etc.

Various adjustments to the baseline cadence and MOPS can boost the completeness for $H\le22$ PHAs.
By improving MOPS and increasing the MOPS linking window from 15 to 30 days we can boost completeness
by about 3\%. By running the survey for an additional two years, we can boost completeness by another 4\%.
Considering this `extended' LSST in the context of existing/ongoing surveys would result in a system-wide cumulative completeness of 86\% for PHAs (77\% for NEOs), approaching the 90\% required by the Congressional mandate (see
lower panels of Fig.~\ref{Fig:Cneo}).

\subsubsection{The Expected Accuracy of Trigonometric Parallax and Proper Motion Measurements }
\label{sec:astrom}

To model the astrometric errors, we need to consider both random and
systematic effects.
Random astrometric errors per
visit for a given star are modeled as $\theta/SNR$, with $\theta=700$ mas and $SNR$ determined using
eq.~\ref{m5}.
Systematic
errors of 10 mas are added in quadrature, and are assumed to be {uncorrelated}
between different observations of a given object. Systematic and random
errors become similar at about $r=22$, and there are about 100 stars per LSST
sensor (0.05 deg$^2$) to this depth (and fainter than the LSST saturation limit at
$r\sim16$) even at the Galactic poles.

HSC data from the Subaru telescope reduced with the LSST software stack indicate that systematic errors of
10 mas on spatial scales of several arcminutes are realistic even at this stage of maturity of the
code; results reported by DES \citep{2017PASP..129g4503B} indicate astrometric residuals of $\sim 7\,$mas for 30\,s exposures
in a 4m, with scope for further improvements from denser astrometric standard grids. Even a drift-scanning
survey such as SDSS delivers uncorrelated systematic errors (dominated by seeing
effects) at the level of 20-30 mas \citep[measured from repeated scans;][]{2003AJ....125.1559P};
the expected image quality for LSST will be twice as good as for SDSS. Furthermore,
there are close to 1000 galaxies per sensor with $r<22$, which will provide exquisite
control of systematic astrometric errors as a function of magnitude, color and other
parameters, and thus enable absolute proper motion measurements.

\begin{deluxetable}{l|c|c|c|c|c}
\tablecaption{The expected proper motion, parallax and accuracy for a 10-year long baseline survey.\label{tab:tenyear}}
\tablehead{
\colhead{$r$} &  \colhead{$\sigma_{xy}$\tablenotemark{a}} & \colhead{$\sigma_\pi$\tablenotemark{b}} & \colhead{$\sigma_\mu$\tablenotemark{c}} & \colhead{$\sigma_1$\tablenotemark{d}} &  \colhead{$\sigma_C$\tablenotemark{e}} \\
\colhead{mag} & \colhead{mas} & \colhead{mas} & \colhead{mas/yr} & \colhead{mag} & \colhead{mag}
}
\startdata
       21 &  11  &  0.6  &  0.2   &   0.01  &   0.005 \\
       22 &  15  &  0.8  &  0.3   &   0.02  &   0.005 \\
       23 &  31  &  1.3  &  0.5   &   0.04  &   0.006 \\
       24 &  74  &  2.9  &  1.0   &   0.10  &   0.009 \\
\enddata
\tablenotetext{a}{Typical astrometric accuracy (rms per coordinate per visit).}
\tablenotetext{b}{Parallax accuracy for 10-year long survey.}
\tablenotetext{c}{Proper motion accuracy for 10-year long survey.}
\tablenotetext{d}{Photometric error for a single visit (two 15-second exposures).}
\tablenotetext{e}{Photometric error for coadded observations (see Table~\ref{tab:baseline}).}
\end{deluxetable}

Given the observing sequence for each sky position in the main survey
as produced by the Operations Simulator (\S~\ref{sec:opsim}), we
generate a time sequence of mock astrometric measurements, with random
and statistical errors modeled as described above.
The astrometric transformations for a given CCD and exposure, and
proper motion and parallax for all the stars from a given CCD, are simultaneously
solved for using an iterative algorithm. The astrometric transformations from
pixel to sky coordinates are modeled using low-order polynomials and standard
techniques developed at the U.S.\ Naval Observatory \citep{2003AJ....125..984M}. The expected
proper motion and
parallax errors for a 10-year long baseline survey, as a function of apparent
magnitude, are summarized in Table~\ref{tab:tenyear}. Blue stars (e.g., F/G stars) fainter than
$r\sim23$ will have about 50\% larger proper motion and parallax errors than
given in the table due to decreased numbers of $z$ and $y$ band detections. The
impact on red stars is smaller due to a relatively small number of observations
in the $u$ and $g$ bands, but extremely red objects, such as L and T dwarfs,
will definitely have larger errors, depending on details of their spectral
energy distributions.  After the first three years of the survey,
{the proper motion errors will be about five times as large, and parallax
errors will be about twice as large,} as the values given in Table~\ref{tab:tenyear}; the errors
scale as $t^{-3/2}$ and $t^{-1/2}$, respectively. This error behavior is
a strong independent argument for a survey lifetime of at least 10 years
(c.f. \S~\ref{Sec:refdesign}).

For comparison with Table~\ref{tab:tenyear}, the SDSS-POSS proper motion measurements have an
accuracy of $\sim$5 mas yr$^{-1}$ per coordinate at $r=20$ \citep{2004AJ....127.3034M}. Gaia
is expected to deliver parallax errors of 0.3 mas and proper motion errors of
0.2 mas yr$^{-1}$ at its faint end at $r\sim20$ \citep{2001A&A...369..339P}. Hence, LSST will smoothly
extend Gaia's error vs.\ magnitude curve 4 magnitudes fainter (for illustration,
see fig.~21 in \citealt{2012ARA&A..50..251I}).

\subsection{             Data Products and Archive Services          }
\label{Sec:dp}

Data collected by the LSST telescope and camera will be automatically processed to \emph{data products} -- catalogs, alerts,
and reduced images -- by the LSST Data Management system
(\S~\ref{sec:dm}). These products are designed to
enable a large majority of LSST science cases, without the need to
work directly with the raw pixels.  We give a high-level overview of
the LSST data products here; further details may be found in the LSST
Data Products Definition Document \citep{LSE-163}.

Two major categories of data products will be produced and delivered by LSST DM:
\begin{itemize}
\item \textbf{  Prompt products\footnote{Historically, these have been referred to as ``Level
1 Data Products'', but going forward we prefer to use the more descriptive
\emph{Prompt Products} designation. Note that the old terminology is still
in use in present-day LSST documents and code; new and updated
documents will gradually transition to the new, descriptive, nomenclature
used in this paper.
}}, designed to support the discovery,
  characterization, and rapid follow-up of time-dependent phenomena
  (``transient science''). These will be generated continuously every
  observing night, by detecting and characterizing sources in images
  differenced against deep templates. They will include alerts to
  objects that were newly discovered, or have changed brightness or
  position at a statistically significant level. The alerts to such
  events will be published within 60   seconds of observation; we
  expect several million alerts per night.\\
\\
In addition to transient science, the prompt products will support
discovery and follow-up of objects in the Solar System. Objects with
motions sufficient to cause trailing in a single exposure will be
identified and flagged as such when the alerts are broadcast. Those
that are not trailed will be identified and linked based on their
motion from observation to observation, over a period of a few
days. Their orbits as derived by MOPS will be published within 24 hours of
identification. The efficiency of linking (and thus the completeness
of the resulting orbit catalog) will depend on the final observing
cadence chosen for LSST, as well as the performance of the linking
algorithm (\S~\ref{Sec:NEOc}).
\item \textbf{Data release products\footnote{These have been referred to as ``Level
2 Data Products'' in the past; as with their ``Level 1'' counterparts, we
will use the more descriptive nomenclature going forward.}} are
  designed to enable systematics- and flux-limited science, and will
  be made available in annual Data Releases\footnote{The first-year
    data will probably be split into two data releases.}. These will include the (reduced and raw) single-epoch images, deep coadds of the observed sky, catalogs of objects detected in LSST data, catalogs of sources (the detections and measurements of objects on individual visits), and catalogs of ``forced sources'' (measurements of flux on individual visits at locations where objects were detected by LSST or other surveys). LSST data releases will also include fully reprocessed prompt products, as well as all metadata and software necessary for the end-user to reproduce any aspect of LSST data release processing.\\
\\
A noteworthy aspect of LSST data release processing is that it will largely
rely on \textbf{multi-epoch model fitting}, or \textbf{\emph{MultiFit}}, to
perform near-optimal characterization of object properties. That is,
while the coadds will be used to perform object \emph{detection}, the
\emph{measurement} of their properties will be performed by
simultaneously fitting (PSF-convolved) models to all single-epoch
observations. It is not yet clear to what extent we will be able to make some of these
measurements on suitable linear combinations of input images (with careful propagation
of PSFs and noise).
An extended source model -- a constrained linear
combination of two S\'ersic profiles -- and a point source model with
proper motion -- will generally be
fitted to each detected object\footnote{For performance reasons, it is
  likely that only the point source model will be fitted in the most
  crowded regions of the Galactic plane.}.\\
\\
Secondly, for the extended source model fits, the LSST will
characterize and store the shape of the associated likelihood surface
(and the posterior), and not just the maximum likelihood values and
covariances. The characterization will be accomplished by sampling,
with up to $\sim$200 (independent) likelihood samples retained for
each object. For storage cost reasons, these samples
may be retained only for those bands of greatest interest for
weak lensing studies.

\end{itemize}

As described in \S~\ref{Sec:minisurveys}, approximately 10\% of the
observing time will be devoted to mini-surveys that do not follow the
LSST baseline cadence. The data products for these programs will be
generated using the same processing system and will be released on the
same timescale as the rest of the survey; any specialized processing
that these require will be the responsibility of the community.

While a large majority of science cases will be adequately served by prompt
and data release products, more specialized investigations may benefit from
custom, user-created, products derived from the LSST data.  These could be
new catalogs created by simple post-processing of the LSST data release
catalogs, entirely new data products generated by running custom code on raw
LSST imaging data, or something in-between.  We will make it possible for the
end-users to create (or use) such \textbf{user-generated\footnote{Formerly known as
``Level 3 Data Products''.}} products at the LSST Data Facility,
using the services offered within the LSST Science Platform (\S~\ref{sec:lsp}).

\subsubsection{The LSST Science Platform}
\label{sec:lsp}

The LSST Science Platform \citep{LSE-319} represents LSST's vision for a
large-scale astronomical data archive that can enable effective research
with datasets of LSST size and complexity.  It builds on recent trends in
remote data analysis, and practical experiences in the astronomical context
gathered by projects such as the JHU
SciServer \citep{2017AAS...22923615R}, Gaia GAVIP \citep{2016SPIE.9913E..1VV}, or NOAO Datalab \citep{2016SPIE.9913E..0LF}.

The LSST Science Platform will be a set of web applications (portals) and
services through which the users will access the LSST data products and, if
desired, conduct remote analyses or create user generated products.  The
platform makes this possible through three user-facing \emph{aspects}:
\begin{itemize}

\item The web \textbf{Portal}, designed to provide the essential data access and
visualization services through a simple-to-use website.  It will enable
querying and browsing of the available datasets in ways the users are
accustomed to at archives such as IRSA, MAST, or the SDSS archive.

\item The \textbf{JupyterLab} aspect, that will provide a
Jupyter\footnote{\url{http://jupyter.org/}} Notebook-like
interface and is geared towards enabling next-to-the-data remote analysis.
A large suite of commonly used astronomical software, including the LSST
software stack (\S~\ref{sec:dmstack}), will be made available through this
interface.  The user experience will be nearly identical to working with
Jupyter notebooks locally, except that computation and analysis will occur
with resources provided at the LSST Data Access Center.  This is an
implementation of the ``bringing computation to the data'' paradigm: rather
than imposing the burden of downloading, storing, and processing
(potentially large) subsets of LSST data at their home institutions, we make
it possible for the users to bring their codes and perform analyses at the
LSST DAC.  This reduces the barrier to entry and shortens the path to
science for the LSST science community.

\item The \textbf{Web API} aspect will expose the LSST DAC services to other
software tools and services using commonly accepted formats and protocols\footnote{For
example, industry-standard protocols such as WebDAV may be used to expose
file data, or Virtual Observatory protocols such as TAP and SIAP may
be used for access to catalogs and images respectively.}.  This interface will open the
possibility for remote access and analysis of the LSST data set using
applications that the users are already comfortable with such as
TOPCAT \citep{2005ASPC..347...29T}, or libraries such as Astropy \citep{2013A&A...558A..33A,2016SPIE.9913E..0GJ}.  Furthermore, the offered APIs will allow
for federation with other astronomical archives, bringing added value to the
LSST dataset.
\end{itemize}

Approximately 10\% of the total budget for the LSST Data Facility
compute and storage capacity has been reserved for the LSST Science Platform
needs, and to be shared by all LSST DAC users.  Based on the current plans and
technology projections, these equate to approximately 2,400 cores, 4 PB of
file storage, and 3 PB of database storage at the beginning of LSST
operations (in 2022).

\begin{figure}
\begin{center}
\includegraphics[width=0.47\textwidth,clip]{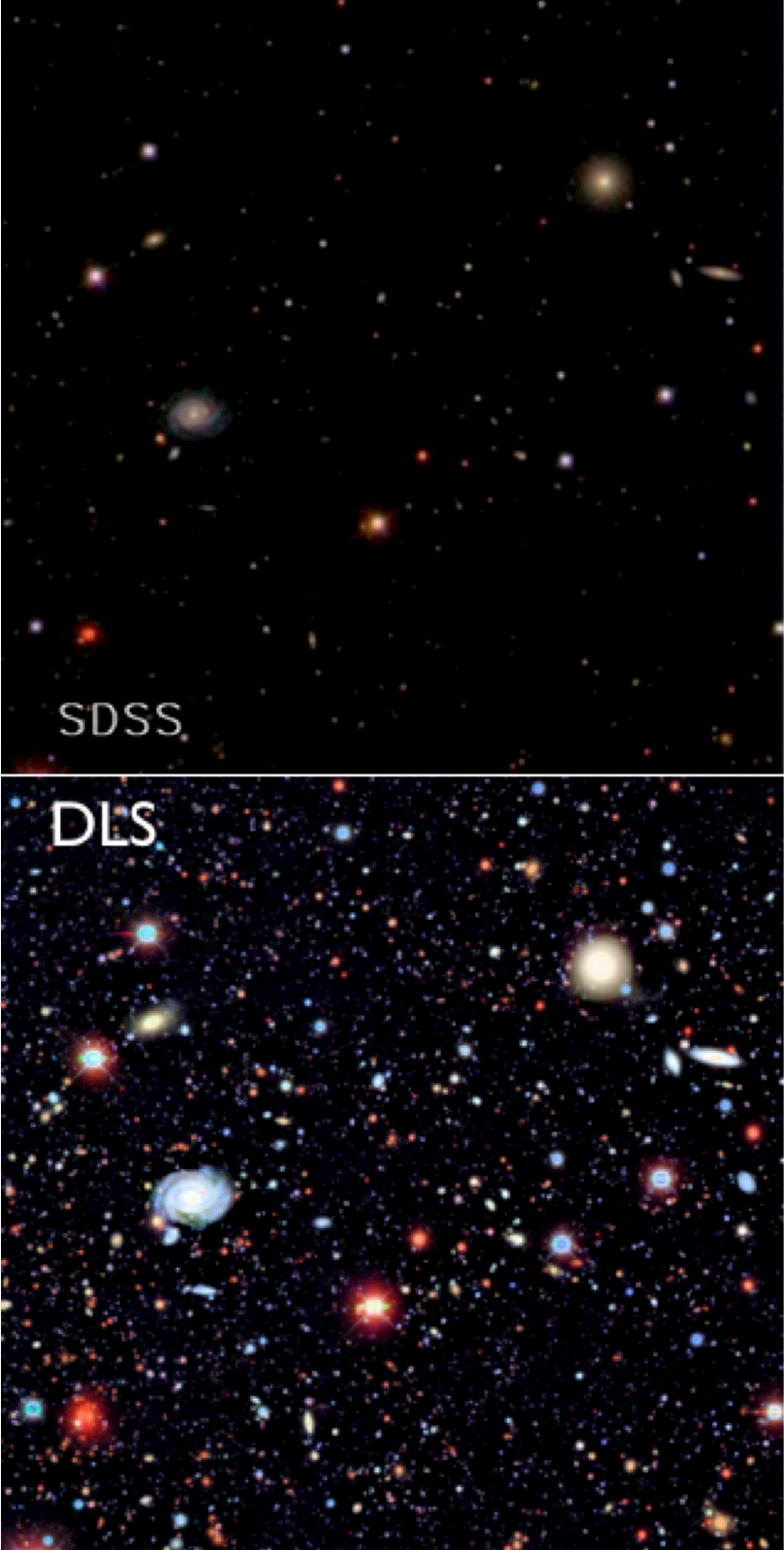}
\end{center}
\caption{A comparison of $\sim7.5\times7.5$ arcmin$^2$ images of
the same area of sky (centered on $\alpha$=9$^h$ 20$\arcmin$ 47$\arcsec$ and
$\delta$=30$^\circ$ 8$\arcmin$ 12$\arcsec$) obtained by the SDSS (top, $r<22.5$) and
the Deep Lens Survey (bottom, $r<24.5$). These are gri composites,
colorized following \citet{2004PASP..116..133L}.  The depth gain for the bottom image
is mostly due to the lower surface brightness limit, which is also responsible
for the apparent increase of galaxy sizes. LSST will obtain $\sim$100 $gri$
color images to the same depth ($\sim$200 for the $riz$ composites) of each point
over half the Celestial sphere (18,000 deg$^2$, equivalent to 1.15 million $\sim7.5\times7.5$
arcmin$^2$ regions), and with better seeing. After their coaddition, the final
image will be another $\sim3$ mag deeper (a faint limit of $r=27.5$ for point
sources).}
\label{Fig:panels1}
\end{figure}

\begin{figure}
\begin{center}
\includegraphics[width=0.47\textwidth,clip]{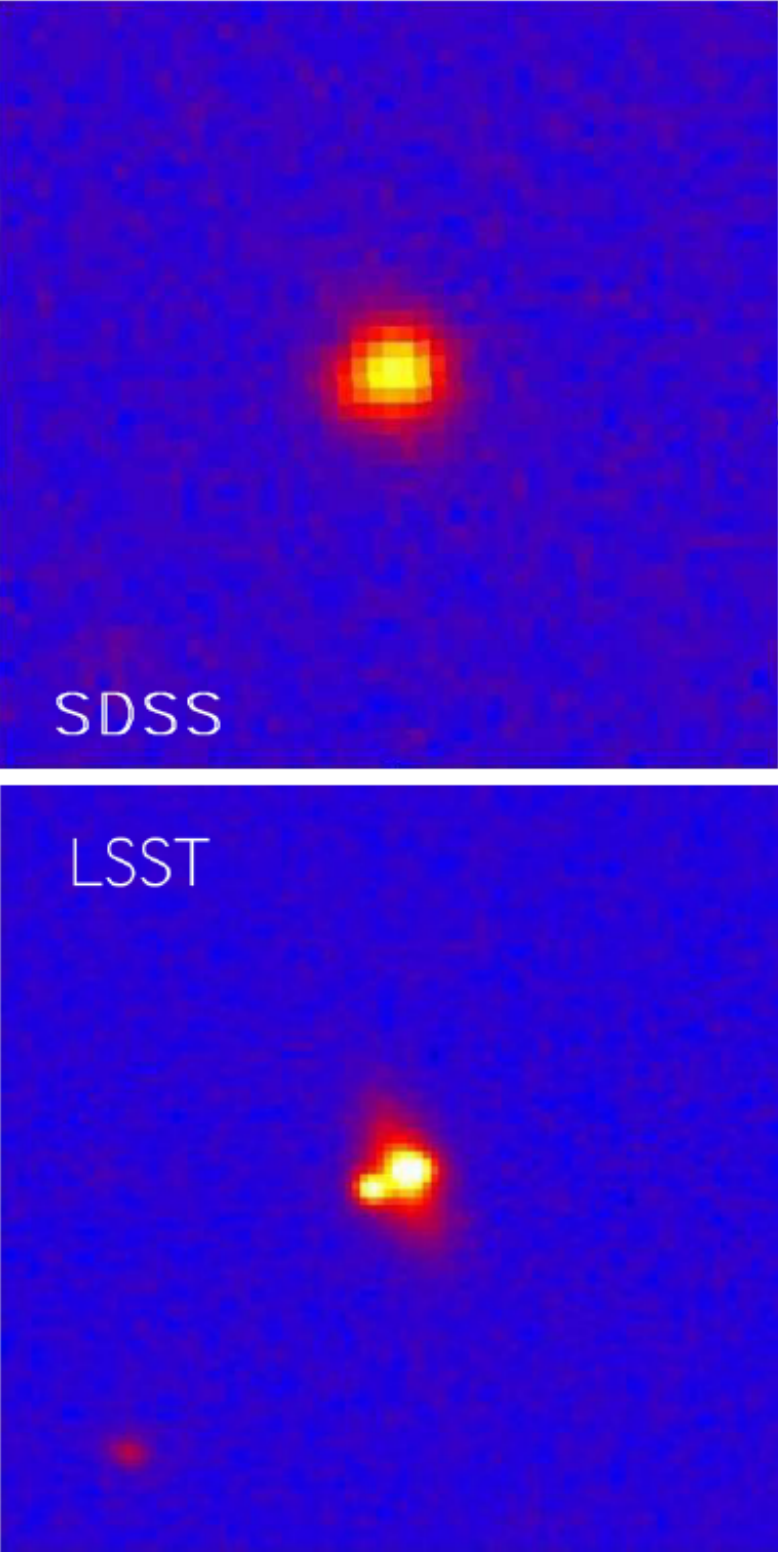}
\end{center}
\caption{A comparison of angular resolution for $20\times20$ arcsec$^2$ images obtained
by the SDSS (top, median seeing of 1.5 arcsec) and expected from LSST (bottom,
seeing of 0.7 arcsec). The images show a lensed SDSS quasar
\citep[SDSS J1332+0347,][]{2007AJ....133..214M}; the bottom image was taken with Suprime-cam at Subaru.
Adapted from \citet{2007AAS...21113707B}.}
\label{Fig:panels2}
\end{figure}

\section{EXAMPLES OF LSST SCIENCE PROJECTS}
\label{Sec:science}

The design and optimization of the LSST system leverages its unique capability
to scan a large sky area to a faint flux limit in a short amount of time.
The main product of the LSST system will be a multi-color $ugrizy$ image of about
half the sky to unprecedented depth ($r\sim27.5$). For a comparison,
one of the best
analogous contemporary datasets is that of SDSS, which provides $ugriz$ images
of about a quarter of the sky to $r\sim22.5$, with twice as large seeing
(see Figs.~\ref{Fig:panels1} and \ref{Fig:panels2}). A major advantage of LSST
is the fact that this deep sky map will be produced by taking hundreds of
shorter exposures (see Table~\ref{tab:baseline}). Each sky position within the main survey area
will be observed close to 1000 times, with time scales spanning seven orders of
magnitude (from 30 sec to 10 years), and produce roughly \textit{thirty
trillion photometric measures} of celestial sources.

It is not possible to predict all the science that LSST data will enable.
We now briefly discuss a few projects to give a flavor of anticipated studies,
organized by the four science themes that drive the LSST design
(although some projects span more than one theme).
For an in-depth discussion of LSST science cases, we refer the reader to the
LSST Science Book, and more specialized documents discussing cosmology
\citep{2012arXiv1211.0310L, 2018RPPh...81f6901Z}, galaxy science
\citep{2017arXiv170801617R}, and synergy with other ground-based and
space-based facilities
\citep{2016arXiv161001661N,2015arXiv150107897J,2017ApJS..233...21R}.

\subsection{Probing Dark Energy and Dark Matter }

\begin{figure}
\includegraphics[width=1.0\hsize,clip]{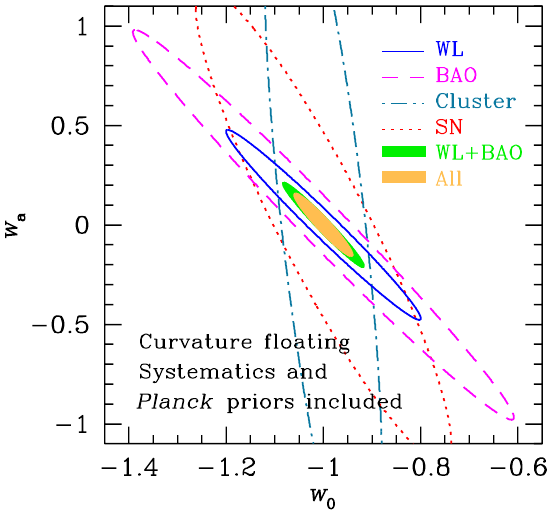}
\caption{
Constraints on the dark energy equation of state ($w = w_0 +
w_a(1-a)$) from LSST cosmological probes.  The various ellipses assume
constraints from BAO (dashed line), cluster counting (dash-dotted line),
supernovae (dotted line), WL (solid line), joint BAO and WL
(green shaded area), and all probes combined (yellow shaded area).
The BAO and WL results are based on galaxy--galaxy, galaxy--shear,
and shear--shear power spectra only.
Adding other probes such as strong lensing time delay
and higher-order galaxy and shear statistics will further improve
the constraints.  While the details of the contours will change
slightly as the survey parameters are updated,
the key point remains that
this combination of dark energy probes results in contours with different degeneracy directions, and
hence their combination results in tight constraints on the dark energy equation of state. }
\label{Fig:DEellipses}
\end{figure}

A unique aspect of LSST as a probe of dark energy and dark matter is
the use of multiple cross-checking probes that reach unprecedented
precision (see Fig.~\ref{Fig:DEellipses}). Any given probe constrains
degenerate combinations of cosmological parameters, and each probe is
affected by different systematics, thus the combination of probes
allows systematics to be calibrated and for degeneracies to be
broken.  Dark energy manifests itself in two ways.  The first is the
relationship between redshift and distance (the Hubble diagram), or
equivalently the expansion rate of the Universe as a function of
cosmic time.  The second is the rate at which matter clusters with
time.
Structure formation involves a balance between
gravitational attraction of matter over-densities and the rapid
expansion of the background.  Thus, quantifying the rate of growth of
structures from early times until the present provides additional
tests of the energy contents of the Universe and their interactions.

The joint analysis of LSST weak lensing and galaxy clustering is
particularly powerful in constraining the dynamical behavior of dark
energy, i.e., how it evolves with cosmic time or redshift \citep{2004PhRvD..70D3009H,2006JCAP...08..008Z}.  By
simultaneously measuring the growth of large-scale structure, and
luminosity and angular distances as functions of redshift (via weak
lensing, LSS, SN, and cluster counting), LSST data can reveal whether
the recent cosmic acceleration is due to dark energy or modified
gravity \citep{2004PhRvD..69D4005L,2006PhRvD..74B3512K,2006PhRvD..74d3513I,2008PhRvD..78f3503J,2011PhRvD..83b3008O,2013ApJ...779...39J,2013PhR...530...87W}.
The Dark Energy Survey \citep[see e.g.,][and references therein]{2017arXiv170801530D} provides a compelling proof of concept for this program.

Over a broad range of accessible redshifts, the simple linear model
for the dark energy equation of state ($w = w_0 + w_a(1-a)$) is a poor representation of more
general dark energy theories. \citet{2008PhRvD..78d3528B} showed that in a high-dimensional dark energy model space,
LSST data could lead to a hundred- to thousand-fold increase in precision over its
precursor experiments, thereby confirming its status as
a premier Stage IV experiment in the sense of \citet{2006astro.ph..9591A}.

The power and accuracy of LSST dark energy and dark matter probes are
a result of the enormous samples that LSST will have, including
several billion galaxies and millions of Type Ia
supernovae. At $i < 25.3$ (SNR${}>20$ for point sources), the
photometry of galaxies will be of high enough quality to provide
photometric redshifts with an RMS accuracy ($\sigma/(1+z)$) of 2\%
over the range $0.3 < z < 3.0$ (only 10\% of the sample will have redshift errors larger than 6\%).
This number represents a requirement on the accuracy of the photometry at delivering photometric
redshifts given known templates for the SEDs.  The degradation in photometric redshift quality associated with
requiring more training data than currently exists to define the template set increases the
expected $\sigma/(1+z)$ to $\sim$0.05 \cite[e.g.][]{2015APh....63...81N,2018AJ....155....1G}, which is still well within the
expected range for a Stage IV dark energy experiment.  The
sample to $i=25.3$ will include several billion galaxies.  At a
slightly brighter cut, there will be around 30 galaxies arcmin$^{-2}$
with shapes measured well enough for weak lensing measurements \citep{2013MNRAS.434.2121C,2015MNRAS.447.1746C},
with the number realized in practice being dependent on
the performance of the deblending and shape measurement algorithms.
The median redshift for
this sample will be $z\sim$1.2, with the third quartile at $z\sim2$.
It will be possible to further improve photometric redshift calibration
by cross-correlating the photometric sample with redshift surveys of
galaxies and quasars in the same fields
\citep{2008ApJ...684...88N,2010ApJ...721..456M,2013arXiv1303.4722M,2017arXiv171002517D}.

\begin{figure}
\includegraphics[width=1.0\hsize,clip]{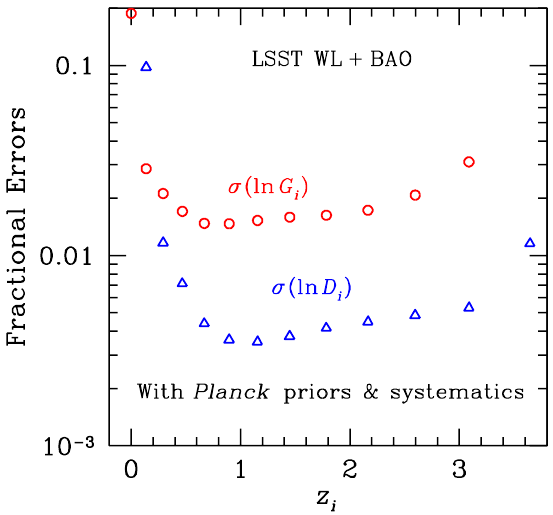}
\caption{Marginalized $1\sigma$ errors on the comoving distance
(open triangles) and growth factor (open circles) parameters from
the joint analysis of LSST LSS and WL (galaxy--galaxy, galaxy--shear,
and shear--shear power spectra) with a
conservative level of systematic uncertainties in the photometric redshift error
distribution and additive and multiplicative errors in the shear and
galaxy power spectra. The maximum multipole used for WL is
2000, and that for LSS is 3000 [with the additional requirement that
$\Delta_\delta^2(\ell/D_{A};z) < 0.4$].
The growth parameters
are evenly spaced in
$\log(1+z)$ between $z = 0$ and 5, and the distance parameters
start at $z_1 = 0.14$.
The error of each distance (growth) parameter is marginalized
over all the other parameters including growth (distance) parameters. The joint constraints on
distance are relatively insensitive to the assumed systematics
\citep{2009ApJ...690..923Z}.}
\label{Fig:bao2}
\end{figure}

The main LSST observables in the context of dark energy and matter are described below.

\begin{itemize}
\item The
joint analysis of shear--shear, galaxy--shear, and galaxy--galaxy
correlation functions has become standard in analyses of precursor datasets
\cite[e.g.][]{2017arXiv170801530D,2018MNRAS.474.4894J}. WL and LSS are highly complementary probes, and the combination
of their auto- and cross-correlations will constrain the properties of the late-time accelerated expansion while providing
internal cross-checks for marginalizing over systematic uncertainties \cite[e.g.,][]{2017arXiv171003235M}.
These measurements consist of the two-point auto- and cross-correlations of shear and positions for billions of galaxies across $\sim 10$ redshift bins.
As described in the following two items, the galaxy-galaxy and galaxy-shear correlation functions provide additional probes of dark energy and dark matter.
\item The galaxy--shear correlation function probes the growth of dark matter large-scale structure and is a
diagnostic of the underlying cosmology. The combination with the
galaxy--velocity correlation function estimated from currently planned
spectroscopic surveys
could test General Relativity and its variants at high redshift \citep{2010Natur.464..256R}.
\item The galaxy--galaxy correlation function is vital to constrain the galaxy bias impacting the galaxy-shear correlation and is therefore
an essential component in the joint analysis of LSS and WL. In addition, the presence of
Baryon Acoustic Oscillations in the galaxy angular correlation functions is a strong cosmological
probe on its own. The sound horizon at decoupling, which is imprinted on the
mass distribution at all redshifts and calibrated with the CMB, provides a standard ruler to measure the angular diameter
distance as a function of redshift
\citep{1998ApJ...504L..57E,2001ApJ...557L...7C,2003ApJ...594..665B,2003PhRvD..68f3004H,2003PhRvD..68h3504L,2003ApJ...598..720S}.
LSST photo-$z$ BAO will achieve percent-level precision on the angular
diameter distance at $\sim$10 redshifts logarithmically spaced between $z = 0.4$ to 3.6. The combination with CMB
and weak lensing (WL) shear yields tight constraints on the
dynamical behavior of dark energy (Fig.~\ref{Fig:bao2}). In particular, high-redshift BAO data can break
the degeneracy between curvature and dark energy, constraining $\Omega_k$ to within
0.001.
\item Higher-order shear and galaxy statistics and shear peak counts can improve dark energy
constraints and provide self-calibration of various systematics
\citep{2004MNRAS.348..897T,2006MNRAS.366..884D,2006MNRAS.366..101H,2016PhRvD..94f3534P}. They are also probes of both
primordial non-Gaussianities and those caused by non-linear structure.
\item Primordial non-Gaussianity is also probed by the large-scale power of any biased tracer of the matter
overdensities \citep{2008PhRvD..77l3514D}. Although measurements of the galaxy power spectrum on very large scales
are challenging due to sky systematics \citep{2014PhRvL.113v1301L} and cosmic variance, the prospect of using
multiple tracers of the same field could significantly improve the constraining power for this observable
\citep{2009PhRvL.102b1302S}. Similar measurements of the large-scale power will also be used to test phenomenological
models of clustering dark energy \citep{2006PhRvD..74d3505T}.
\item Similarly, weak lensing magnification tomography \citep{2012MNRAS.426.2489M} offers a
complementary probe of a mix of cosmic geometry and growth of dark matter structure.
\item The two LSST observing programs are complementary in the supernova samples they will provide. The main survey will
obtain light curves in six bands and photometric redshifts of about 400,000 photometrically-classified Type
Ia supernovae that can be used for cosmological distance measurements, with further spectroscopic
follow-up of a sub-sample of their host galaxies.
Such a sample will not only provide larger statistics
for the study of the Type Ia population in the universe, but will also
be spread across the full 18,000 deg$^2$ LSST main survey footprint, allowing different probes of the large scale structure of the low redshift
universe. This sample of supernovae can be used as a tracer of large scale structure by directly probing the
gravitational potential of structure through inferences of their peculiar velocities
\citep{2007PhRvL..99h1301G,2011PhRvD..83d3004B,2017ApJ...847..128H}, weak lensing of supernova brightnesses
\citep{2006PhRvD..74f3515D,2014PhRvD..89b3009Q,2017MNRAS.467..259M,2017MNRAS.465.2862S}, and the local bulk flow
\citep{2000ASPC..201...80R,2011JCAP...04..015D,2012MNRAS.420..447T,2013A&A...560A..90F,2015JCAP...12..033H},
as well as low redshift constraints on the isotropy of the universe
\citep{2010JCAP...12..012A,2011MNRAS.414..264C,2011PhRvD..83j3503C,2013PhRvD..87l3522C,2015ApJ...810...47J}.
The rapidly sampled deep drilling fields, possibly coadded over short time scales,  will yield well-sampled
light curves of tens of thousands of supernovae to redshifts peaking around $z\sim 0.7$ and reaching beyond a
redshift of 1.0, limited by the systematics related to the limits of our astrophysical understanding of supernovae
populations and relative photometric calibration. In addition to the
usual use of Type Ia supernovae to probe the redshift-distance
relation to high redshift, the luminosities will be magnified by
lensing from foreground structure, a correlation which can be probed
with these data.
The ultimate promise of such supernova surveys will be linked to the observing strategy employed by the LSST.
\item  Cosmological analyses can be carried out using SN, WL, and LSS
  in subsets of the LSST data in different regions of the sky, testing
  fundamental cosmological
assumptions of homogeneity and isotropy
\citep[e.g.,][]{2009ApJ...690..923Z}.
\item The shape of the power spectrum of dark matter fluctuations measured by LSST weak lensing
will constrain the sum of neutrino masses with an accuracy
of 0.04 eV or better
\citep{1999A&A...348...31C,2004PhRvD..70f3510S,2006JCAP...06..025H}.
Given the current constraints on neutrino mass mixing, this is at the
level to determine whether there is an inverted neutrino mass hierarchy, a
fundamental question in particle physics.
\item Tens of thousands of galaxy-galaxy lenses will provide the needed statistics to probe dark matter
halo profiles and substructure \cite[e.g.,][]{2006MNRAS.368..715M,2012Natur.481..341V}. The image fluxes in several thousand well-measured
strongly lensed quasars will enable constraints of the dark matter mass function on small scales \citep{2002ApJ...572...25D}.
\item The abundance of galaxy clusters as a function of mass and redshift is sensitive to cosmological parameters
\citep[SciBook, Ch.~13;][]{2014MNRAS.443.1973V}. LSST will produce a large catalog of clusters detected through their member galaxy population
to redshift $z\sim 1.2$.  In addition, LSST will identify optical counterparts and provide deep optical
imaging for clusters detected in other wavebands \citep[e.g.,][]{2009ApJ...701...32S}.
\item The clustering properties of those same galaxy clusters will also be used to constrain
  cosmological parameters \citep{1996MNRAS.282.1096M,2013MNRAS.434..684M}, to marginalize over uncertainties in
  the mass-observable relation and photometric redshift uncertainties \citep{2011PhRvD..83b3008O}, and to constrain the effects of super-sample covariance in the two-point functions of WL and LSS \citep{2003ApJ...584..702H,2014MNRAS.441.2456T}.
\item LSST will discover several hundred galaxy clusters that produce multiple-image lenses of background objects.
Cluster mass reconstruction based on the multiple image positions
 can probe the cluster inner mass profile, and can provide a separate test of cosmology, especially
in cases with strongly lensed background objects at different redshift \citep{2000ApJ...532..679P,2003MNRAS.338L..25O}.
\item Time delays of galaxy-scale lensed quasars will allow one to measure Hubble's constant
\citep[e.g.,][]{2010ApJ...711..201S,2017MNRAS.465.4914B} in hundreds of systems; sub-percent level precision in
$H(z)$ should be achievable \citep{2009ApJ...706...45C,2016A&ARv..24...11T}, providing a further independent dark energy probe.
LSST will also discover between 500 and 1000 strongly lensed Type Ia supernovae \citep{2017ApJ...834L...5G,2017arXiv170800003G}, which will provide hundreds of additional high-quality time delays.
Time delays for quasars multiply lensed by clusters as a function of redshift are an independent test
of dark energy \citep{1997ApJ...482...75K}. The natural timescale (many months to years) is well matched
to the LSST survey \citep{2010MNRAS.405.2579O}.
\item Standard sirens are a new cosmological probe, demonstrated by the recent discovery of a binary
  neutron star merger by LIGO with an electromagnetic counterpart \citep{2017ApJ...848L..12A}, which was
  used to constrain the Hubble parameter to roughly 15\% precision \citep{2017Natur.551...85A}.
  Constraints from standard sirens are independent of the local distance ladder, with the primary
  uncertainties being the local velocity field and the inclination angle of the system.  \citet{2018ApJ...852L...3S}
  estimate of order 70 such systems could be found with LSST.
\end{itemize}

\subsection{Taking an Inventory of the Solar System}

The small bodies of the Solar System, such as main-belt asteroids,
the Trojan populations of the giant planets and the Kuiper Belt objects,
offer a unique insight into its early stages because they provide
samples of the original solid materials of the solar nebula.
Understanding these populations, both physically and in their number
and size distribution, is a key element in testing various theories of
Solar System formation and evolution.

The baseline LSST cadence will result in orbital parameters for several
million objects; these will be dominated by main-belt asteroids, with
light curves and multi-color photometry for a substantial fraction of detected objects.
The LSST sample of asteroids with accurate orbits and multi-color light curves
will be 10 to 100 larger than currently available sample.
LSST will make a significant contribution to the Congressional target
completeness of 90\% for PHAs larger than 140\,m (\S~\ref{Sec:NEOc}), and will detect over 30,000 TNOs brighter than $r\sim24.5$ using its baseline cadence. LSST will be capable
of detecting objects like Sedna to beyond 100 AU, thus enabling \textit{in situ} exploration
far beyond the edge of the Kuiper belt at $\sim$50 AU. Because most of these
objects will be observed several hundred times, accurate orbital elements,
colors, and variability information will also be available.

\begin{figure}
\includegraphics[width=1.0\hsize,clip]{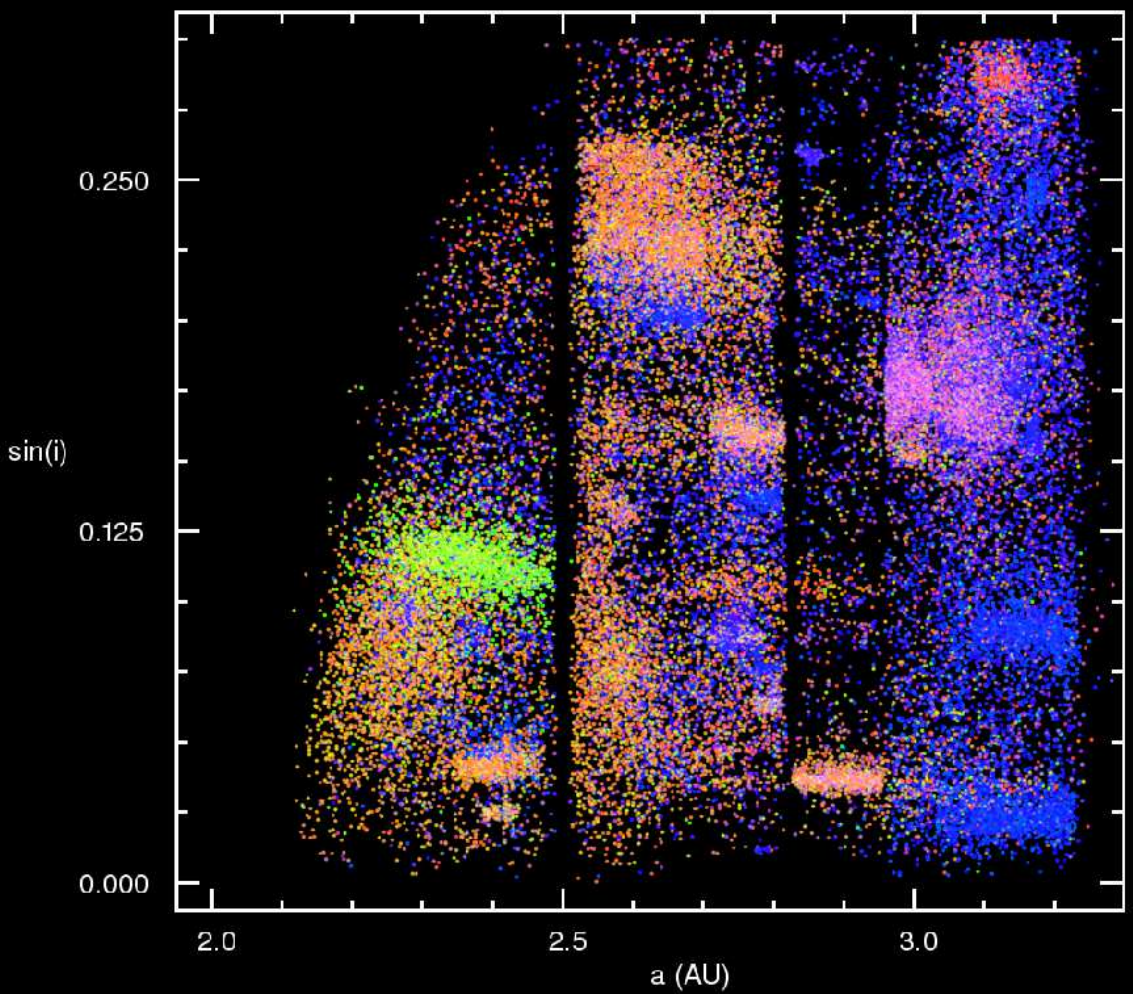}
\caption{An example of color-based asteroid taxonomy. The figure
shows the distribution of asteroids in the proper semi-major axis vs. $\sin(i)$
plane for 45,000 asteroids with colors measured by SDSS \citep{2008Icar..198..138P}.
The color of each dot is representative of the object's color.
Note the strong correlation between asteroid families (objects with
similar orbital elements) and colors. There are
at least five different taxonomic types distinguishable with SDSS measurements;
LSST color measurements of asteroids will be more than twice as accurate
and will increase the number of objects by roughly two orders of magnitude.}
\label{Fig:asteroids}
\end{figure}

\begin{figure}
\includegraphics[width=1.0\hsize,clip]{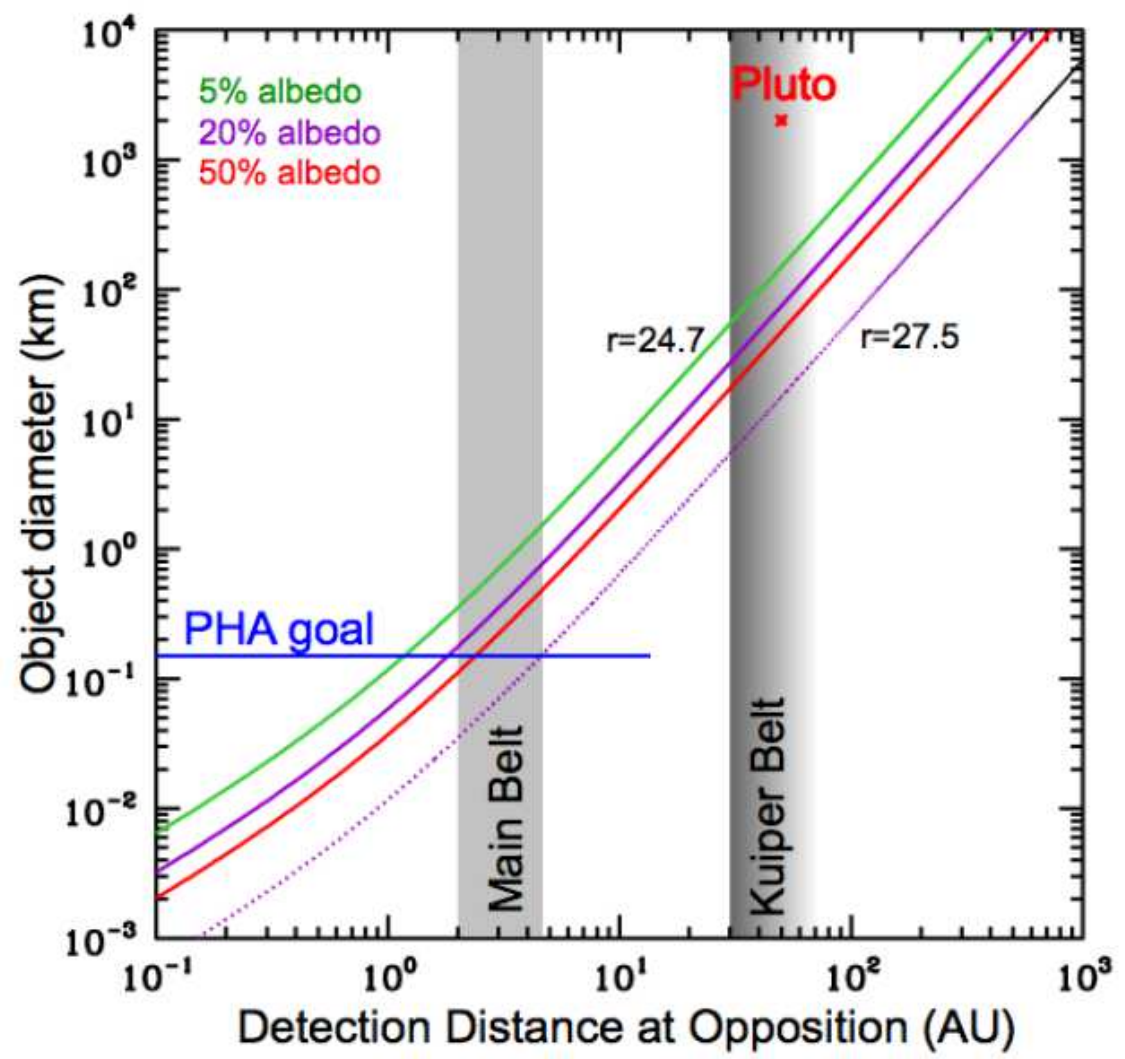}
\caption{The LSST detection limits for distant Solar System objects as
  a function of distance.
Moving objects with diameters as small as 100\,m in the main asteroid belt and
100\,km in the Kuiper Belt (TNOs) will be detected in individual visits,
depending on the albedo. Specialized deeper observations
(see \S~\ref{Sec:minisurveys}) will detect TNOs as small as 10 km. Adapted from
\citet{2007AAS...21113714J}.}
\label{Fig:Af9}
\end{figure}

The following are some examples of the LSST science opportunities in
Solar System science:
\begin{itemize}
\item Studies of the distribution of orbital elements for over 5 million main-belt
asteroids as a function of color-based taxonomy (see Fig.~\ref{Fig:asteroids})
and size; size distributions of asteroid families \citep{2008Icar..198..138P} and their
correlations with age \citep{2004Natur.429..275J,2005Icar..173..132N}; dynamical
effects \citep{2001Sci...294.1693B}; and studies of object shapes and spin states using
light curve inversion techniques \citep{2000Icar..148...12P,2009A&A...493..291D}.
\item Studies of transient mass loss in asteroids (active asteroids or main belt comets,
\citealt{2006Sci...312..561H}); such objects will appear extended in the
sensitive LSST images. Only a few such objects are currently known
\citep{2011AJ....142...28J,2012AJ....143...66J};
LSST
will increase the sample of such objects to $\sim$100.
\item Studies of the distribution of orbital elements for about 100,000 NEOs as a
function of color and size \citep{1993ApJ...407..412R,2003Icar..163..363D};
correlations with the analogous distributions for
main-belt objects, and studies of object shapes and structure using light curves.
\item Studies of the distribution of orbital elements for close to 300,000 Jovian Trojan
asteroids as a function of color and size \citep{2000AJ....120.1140J,2005AJ....130.2900Y,2007MNRAS.377.1393S};
the search for dynamical families
\citep{2005HiA....13..758K}; studies of shapes and structure using light curves.
\item Studies of the distribution of orbital elements for about 30,000 TNOs (see
Fig.~\ref{Fig:Af9}) as a function of color and size; the search for dynamical families
\citep{2011ApJ...733...40M}; studies of shapes and structure using light curves
\citep{1995AJ....110.3073D,2001AJ....122..457T,2001AJ....122.1051G,2004AJ....128.1364B,2005AJ....129.1117E,2006Icar..185..508J,2007AJ....134.2186D}.
\item An unbiased and complete census of both Jupiter-family and Oort-cloud
comets. These comets will have detailed six-band high-resolution
images extending to low surface brightness, in multiple points through
their orbits, allowing detailed studies of activity as a function of
distance from the Sun \citep{1999A&A...349..649L,2004come.book...17A}.
LSST will
discover an unprecedentedly large number of comets with typically 50
observations per object spread throughout their orbits during the 10-year
survey, and will help us to constrain models of the origin of comets
\citep{2010PhDT.......241S,2016AJ....152..103S}.  Combining the CN production
rates determined from observations in the $u$ bandpass, as a proxy for overall
gas activity, with the non-volatile production rate calculated from the
continuum-sensitive $r$, $i$, and $z$ bands allows for the determination of the
gas-to-dust ratio.  The relationship between the gas-to-dust ratio in comets
and their dynamical class (and places of formation) is a fundamental, and still
unresolved, question in cometary science \citep[see
e.g.,][]{1995Icar..118..223A,2017RSPTA.37560252B}.
\item Searching for objects with perihelia out to several hundred AU. For example, an object
like Sedna \citep{2004ApJ...617..645B} would be detectable at 130 AU. This will result
in a much larger, well-understood sample of inner Oort Cloud objects like Sedna and 2012 VP113
 \citep{2014Natur.507..471T}.  Studying the distribution of their orbits (in particular including any
clustering in the argument of perihelion) will test models predicting the existence of a planetary-mass object beyond Neptune, a proposed Planet 9 \citep{2014Natur.507..471T,2016AJ....151...22B,2016ApJ...824L..23B,2016AJ....152..221S,2017AJ....154...65B}. Depending on the proposed Planet 9's on-sky location and brightness, it may be possible for LSST to directly detect it in the wide survey images \citep{2016AJ....151...22B,2016ApJ...824L..23B,2016AJ....152..221S,2017AJ....154...65B}.

\item Mapping the propagation of coronal mass ejections through the
  Solar System using induced
 activity in a large sample of comets at different heliocentric distances
(SciBook Ch.~5).
\item Probing the inventory and frequency of  interstellar
  asteroids/comets. The recent Pan-STARRS1 discovery of the interstellar object
1I/2017 U1 (`Oumuamua)
\citep{2017MPEC....U..181B} has shown the power
of large, complete all-sky surveys to unearth
rare and exciting classes of objects. LSST will be some three magnitudes more sensitive
than  current NEO surveys (like Pan-STARRS1), and will cover more sky more
often. Therefore, LSST is likely to find more interstellar objects, and more frequently.
Estimates from  \citet{2016ApJ...825...51C},  \citet{2017AJ....153..133E}, and \citet{2017ApJ...850L..38T} suggest that LSST will increase the number of  such rare objects by an order of magnitude which, among other outcomes, will help
constrain the frequency and properties of planetary system formation in the solar neighborhood.
\end{itemize}

\subsection{ Exploring the Transient Optical Sky }

\begin{figure}
\includegraphics[width=1.0\hsize,clip]{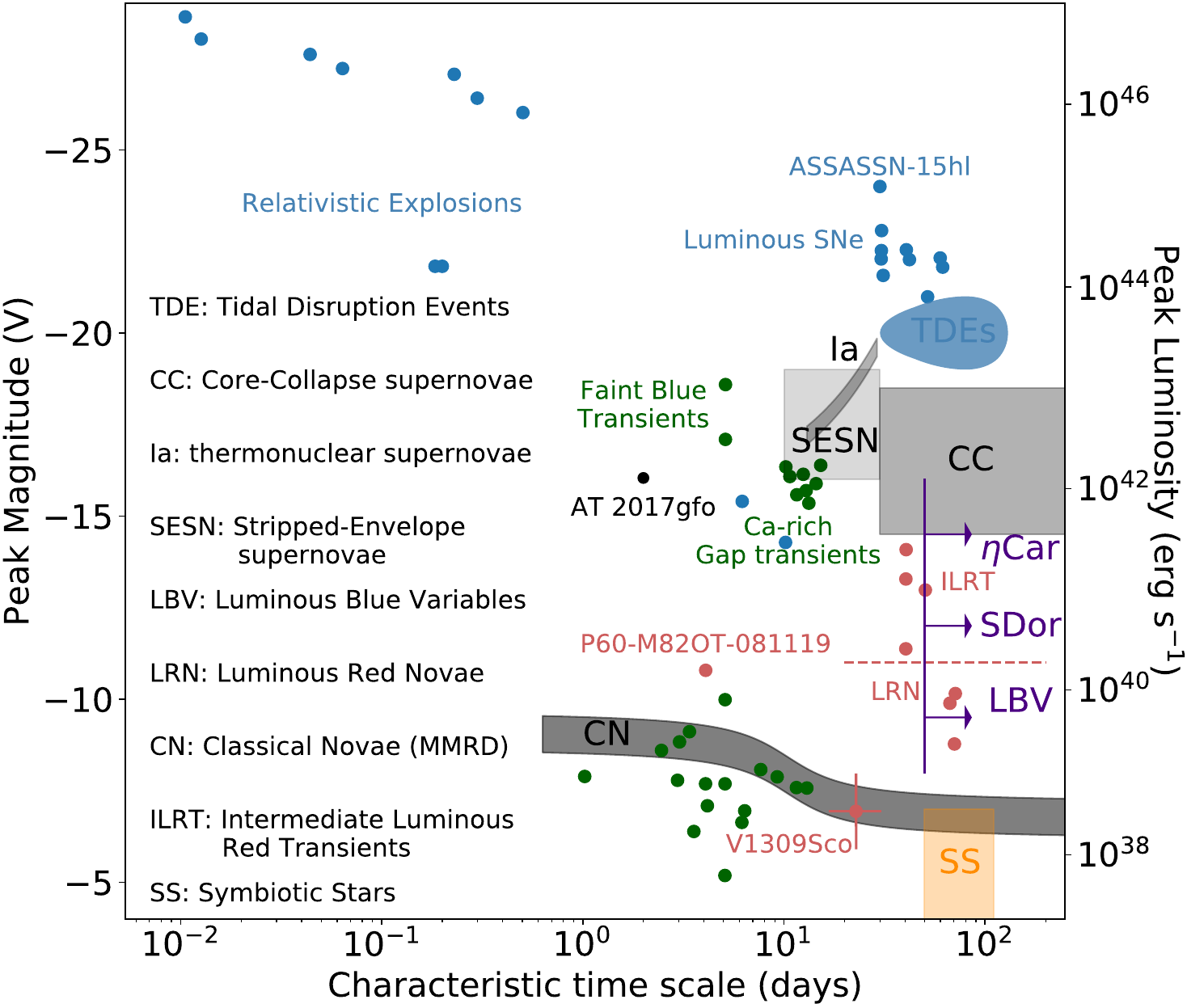}
\caption{The phase space of cosmic explosive and eruptive transients
  as represented by their absolute $V$ band peak brightness and the
  event timescale, defined as the time taken to drop one magnitude in
  V band brightness from peak luminosity (adapted from \citet{2007Natur.447..458K}
  and \citet{2011PhDT........35K}).  The locus of the Classical
  Novae
is as described in \citet{1995ApJ...452..704D}.  LSST
  will open up large regions of this phase space for systematic
  exploration by extending time-volume space more than 100 times over
  existing surveys.}
\label{Fig:shri}
\end{figure}

Time domain science will greatly benefit from LSST's unique capability
to simultaneously provide large area coverage, dense temporal
coverage, accurate color information, good image quality, and rapid
data reduction and classification. Since LSST extends time-volume-color
space 50-100 times over current surveys \citep[e.g.,][]{2013pss2.book..223D}
it will facilitate new population and statistical studies and also the discovery of new classes of
objects.  LSST data products will enable many projects including:

\begin{itemize}

\item Discovery and characterization of thousands of hot Jupiters
  in exoplanetary systems via the transit method \citep{2012ApJ...753..160W}.
LSST will extend the extrasolar planet census to larger distances within the Galaxy, thus enabling detailed studies
of planet frequency as a function of stellar metallicity and parent population \cite[e.g.,][]{2009ApJ...695..336H,2011ApJ...743..103B}.
The out-of-transit variability of exoplanet host stars will also provide characterization of the system
via flaring behavior and stellar age via gyrochronology, the latter helping constrain theories of tidal evolution and
migration in giant planets.

\item Gravitational microlensing in the Milky Way \cite[see][]{2008ApJ...681..806H} as well as in the Local Group and beyond \citep{2008A&A...478..755D}.

\item Studies of dwarf novae, including their use as probes of stellar populations and
      structure in the Local Group
      \citep{2005AJ....129.1873N,2006AJ....131.2980S,2009ApJ...692..324S}. Population
      studies of the end points of binary evolution, mapping the
      distribution and quantifying the demographics of
      long and short orbital period dwarf novae, and distinguishing
      recurrent from normal novae. Regular cadence, long term color
      observation on a large sample of galactic sources will enable
      the identification of CVs containing highly magnetic white
      dwarfs, that are red due to cyclotron emission from the magnetic
      accretion column and in the low state for the majority of the 10-year survey.

\item Studies of  transients from poorly-constrained stages of stellar evolution including
stellar eruptions,  luminous blue variable (LBV), stellar mergers, and
helium core flashes leading to white dwarf formation.  We will be able
to identify the progenitors of eruptive transients in the
deep LSST stacks and even look for faint precursor eruptions.  We will
also constrain the rates of individual eruption subclasses
\citep{2014ARA&A..52..487S} by detecting them in galaxies out to tens
of Mpc.

\item A census of light echoes of historical explosive and eruptive
  transients in the Milky Way and Local Group through high resolution
  time series.

\item Studies of known and unusual SN populations and parameterization of their light curves \citep[e.g.,][]{1998ApJ...495..617H,2003ApJ...590..944W,2007ApJ...667L..37H,2008ApJ...686..749K,2009ApJ...700.1097H,2012ApJ...748..127F,2014ApJS..213...19B,2017Natur.551..210A}, including late-time observations of rapidly-evolving transients to deep limits, critical for ascertaining their nature. Measurements of intrinsic rates for both peculiar transients \cite[e.g.,][]{2014ApJ...794...23D} and for SN as a function of sub-type and host environment properties \citep[e.g., metallicity;][]{2017ApJ...837..120G}.

\item A deep search for new populations of novae and supernova progenitors
      \citep[][see Fig.\ref{Fig:shri}]{2009ARA&A..47...63S,2009ApJ...705.1364T,2011MNRAS.415..773S} both through direct imaging and through the detection of SN precursor events \citep{2013Natur.494...65O}, characterization of pre-SN variability of SN progenitors and the frequency of pre-SN outbursts.

\item The discovery of strongly lensed SNe; $500-1000$ multiply imaged SN~Ia \citep{2017ApJ...834L...5G,2017arXiv170800003G}
and at least several hundred strongly lensed core-collapse SNe \citep{2010MNRAS.405.2579O}
are expected to be discovered by LSST. Time delays between the multiple images of strongly lensed core-collapse
SNe can be used to observe the elusive shock breakout phase of the light curve, providing an unprecedented look
at the earliest emission from these transients \citep{2018MNRAS.474.2612S}.

\item A large, well characterized sample of super luminous supernovae
including object at redshift as high as $z=2.5$, a sample large enough to be leveraged for cosmology  improving constraints on $w$ and $\Omega_m$ \citep{2016MNRAS.456.1700S}.

\item Studies of optical bursters (those varying faster than 1 mag hr$^{-1}$) to $r\sim25$ mag.

\item Detection and measurement of gamma-ray burst afterglows and transients
      \cite[e.g.,][]{2004IJMPA..19.2385Z,2006ApJ...642..354Z,2010ApJ...720.1513K} to high redshift ($\sim$7.5).

\item Large scale studies of stellar tidal disruptions by nuclear supermassive
  black holes \cite[e.g.,][]{1989ApJ...346L..13E,2008ApJ...676..944G,2009MNRAS.400.2070S,2011Sci...333..203B,2012EPJWC..3903001G,2015JHEAp...7..148K}, as well as binary
  supermassive black holes in the in-spiral phase \cite[e.g.,][]{2009MNRAS.393.1423C,2017MNRAS.465.3840C}.
  Persistent observations leading to complete lightcurves (other than
  the seasonal gaps) of long duration events like
  TDEs. Measurements of rates as function of galaxy type, redshift,
  and level of nuclear activity. An assessment of the diversity
  of these events in terms of total power, effective temperature, and
  jet launching efficiency.

\item A study of quasar variability using accurate, multicolor light
  curves for a few million
quasars, leading to constraints on the accretion physics and nuclear environments \citep{2003AJ....126.1217D,2004ApJ...601..692V,2010ApJ...721.1014M,2017ApJ...836..186J}.
      Relations between quasar variability
      properties and luminosity, redshift,
      rest-frame wavelength, time scale, color, radio-jet emission, black-hole
      mass, and Eddington-normalized luminosity will be defined with massive
      statistics, including the potential to detect rare but important events such as
      jet flares and obscuration events. Microlensing events will also be monitored in the $\sim$4000 gravitationally-lensed
      quasars discovered by LSST and used to measure the spatial structure of quasar accretion disks.

\item The superb continuum light curves of AGN will enable economical ``piggyback''
      reverberation-mapping efforts using spectroscopy of emission lines
      \cite[e.g.,][]{2012ApJ...747...62C,2015ApJS..216....4S,2017ApJ...851...21G}. These results
      will greatly broaden the luminosity-redshift plane of reverberation-mapped AGNs
      with black-hole mass estimates. For LSST data alone, the inter-band continuum lags
      will provide useful structural information.

\item Optical identification of transients and variables detected in
  other electromagnetic wavebands, from gamma rays to radio. Examples
  include optical and gamma ray variability in blazars \citep{2014MNRAS.439..690H},
  radio transients associated with tidal disruption flares
  \citep{2011MNRAS.416.2102G}, and radio counterparts to supernovae and
  GRBs \citep{2006ApJ...639..331G}. Deep optical observations with LSST may
  also help illuminate the nature of fast radio bursts \cite[FRBs,][]{2007Sci...318..777L,2013Sci...341...53T}.

\item Optical identification of counterparts to non-electromagnetic
  sources, such as gravitational waves (GW) and neutrino events
  (LIGO\footnote{\url{http://www.ligo.caltech.edu}},
  ICECUBE\footnote{\url{http://icecube.wisc.edu}}).  LSST's unique ability
  to characterize the faint variable sky over large areas will be
  important for the detection of GW associated sources, with an estimate
  of $\sim 7$ discoveries per year \citep{2018ApJ...852L...3S}.  The power
  of the Advanced LIGO
  (aLIGO)/Virgo\footnote{\url{http://public.virgo-gw.eu/language/en/}}
  experiment has led to the discovery
  of four GW events in less than a year. The binary
  neutron star merger event GW170817 was accompanied by emission
  detected across the entire electromagnetic spectrum \citep{2017ApJ...848L..12A}.
The optical/NIR emission had two distinct components, a blue emission
(which peaked and then faded away
on a time scale of a few days) and a redder component  that persisted
for $\sim 15$ days.  This longer-lasting component arose from the radioactive decay of heavy elements
synthesized during the NS merger, a ``kilonova'' (AT 2017gfo).
While both these components had been predicted \citep{2017LRR....20....3M},
the $\sim  100$ kilonova sample that LSST is expected to generate will
enable comparative studies of these transients, allowing us to
understand how  the presence and relative luminosity of the two
components varies to the properties of the binary system (e.g., mass)
and its remnant.
  Furthermore, LSST will be important for identifying the optical transient
  corresponding to LIGO events in the first place, eliminating false
  positives \citep{2013ApJ...767..124N,2012ApJ...746...48M,2015ApJ...814...25C, 2017ApJ...849...12C}. 
  At 24th
  mag, rejecting thousands of false positives from other new
  transients appearing during the imaging of the GW event area
  requires a strategy of multiple passes in different filters.

\end{itemize}

\subsection{Mapping the Milky Way }

The LSST will map the Galaxy in unprecedented detail, and by doing so revolutionize the fields of Galactic
Astronomy and Near-field Cosmology. The great detail with which the Milky Way can be studied complements
the statistical power of extra-galactic observations.  The overarching goal of near-field cosmology is to use
spatial, kinematic, and chemical measurements of stars to reveal the structure and evolution history of the Milky Way
and its environment. LSST will reveal this fossil record in great detail and provide a Rosetta Stone for extragalactic
astronomy by setting the context within which we interpret these
broader datasets. Moreover, different candidate supersymmetric
particle dark matter models predict different mass clustering on small
scales, and thus different mass functions for subhalos of the Milky
Way.  Thus the LSST census of faint satellites and stellar streams in
the halo will offer a unique means to constrain the
particle nature of dark matter.

\begin{figure}
\includegraphics[width=1.0\hsize,clip]{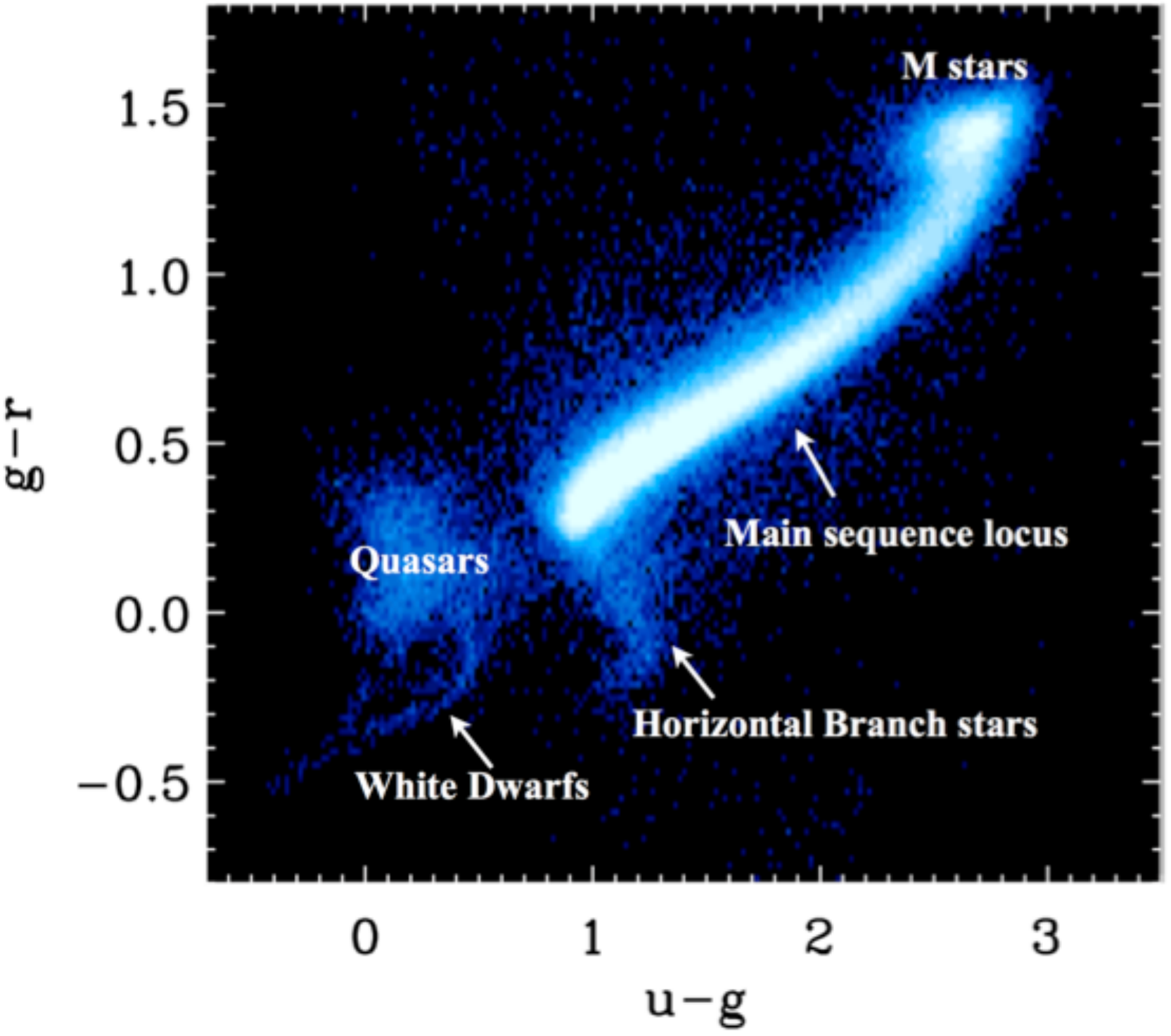}
\caption{The $g-r$ vs. $u-g$ color-color diagram for about a million point sources
from the SDSS Stripe 82 area. Accurate multi-color photometry
contains information that can be used for source classification and determination of
detailed stellar properties such as effective temperature and metallicity. LSST will
enable such measurements for billions of stars.}
\label{Fig:FeH}
\end{figure}

\begin{figure}
\includegraphics[width=1.\hsize,clip]{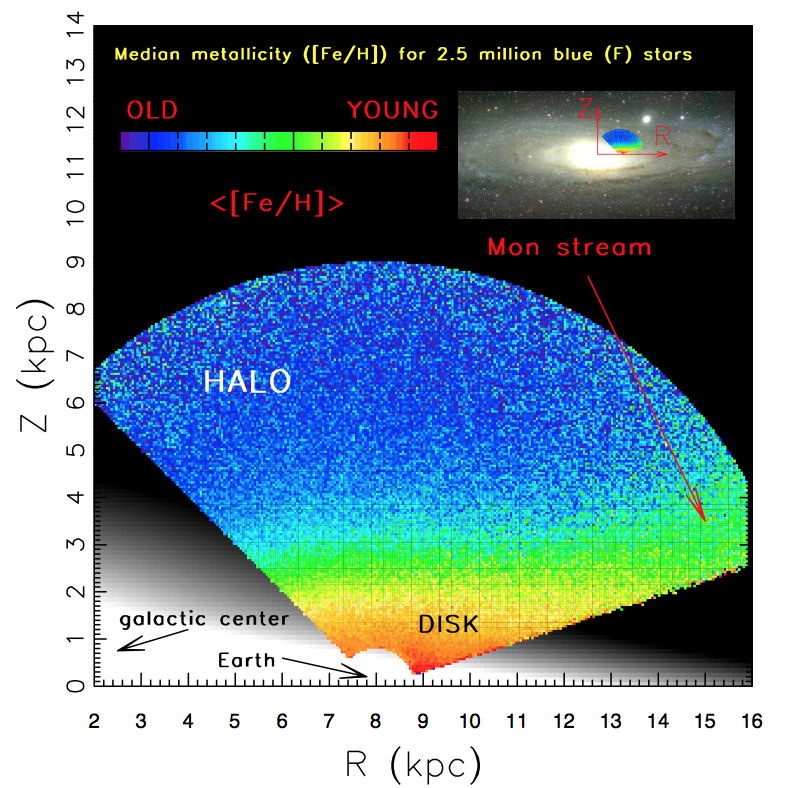}
\caption{
The median metallicity map for 2.5 million main-sequence F-type stars within 10 kpc
from the Sun \citep[adapted from][]{2008ApJ...684..287I}. The metallicity is estimated using
$u-g$ and $g-r$ colors measured by SDSS. The position and size of the mapped
region, relative to the rest of the Milky Way, is illustrated in the top right
corner, where the same map is scaled and overlaid on an image of the Andromeda
galaxy. The gradient of the median metallicity is essentially parallel
to the $Z$ axis, except in the Monoceros stream region, as marked. LSST
will extend this map out to 100 kpc, using a sample of over 100 million
main-sequence F stars.}
\label{Fig:FeH3}
\end{figure}

\begin{figure}
\includegraphics[width=1.\hsize,clip]{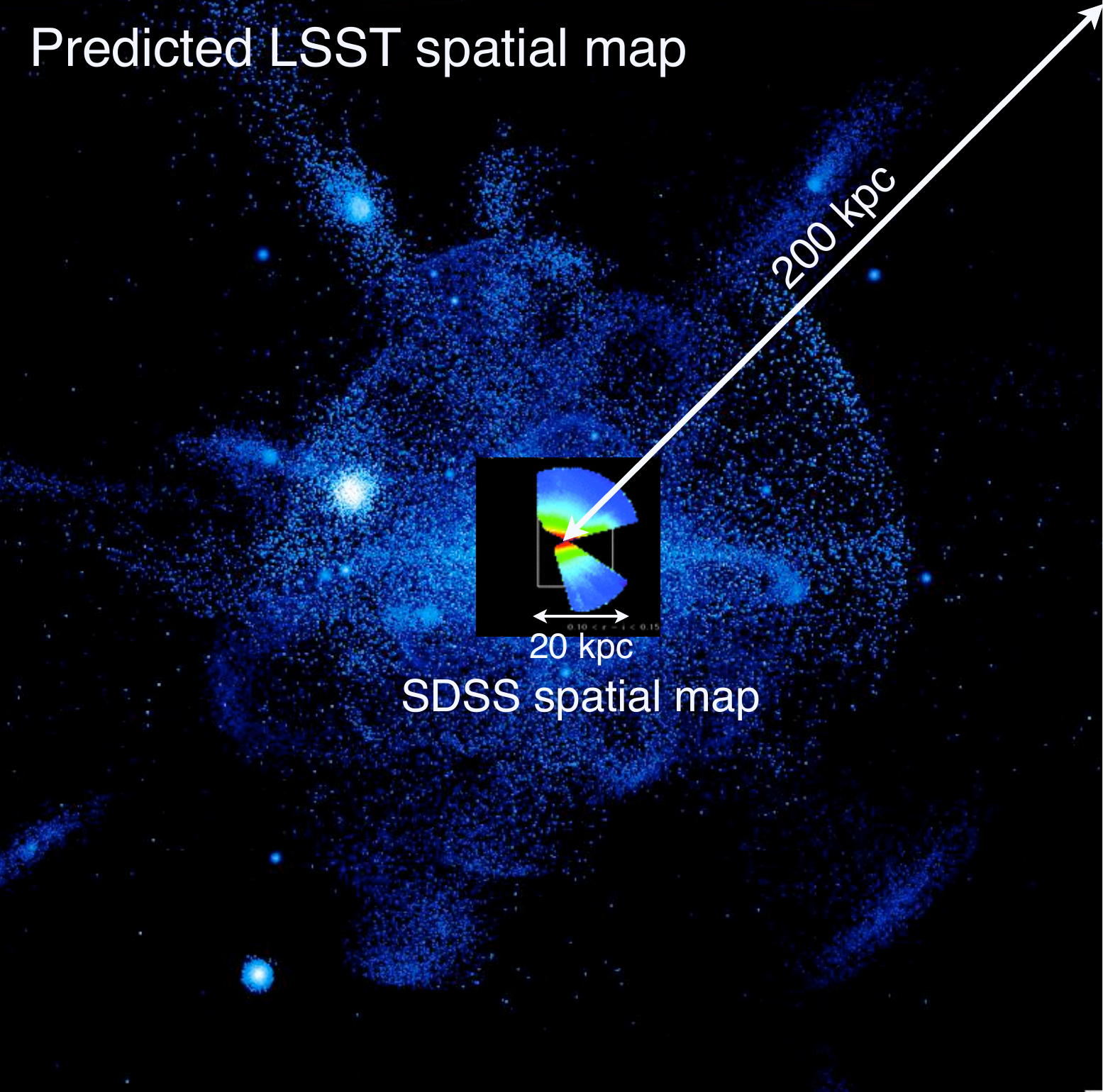}
\caption{A predicted spatial distribution of stars out to 150 kpc from the center of the Milky Way,
from \citet{2005ApJ...635..931B}.  LSST will resolve main sequence turnoff stars out to 300 kpc, ten times
more volume than shown here, enabling a high-fidelity spatial map over
the entire observed virial volume. (Note that this is significantly larger than the
100 kpc probed by {\em metallicity} measurements in
Figure~\ref{Fig:FeH3}, which is limited by the depth of the $u$-band observations.)
Overlaid on this prediction is the observed SDSS stellar number density map based on main sequence stars
with $0.10 < r-i < 0.15$ \citep{2008ApJ...673..864J}.  This map extends up to $\sim$ 20 kpc from the Sun, with
the white box showing a scale of 20 kpc across and the left side aligned with the Galactic center.
The revolutionary Galaxy map provided by SDSS is only complete to $\sim$40 kpc, or only $\sim$1\% of
the virial volume.  However, the outermost reaches of the stellar halo are predicted to bear the most unique
signatures of our Galaxy's formation \citep{2008ApJ...689..936J,2010MNRAS.406..744C}.   LSST will be the only survey
capable of fully testing such predictions.}
\label{Fig:halo}
\end{figure}

The LSST will produce a massive and exquisitely accurate photometric and astrometric dataset for about 20 billion
Milky Way stars. The coverage of the Galactic plane will yield data for numerous star-forming
regions, and the $y$ band data will penetrate through the interstellar dust layer. Photometric metallicity
measurements (see Figs.~\ref{Fig:FeH} and \ref{Fig:FeH3}) will be available for about 200 million main-sequence
F/G stars which will sample the halo to distances of 100 kpc
\citep{2008ApJ...684..287I,2013ApJ...763...65A} over a solid angle of
roughly 20,000 deg$^2$. No other
existing or planned survey will provide such a powerful dataset to
study the outer halo: Gaia
is flux limited at $r=20$, and the Dark Energy Survey \citep{2011AJ....141..185R} and Pan-STARRS both
lack observations in the $u$ band, necessary for estimating metallicity. The LSST in its standard surveying mode will
be able to detect RR Lyrae variables (pulsating stars and standard candles) and classical novae (exploding stars
and standard candles) at a distance of 400 kpc and hence explore the extent and structure of our  halo out to
half the distance to the Andromeda galaxy. Thus, the LSST will enable studies of the distribution of main-sequence
stars beyond the presumed edge of the Galaxy's halo (see Fig.~\ref{Fig:halo}), of their metallicity distribution
throughout most of the halo, and of their kinematics beyond the thick disk/halo boundary. LSST will also obtain
direct distance measurements via trigonometric parallax below the hydrogen-burning limit for a representative
thin-disk sample.

In addition to the study of hydrogen burning stars, LSST will uncover the largest sample of stellar remnants to date.
Over 97\% of all stars eventually exhaust their fuel and cool to become white dwarfs. Given the age of the Galactic
halo, a significant fraction of the mass in this component may reside in these remnant stars
\citep[e.g.,][]{2000ApJ...542..281A,2007A&A...469..387T}
and therefore their discovery directly constrains the Galactic mass budget.  These large
populations of disk and halo white dwarfs will provide unprecedented constraints on the luminosity function of
these stars, which will directly yield independent ages for the Galactic disk and halo (e.g., through the initial-final mass
relation, \citet{2008ApJ...676..594K}).

The sky coverage of LSST naturally targets both field stars and star clusters.  To date, no systematic survey of the stellar
populations of Southern hemisphere clusters has been performed (e.g., such as the CFHT Open Star Cluster Survey, or
the WIYN Open Star Cluster Survey in the North; \citealt{2001AJ....122..257K,2000ASPC..198..517M}).  Multiband imaging of these co-eval,
co-spatial, and iso-metallic systems will provide vital insights into fundamental stellar evolution.  For example, the depth
of LSST will enable construction of  luminosity and mass functions for nearby open clusters down to the hydrogen burning
limit and beyond.  Variations in the initial mass function will be studied as a function of environment (e.g., age and metallicity).
The wide-field coverage will also allow us to track how the stellar populations in each cluster vary as a function of radius,
from the core to beyond the tidal radius. Fainter remnant white dwarfs will be uncovered in both open and globular clusters
(the nearest of which are all in the South), thereby providing a crucial link to the properties of the now evolved stars in each
system.

In summary, the LSST data will revolutionize studies of the Milky Way and the entire Local Group. We list a few specific
Galactic science programs that LSST will enable:

\begin{itemize}
\item High-resolution studies of the distribution of stars in the outer halo
          in the six-dimensional space spanned by position, metallicity and proper
          motions \citep[e.g.,][]{2006AJ....132.1768G,2008ApJ...680..295B,2008ApJ...673..864J,2008ApJ...684..287I,2010ApJ...716....1B}.
\item The most complete search possible for halo streams, Galaxy satellites and intra-Local Group
          stars \citep[e.g.][]{2007ApJ...654..897B,2009AJ....137..450W,2014AJ....147...76B}.
\item Deep and highly accurate color-magnitude diagrams for over half of the known
          globular clusters, including tangential velocities from proper motion
          measurements \citep{2008ApJS..179..326A,2007AJ....134..195C}.
\item Mapping the metallicity, kinematics and spatial profile of the Saggitarius dwarf tidal
          stream \citep[e.g.,][]{2001ApJ...547L.133I,2003ApJ...599.1082M,2005ApJ...619..807L,2014MNRAS.437..116B}
          and the Magellanic stream \citep{2004AJ....128.1606Z}.
\item The measurement of the internal motions of Milky Way dwarf
          galaxies via proper motions, thereby constraining their density profiles and
	  possibly the nature of dark matter \citep[e.g.,][]{2011ApJ...742...20W}.
\item Detailed constraints on the formation and evolution of the populations within the Galactic Bulge, as traced by the spatial
          distribution, motion, and chemistry of $\sim$10$^{7-8}$ of its stars
          \citep[e.g.][]{2011A&A...534A..80H,2014ApJ...787L..19N}.
\item Studies of the clumpiness of the gravitational potential in the Galaxy using
          fragile wide-angle binaries selected with the aid of trigonometric and
          photometric parallaxes, and common proper motion \cite[e.g.,][]{2004ApJ...601..311Y,2010A&A...509A..46L}.
\item Detailed studies of variable star populations; 2\% or better accurate
          multicolor light curves will be available for a sample of at least 50
          million variable stars \citep{2007AJ....134.2236S}, enabling studies of
          cataclysmic variables, eclipsing binary systems, and rare types of variables.
\item Discovery of rare and faint high proper motion objects: probing the
          faint end of the stellar mass function \citep{2008AJ....135.2177L,2010AJ....140..844F}, and searching for
          free-floating planet candidates \citep{2000MNRAS.314..858L,2014ApJ...786L..18L}.
\item Direct measurement of the faint end of the stellar luminosity function
          using trigonometric parallaxes \citep{2002AJ....124.2721R} and a complete census of the
          solar neighborhood to a distance of 100 pc based on trigonometric parallax measurements for objects as faint as
          $M_r=17$ ($\sim$L5 brown dwarfs). For example, LSST will deliver 10\% or better distances for a sample of about 2,500 stars
          with 18$<M_r<$19.
                                         
\item The separation of halo M sub-dwarfs from disk M dwarfs, using the $z-y$ color which is sensitive to their rich molecular band
          structure \citep{2011ASPC..448..531W,2013AJ....145...40B}.
\item Studies of white dwarfs using samples of several million objects, including the determination of the halo white dwarf luminosity
          function (SciBook Ch.~6).
\item Measurements of physical properties of stars using large samples of eclipsing binary stars \citep{2013AAS...22111601S}.
\item High-resolution three-dimensional studies of interstellar dust using 5-color
          SEDs of main sequence stars \citep{2011A&A...536A..23P,2012ApJ...757..166B,2014ApJ...783..114G}.
\item A census of AGB stars in the Galaxy by searching for resolved envelopes and optical  identifications of IR counterparts
         (e.g., from the WISE survey), and by using long-term variability and color selection \citep{2007ASPC..378..485I}.
\item A complete census of faint populations in nearby star forming regions using
          color and variability selection \citep[e.g.][]{2005AJ....129..907B}.
\end{itemize}

\begin{figure}
\includegraphics[width=1.0\hsize,clip]{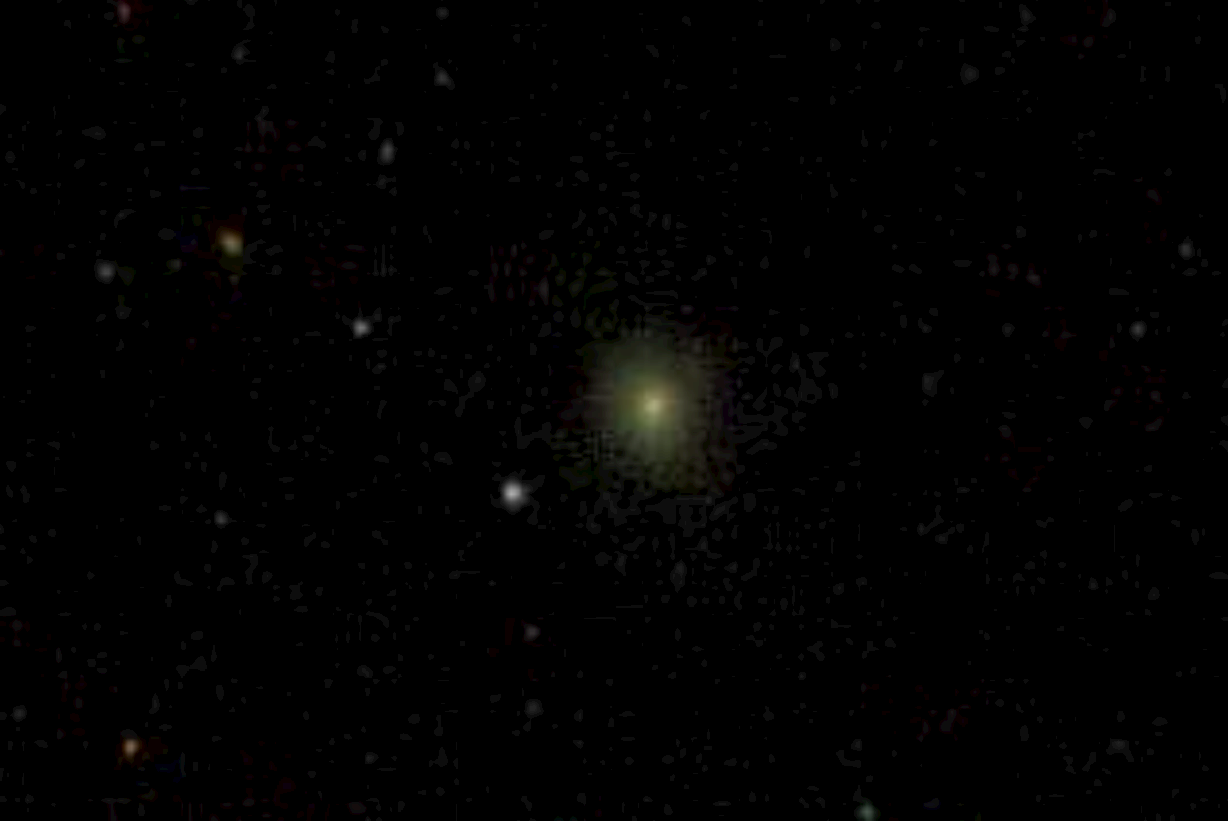}
\includegraphics[width=1.0\hsize,clip]{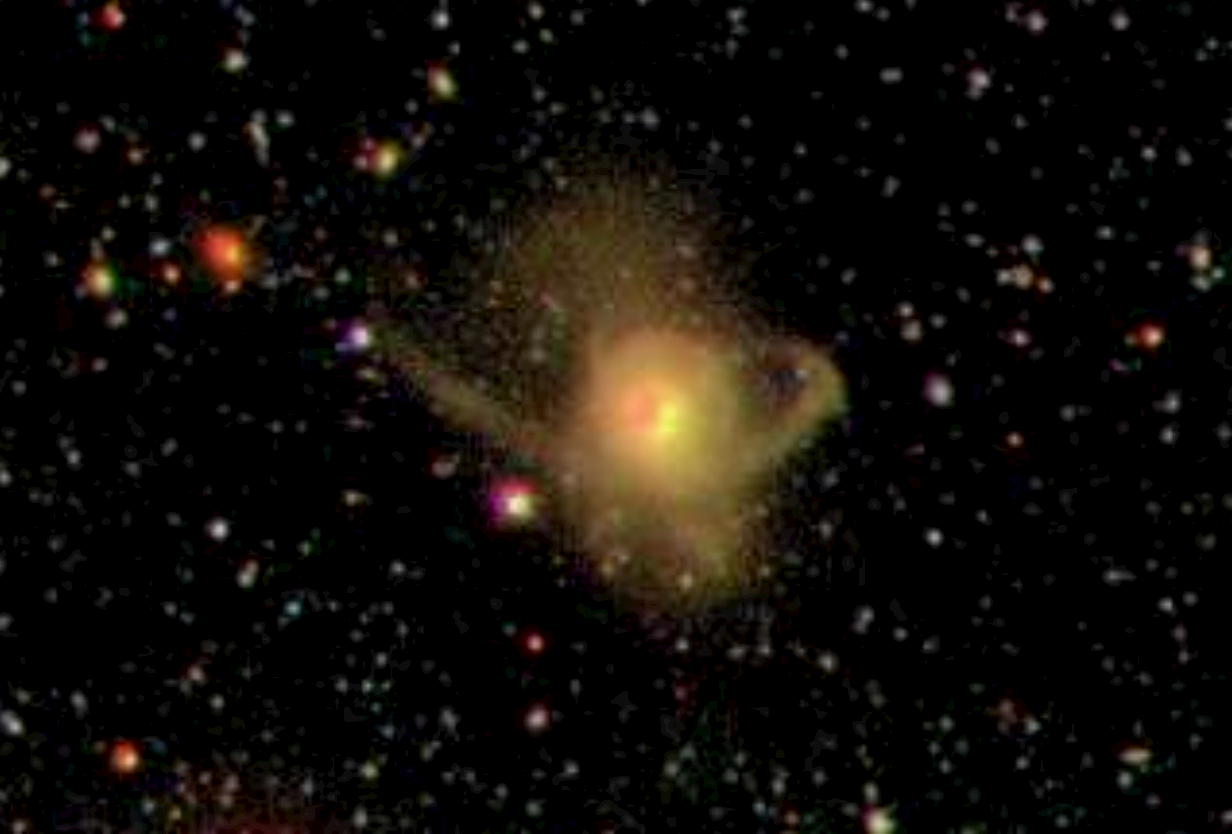}
\caption{
A comparison of an SDSS image (2$\times$4 arcmin$^2$ $gri$ composite) showing a typical galaxy at
a redshift of $\sim$0.1 (top) with a similar $BVR$ composite image of the same field obtained by the MUSYC survey
\citep[bottom;][]{2006ApJS..162....1G}. The MUSYC image is about 4 mag deeper than the SDSS image (and about 1 mag less deep
than the anticipated LSST 10-year coadded data). Note the rich surface brightness structure seen in the MUSYC
image that is undetectable in the SDSS image.}
\label{Fig:musyc}
\end{figure}

\subsection{  Additional Science Projects}

The experience with any large survey (e.g., SDSS, 2MASS, VISTA, WISE, GALEX, to name but a
few) is that much of their most interesting science is often unrelated to
the main science drivers, and is often unanticipated at the time the survey is
designed. LSST will enable far more diverse science than encompassed by the
four themes that drive the system design. We list a few anticipated major
programs.

\begin{figure}
\includegraphics[width=1.0\hsize,clip]{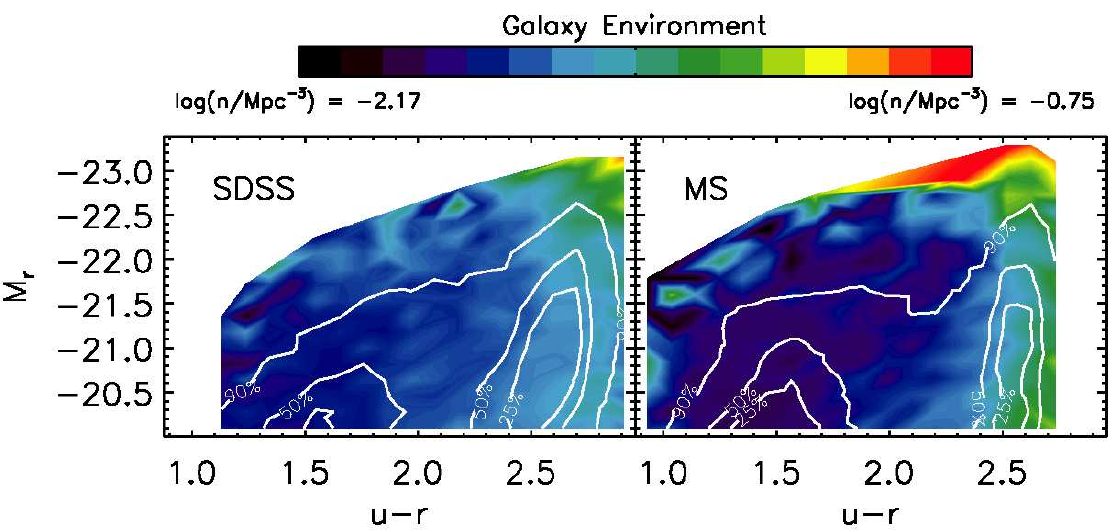}
\caption{A comparison of the distribution of galaxies in
luminosity--color--density space measured by SDSS (left) and a model based
on the Millennium simulation (right). The linearly-spaced contours outline
the distribution of a volume-limited sample of galaxies in the plotted diagram, and
the color-coded background shows the median environmental density (computed
using the ten nearest neighbors) for galaxies
with the corresponding luminosity and color. Such multi-variate distributions
encode rich information about formation and evolution of galaxies. Galaxies
detected by SDSS are representative of the low-redshift Universe (the median
redshift is $\sim$0.1). The LSST will enable such studies as a function of
redshift, to $z\sim$2. Adapted from \citet{2008ApJ...674L..13C}.}
\label{Fig:cowan}
\end{figure}

\begin{figure}
\includegraphics[width=1.0\hsize,clip]{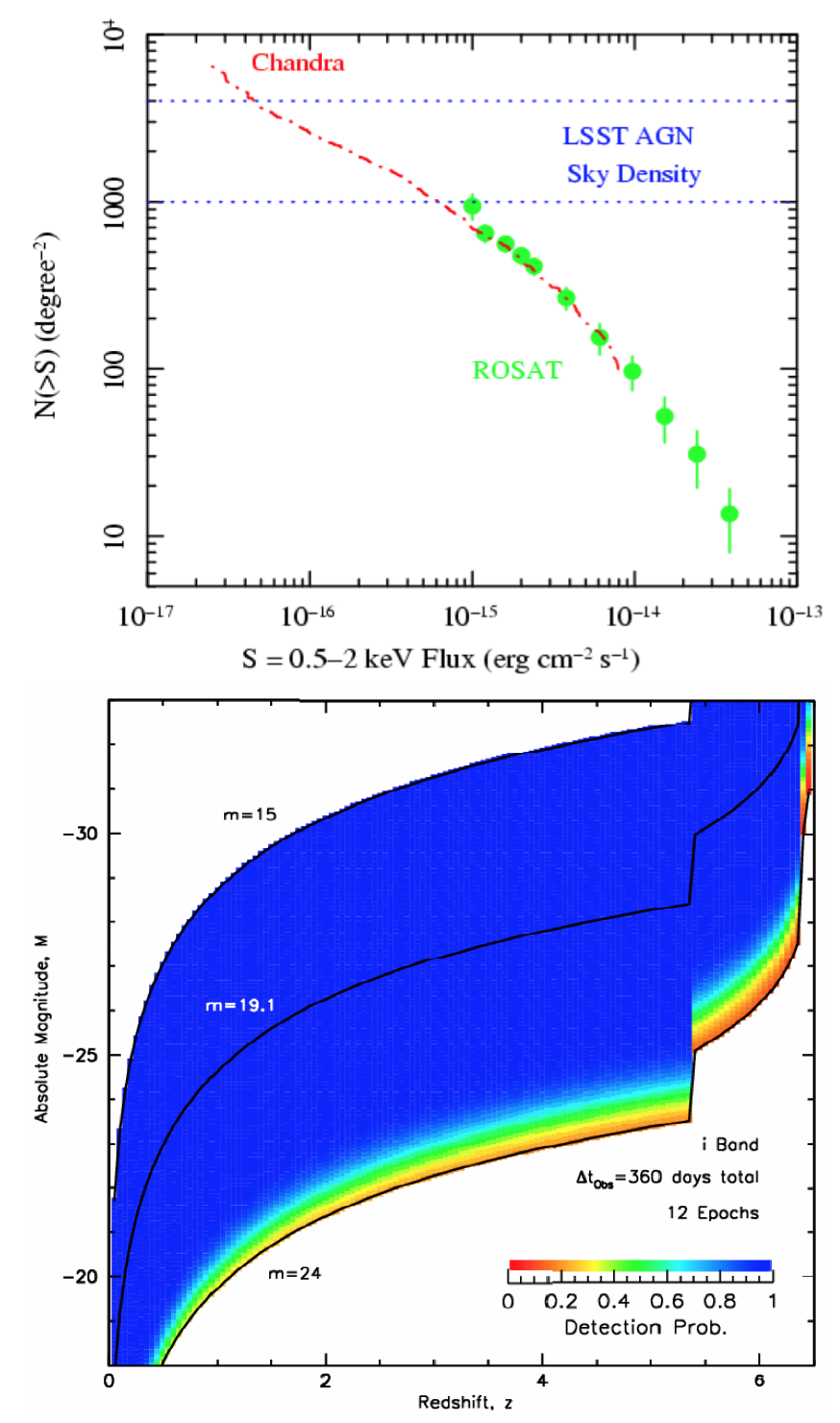}
\caption{The LSST will deliver AGN sky densities of 1000--4000\,deg$^{-2}$ (top panel);
The total LSST AGN yield, selected using colors and variability, should be well over
10 million objects, especially once multiwavelength data are also utilized.
The bottom panel shows the expected distribution of these objects in the
absolute magnitude vs. redshift plane, color-coded by the probability for
an object to be
detected as variable after 1 year of observations. Note that quasars will
be detected to their formal luminosity cutoff ($M< -23$) even at redshifts
of $\sim$5. Adapted from \citet{2007AAS...21113709B}.}
\label{Fig:panels3}
\end{figure}

\begin{itemize}
\item Detailed studies of galaxy formation and evolution using their distribution in
luminosity-color-morphology space as a function of redshift. For example, LSST will
enable studies of the rest-frame UV emission, similar to those based on GALEX data
for local galaxies, to a redshift of $\sim$2 for an unprecedentedly large number of
galaxies. These studies project onto many axes:
\begin{itemize}
  \item the evolution of the galaxy luminosity function with redshift, as a function of
        morphology and color;
  \item the evolution of the galaxy color distribution over a wide range of rest-frame
        wavelengths, and as a function of luminosity and morphology;
  \item bulge-disk decomposition, as a function of luminosity and color, over
        a large redshift range;
  \item detailed distribution of satellite galaxies in luminosity-color-morphology space
        as a function of luminosity, color, and morphology of the primary galaxy;
  \item correlations of luminosity, color and morphology with local environment from
           kpc to Mpc scales, and as a function of redshift (see  Figs.~\ref{Fig:musyc} and \ref{Fig:cowan});
  \item the properties of galaxy groups and clusters as a function of cosmic time.
\end{itemize}
\item AGN census to very faint luminosity and a large redshift limit
  \citep{2014IAUS..304...11I}, yielding 20 million objects from LSST
  data alone, and the ability to identify up to $\sim 100$ million objects once multiwavelength
      data are used to aid AGN selection (see Fig.~\ref{Fig:panels3}). By reaching substantially further
      down the AGN luminosity function than has been possible before over a very large solid angle, LSST data
      will test evolutionary cosmic downsizing scenarios across the full range of cosmic environments,
      and lead to a much clearer understanding of black-hole growth during the first Gyr. For
      example, LSST should discover several thousand AGNs at $z\sim6-7.5$,
      representing a dramatic increase over present samples
      \citep[][see also SciBook Ch.~10]{2007AAS...21113709B}.

\item The combination of LSST, Euclid and WFIRST data should allow discovery of at least
       tens of quasars at $z>7.5$ (R. Barnett 2017, priv. comm).
\item LSST data will provide good constraints on AGN lifetimes, or at least the timescales over which
         they make distinct accretion-state transitions \citep{2016MNRAS.457..389M}, due to large sample
         size and survey lifetime \citep[e.g.][]{2003ApJ...597L.109M}.
\item The first wide field survey of ultra low surface brightness galaxies, with
      photometric redshift information. The currently available
      samples \citep[e.g.][]{2018ApJ...857..104G} are highly
      incomplete, especially in the Southern Hemisphere \citep[see Fig.~7 in][]{2007ApJ...654..897B}.
\item Search for strong gravitational lenses to a faint surface
  brightness limit \citep[e.g.][]{1998A&A...330....1B,1998ApJ...498L.107T,2007ApJ...671L...9B}, which can be used to
  explore the dark matter profiles of galaxies \citep[e.g.,][]{2006ApJ...640..662T}.
\end{itemize}

\subsubsection{  Synergy with other projects }

LSST will not operate in isolation and will greatly benefit from other precursor and coeval
data at a variety of wavelengths, depths, and timescales. For example,
in the visual wavelength range, most of the Celestial
Sphere will be covered to a limit several magnitudes fainter than LSST saturation
($r\sim16$), first by the combination of SDSS, Pan-STARRS1 (PS1), the
Dark Energy Survey \citep{2016MNRAS.460.1270D} and SkyMapper
\citep{2007PASA...24....1K},
and then by the Gaia survey. The SkyMapper survey will obtain imaging data in the southern
sky to similar depths as SDSS, the PS1 surveys provides multi-epoch
data somewhat deeper
than SDSS in the northern sky, and the Dark Energy Survey
 is scanning
$\sim$5000 deg$^2$ a magnitude deeper still in the southern sky. Despite the lack of
the $u$ band data and its relatively shallow imaging, the Pan-STARRS surveys
represents a valuable complement to LSST in providing Northern sky coverage to a limit
fainter than that of SDSS and SkyMapper. LSST and Gaia will
be highly complementary datasets for studying the Milky Way in the multi-dimensional space of
three-dimensional positions, proper motions and metallicity \citep{2012ARA&A..50..251I}.
The Gaia survey will provide calibration checks at the bright end for proper
motions and trigonometric parallax measurements by LSST, and LSST will extend the
Gaia survey by four magnitudes. The upcoming Zwicky Transient Facility \cite[e.g.,][]{2017arXiv170801584L},
with its 600 Megapixel camera and a 47 deg$^2$ large field of view, will generate the largest
optical transient stream prior to LSST (at about one tenth of LSST rate) and thus provide
an early insight into astrophysical surprises and technical challenges awaiting LSST.
The LSST data stream will invigorate subsequent investigations by numerous other telescopes
that will provide additional temporal, spectral and spatial resolution coverage.

WFIRST and Euclid will carry out wide-field imaging surveys in the
near-infrared, giving highly complementary photometry to LSST.  The
resulting galaxy SEDs should give rise to even better photometric
redshifts, as well as tighter constraints on stellar masses and star
formation histories crucial for galaxy evolution studies.  The weak
lensing analyses from space and from the ground will also be highly
complementary, and will provide crucial cross-checks of one another.
LSST also presents the opportunity to conduct simultaneous observations
of WFIRST's Galactic Bulge survey fields, from which it will be possible to
measure the parallax and hence the lens masses for most microlensing
events, as well as providing valuable lightcurve coverage during the large
data gaps between WFIRST survey `seasons'.

LSST will also enable multi-wavelength studies of faint optical
sources using gamma-ray, X-ray, IR and radio data.  For example, the
SDSS detected only 1/3 of all 20cm FIRST sources \citep{1995ApJ...450..559B}
because it was too shallow by $\sim$4 mag for a complete optical
identification. Similarly, deep optical data are required for
identification of faint X-ray sources \citep{2005ARA&A..43..827B,2017AN....338..241B}.

LSST will provide a crucial complementary capability to space
experiments operating in other wavebands, such as the
NuSTAR Mission \citep{2013ApJ...770..103H},
eROSITA \citep{2012arXiv1209.3114M},
and the \textit{Fermi}
Gamma-ray Space Telescope \cite[e.g.,][]{2009ApJ...697.1071A}.
The Laser Interferometer Gravitational
Wave Observatory (LIGO) has now detected ultracompact binaries and black-hole mergers through the
gravitational wave outbursts that are emitted during the final stages of such events.
LSST will aid studies of  the electromagnetic signal that accompanies the gravitational wave emission,
thereby providing an accurate position on the sky for the system, which is
crucial for subsequent observations. LSST will also add new value to the archives for
billion-dollar class space missions such as Chandra, XMM-Newton,
Spitzer, Herschel, Euclid, and WFIRST,
because deep optical multi-color data will enable
massive photometric  studies of sources from these missions.
All areas of the sky -- whether by design or by serendipity -- in which past, present, or future
multiwavelength surveys overlap with LSST sky coverage, will be further promoted by LSST
investigations to become ``optical plus multiwavelength Selected Areas''.
Last but not least, the huge samples of various astronomical source
populations will yield extremely rare objects for investigations by powerful
facilities such as JWST \citep{2006SSRv..123..485G} and the next generation
of 20--40 meter telescopes.

In summary, the diversity of science enabled by LSST will be
unparalleled, extending from the physics of gravity and the
early Universe to the properties of ``killer'' asteroids. While
there are other projects that aim to address some of the same
science goals, no other project matches this diversity and
LSST's potential impact on society in general.

\section{COMMUNITY INVOLVEMENT}
\label{Sec:community}

LSST has been conceived as a public facility: the database that it will
produce, and the associated object catalogs that are generated from that
database, will be made available with no proprietary period to the
U.S.\ and Chilean scientific communities, as well as to those
international partners who contribute to operations funding.  As
described in \S~\ref{Sec:impact}, data will also be made available to
the general public for educational and outreach activities.
The LSST data management
system (\S~\ref{Sec:dp}) will provide user-friendly tools to access this database, support
user-initiated queries and data exploration, and carry out scientific analyses on the
data, using LSST computers either at the archive facility
or at the data access centers.
We expect that many, perhaps even the majority,
of LSST discoveries will come from research astronomers with no formal
affiliation to the project, from students, and from interested amateurs,
intrigued by the accessibility to the Universe that this facility uniquely
provides.

The SDSS provides a good example for how the scientific
community can be effective in working with large, publicly available
astronomical data sets. The SDSS has published a series of large incremental
data releases via a sophisticated database, roughly once per year, together with
a paper describing the content of each data release, and extensive on-line
documentation giving instructions on downloading the catalogs and image data
(see \url{http://www.sdss.org}). The overwhelming majority of the almost
8000 refereed papers based
on SDSS data to date have been written by scientists from outside
the project, and  include many of the most high-profile results that have come
from the survey.

Nevertheless, it is clear that many of the highest priority LSST science
investigations will require organized teams of professionals working together
to optimize science analyses and to assess the importance of systematic
uncertainties on the derived results. To meet this need, a number of science
collaborations have been established in core science
areas. For example, the LSST Dark Energy Science Collaboration includes
members with interests in the study of dark energy and related topics in
fundamental physics with LSST data. As of the time of this contribution, there are
over 800 participants in these collaborations.
The science collaborations are listed on the LSST web page, together
with a description of the application process for each one.
All those at US and Chilean institutions,
as well as named individuals from institutions in other countries
which have signed Memoranda of Agreement to contribute to LSST
operations costs are eligible to apply. As described in
\S\S~\ref{sec:dm} and \ref{Sec:dp}, LSST will make available
substantial computational resources to the
science community to carry out their analyses;
the system has been sized accordingly.

As we design our observing strategies, we are actively seeking and implementing
input by the LSST science community.  The LSST science collaborations
in particular have helped develop the LSST science case and continue
to provide advice on how to optimize their science with choices in
cadence, software,
and data systems. A recent example is the publication of a document
entitled ``Science-Driven Optimization of the LSST Observing
Strategy'' \cite{2017arXiv170804058L}, a living document that
quantifies the science returns in different areas for different
observing cadence.  The cadence will continue to be refined, with
input from the science collaborations, during the commissioning, and
the observing strategy will be regularly reviewed, with flexibility
built in, during operations.

The Science Advisory Committee (SAC), chaired by Michael Strauss,
provides a formal, and two-way, connection to the external science
community served by LSST. This committee takes responsibility for
policy questions facing the project and also deals with technical
topics of interest to both the science community and the LSST
Project. The SAC minutes and notes are available
publicly. Current members on this committee are: T. Anguita (Andr\'es
Bello, Chile), R. Bean (Cornell), W.N. Brandt
(Penn State), J. Kalirai (STScI), M. Kasliwal
(Caltech), D. Kirkby (UC Irvine), C. Liu (Staten Island), A. Mainzer
(JPL), R. Malhotra (U Arizona),
N. Padilla (U. Cat\'olica de Chile), J. Simon (Carnegie), A. Slosar
(Brookhaven), M. Strauss (Princeton), L. Walkowicz (Adler),
and R. Wechsler (Stanford).

\section{EDUCATIONAL AND SOCIETAL IMPACTS}
\label{Sec:impact}

The impact and enduring societal significance of LSST will exceed its direct contributions to advances in physics and astronomy.  LSST is uniquely positioned to have high impact with the interested public, planetariums and science centers, and citizen science projects, as well as middle school through university educational programs. LSST will contribute to the national goals of enhancing science literacy and increasing the global competitiveness of the US science and technology workforce. Engaging the public in LSST activities has been part of the project design from the beginning.

The mission of LSST's Education and Public Outreach (EPO) program is to provide worldwide access to a subset of LSST data through accessible and engaging online experiences so anyone can explore the universe and be part of the discovery process. To do this, LSST EPO will facilitate a pathway from entry-level exploration of astronomical imagery and information to more sophisticated interaction with LSST data using tools similar to what professional astronomers use for their work.

A dynamic, immersive web portal will enable members of the public to explore color images of the full LSST sky, examine objects in more detail, view events from the nightly alert stream, learn more about LSST science topics and discoveries, and investigate scientific questions that excite them using real LSST data in online science notebooks. The portal will also link to numerous citizen science projects using LSST data.

LSST data can become a key part of classrooms emphasizing student-centered research in middle school, high school, and undergraduate settings. Using online science notebooks, teachers will be able to bring real LSST telescope data into their classrooms without having to download, install, and maintain software locally. Educational investigations will be designed to support key aspects of the Next-Generation Science Standards (NGSS) in the USA, and goals of the Explora program through CONICYT in Chile. Educators will be supported through professional development that offers training on the online notebook technology and also relevant science content. Science notebooks will also accommodate access to LSST data for lifelong learners and anyone that visits the portal.

Anyone around the world will be able to participate in a variety of citizen science projects that use LSST data.  The EPO Team will work with the Zooniverse to develop the \textit{Project Builder} to include tools specifically designed to utilize LSST data, allowing LSST principal investigators to create any number of projects to help them accomplish their science goals. EPO anticipates that the number of citizen science projects in the astronomy field will increase dramatically when LSST is operational, giving a whole new generation of citizen scientists the opportunity to deepen their engagement with astronomy using authentic data from LSST.

LSST EPO will produce and maintain a digital library of multimedia
assets including images, video clips, and 3D models.  Multimedia
assets will be aligned to standards such as IMERSA Dome Master and
Astronomy Visualization Metadata, when applicable, allowing full
flexibility for adoption by content creators at planetariums and
science centers. We will also follow the International Planetarium
Society's Data2Dome standard, to maximize the number of platforms that
can use our assets.

The LSST EPO program will rely on a cloud-based EPO Data Center (EDC) to handle the unique needs of the EPO audiences. These needs include, for example, a fast and smooth browsing experience on mobile devices, and the need to handle inevitable spikes and lulls in visitor traffic and data transfers. As such, the EDC will follow best practices popularized by cloud computing, leveraging on-demand computing and auto-scalable architecture. Remaining agile and relevant during the full lifetime of Operations by adjusting to technology trends and education priorities is an important part of the LSST EPO design process.

LSST EPO is committed to engaging with diverse audiences and is undertaking a multi-faceted approach to reaching diverse individuals.  LSST EPO is planning to partner with at least five organizations serving 1) women/girls, 2) individuals from traditionally underrepresented groups in STEM, and 3) individuals in low socioeconomic communities. Representatives from these organizations will be key stakeholders of the EPO program, helping to shape deliverables and a culturally responsive program evaluation. Furthermore, these relationships will allow for co-creation of EPO deliverables to help ensure materials are accessible to, of interest to, and relevant to diverse populations.

LSST EPO is breaking new ground in bringing astronomical data to the public in a timely, engaging, and easy-to-use way. It is not unreasonable to anticipate tens of millions of public users browsing the LSST sky, exploring discoveries as they are broadcast, and monitoring objects of interest. Results of EPO's ongoing evaluation will be made publicly available, allowing us to share lessons learned, insights, and program impacts with the larger science EPO community.

\section{SUMMARY AND CONCLUSIONS}
\label{Sec:conclusions}

Until recently, most astronomical investigations have focused on small
samples of cosmic sources or individual objects. Over the past few decades,
however, advances in technology have made it possible to move beyond the
traditional observational paradigm and to undertake large-scale sky
surveys, such as SDSS, 2MASS, GALEX, Gaia, and others. This observational
progress, based on a synergy of advances in telescope construction, detectors,
and above all, information technology, has had a dramatic impact on nearly all
fields of astronomy, many areas of fundamental physics, and society in
general. The LSST builds on the experience of these surveys and addresses
the broad goals stated in several nationally endorsed reports by the U.S.\
National Academy of Sciences. The 2010 report ``New Worlds, New Horizons
in Astronomy and Astrophysics'' by the Committee for a Decadal Survey of Astronomy and
Astrophysics
ranked LSST as its top priority for large ground-based programs.
The LSST will be unique: the combination of large aperture and large field of view,
coupled with the needed computation power and database technology, will
enable simultaneously fast and wide and deep imaging of the sky, addressing in
one sky survey the broad scientific community's needs in both the
time domain and deep universe.

The realization of the LSST involves extraordinary engineering and
technological challenges: the fabrication of large, high-precision optics;
construction of a huge, highly-integrated array of sensitive, wide-band
imaging sensors; and the operation of a data management facility
handling tens of terabytes of data each day. The design, development
and construction
effort has been underway since 2006 and will continue through the
onset of full survey operations.  This work involves hundreds of
personnel at institutions all over the US, Chile, and the rest of the
world.

In December 2013, LSST passed the NSF Final Design Review for construction,
and in May 2014 the National Science Board approved the project.
The primary/tertiary mirror was cast in 2008, and the polished mirror
was completed in 2015.
In 2014 LSST transitioned from the design and development phase to
construction, and the Associated Universities for Research in
Astronomy (AURA) has formal responsibility for the LSST project since 2011.  
At this writing,  the project is near the peak of the construction
effort, and is preparing for the transition to
commissioning and operations.

The construction cost of LSST is being borne by the US National Science
Foundation, the Department of Energy, generous contributions from several
private foundations and institutions, and the member institutions of the
LSST Corporation. The LSST budget includes a significant Education and
Public Outreach program (\S~\ref{Sec:impact}).
The U.S.\ Department of Energy is supporting the cost of constructing the
camera. LSST has high visibility in the high-energy physics community,
both at universities and government laboratories. The telescope will
see first light with a commissioning camera in late 2019, and the
project is scheduled to begin regular survey operations by 2022.

The LSST survey will open a movie-like window on objects that
change brightness, or move, on timescales ranging from 10 seconds to 10 years.
The survey will have a raw data rate of about 15 TB per night (about the same as one
complete Sloan Digital Sky Survey per night), and will collect about 60 PB
of data over its lifetime, resulting in an incredibly rich and extensive
public archive that will be a treasure trove for breakthroughs in many areas
of astronomy and physics. About 20 billion galaxies and a similar number of stars
will be detected -- for the first time in history, the number of cataloged
celestial objects will exceed the number of living people! About a thousand
observations of each position across half of the Celestial Sphere will
represent the greatest movie of all time.

Alerts of transient, variable, and moving objects will be issued worldwide within
60 seconds of detection.
An extensive public outreach program will provide a new view of the sky to
curious minds of all ages worldwide.
We are working with prospective foreign partners to make all of the LSST science data
more broadly available worldwide.  As of 2017, 34 institutions from 23 countries
have signed Memoranda of Agreement to contribute significantly to
the LSST operating costs, in exchange for participation in the science collaborations
and data access.  The software which processes the pixels
and creates the LSST database is open source.
LSST will be a significant milestone in the globalization of the information revolution.
The vast relational database of 32 trillion observations of 40 billion objects
will be mined for the unexpected and used for precision experiments in astrophysics.
LSST will be in some sense an internet telescope:
the ultimate network peripheral device to explore the Universe, and
a shared resource for all humanity.

\acknowledgments
This material is based upon work supported in part by the National Science Foundation 
through Cooperative Agreement 1258333 managed by the Association of Universities for 
Research in Astronomy (AURA), and the Department of Energy under Contract No. 
DE-AC02-76SF00515 with the SLAC National Accelerator Laboratory. Additional LSST 
funding comes from private donations, grants to universities, and in-kind support from 
LSSTC Institutional Members.

\facility{LSST}

\appendix

\section{Version History}

\vskip 0.06in
Version 1.0 (May 15, 2008): the first posting.
\vskip 0.06in
Version 2.0 (June 7, 2011): acknowledged the Decadal Survey 2010 report; updated construction schedule;
updated expected performance in Table 2; added sections on LSST simulations and data mining;
updated several figures; updated references; expanded author list.
\vskip 0.06in
Version 3.0 (August 26, 2014): acknowledged the start of federal construction;
updated system description and science examples, updated several figures; refreshed references;
expanded author list.
\vskip 0.06in
Version 4.0 (May 15, 2018): updated system description and science examples, updated expected
performance in Table 2; updated several figures; refreshed references; expanded author list.

\end{document}